# Volatile transport on inhomogeneous surfaces:
# II. Numerical calculations (VT3D)

Leslie A. Young

Southwest Research Institute, Boulder, CO, 80302

`layoung@boulder.swri.edu`

**Abstract**

Several distant icy worlds have atmospheres that are in vapor-pressure equilibrium with their surface volatiles, including Pluto, Triton, and, probably, several large KBOs near perihelion. Studies of the volatile and thermal evolution of these have been limited by computational speed, especially for models that treat surfaces that vary with both latitude and longitude. In order to expedite such work, I present a new numerical model for the seasonal behavior of Pluto and Triton which (i) uses initial conditions that improve convergence, (ii) uses an expedient method for handling the transition between global and non-global atmospheres, (iii) includes local conservation of energy and global conservation of mass to partition energy between heating, conduction, and sublimation or condensation, (iv) uses time-stepping algorithms that ensure stability while allowing larger timesteps, and (v) can include longitudinal variability. This model, called VT3D, has been used in Young (2012a,b), Young (2013), Olkin et al. (2015), Young and McKinnon (2013), and French et al. (2015). Many elements of VT3D can be used independently. For example, VT3D can also used to speed up thermophysical models (Spencer et al. 1989) for bodies without volatiles. Code implementation is included in the supplemental materials and is available from the author.

*Keywords:* Pluto, atmosphere; Pluto, surface; Atmosphere, evolution; Trans-neptunian objects

# 1. Introduction

Pluto and Triton have atmospheres whose pressures have been measured by stellar occultations (e.g., Young et al. 2008a, Olkin et al. 1997) and spacecraft (Gurrola 1995, Krasnopolsky et al. 1993, Stern et al. 2015, Gladstone et al. 2016). These measurements reveal atmospheres for Pluto and Triton that are global in extent, almost certainly controlled by vapor-pressure equilibrium of the surface $N_2$ ice, (Spencer et al. 1997, Yelle et al. 1995), similar to the role of $CO_2$ on Mars (Leighton and Murray 1996).

Vapor pressure is an exceedingly sensitive function of temperature, and early models predicted that the surface pressures of Pluto and Triton would vary by orders of magnitude over their years (e.g., Hansen and Paige 1992, 1996; Moore and Spencer 1990; Spencer and Moore 1992). Those early models were based on a single observation of the atmospheric pressure, either the Triton flyby in 1989 or the definitive discovery Pluto occultation in 1988 (e.g., Elliot and Young 1992). Since that time, further occultations have shown a large increase in the atmospheric pressures of both Pluto and Triton since the late 1980's (Elliot et al. 1998; Elliot et al. 2003). Other advances in the past decade include an improved understanding of the surface compositions of Pluto and Triton (Grundy and Buie 2001, Grundy et al. 2010). It is time for new models (Young 2012a, Young 2013, Olkin et al. 2015, Hansen et al. 2015). This work describes the model used by Young (2012a, 2013) and Olkin et al. (2015).

Since the rash of models in the 1990's, the large, volatile-covered ice worlds Pluto and Triton have been joined by other large, volatile-covered bodies in the outer solar system, including the large Kuiper Belt Objects (KBOs) Eris, Sedna, Makemake, Haumea, Quaoar (Schaller and Brown 2007), and 2007 OR10 (Brown et al. 2011). Some of these should have atmospheres at some time in their orbit (Stern & Trafton 2008). In particular, the 98% albedo of Eris argues for a perihelion atmosphere that collapses near aphelion, freshening Eris's surface (Sicardy et al. 2011).

I present a new model for volatile transport on Pluto, Triton, and other volatile-covered bodies in the outer solar system. As with previous models (Hansen and Paige 1992, 1996; Moore and Spencer 1990; Spencer and Moore 1992), this model includes transport of volatile mass and latent heat, the thermal inertia of a volatile slab, internal heat flux, and thermal conduction to and from the substrate. The major improvements over previous models are improved numerical stability, strict mass conservation including atmospheric escape, a new method for handling the transition between a global and non-global atmosphere, and longitudinal variability. The model is designed to be flexible in order to easily accommodate details such as a change in $N_2$ emissivity with crystalline phase (Stansberry and Yelle 1999).

Although motivated by volatile-covered bodies, the speed improvements make this of interest for other computationally difficult problems of thermal evolution, such as Mercury or Mimas. The intent is that this model will find wide utility in the community. Examples of possible applications include (i) comparison of modeled pressures with occultation results, (ii) comparison of modeled thermal emission with Spitzer observations, and (iii) the exploration of volatile transport on other large, volatile-covered KBOs. Therefore, this paper takes an approach similar to that used in *Numerical Recipes* (Press et al. 2007), which has found wide adoption within the planetary science community. In particular, it presents a description of the numerics in enough detail for the reader to implement all equations. Many of the equations and figures are implemented in the layoung IDL library, found at http://www.boulder.swri.edu/~layoung/. These are indicated in the text, listed in Appendix B, and included in the supplementary materials.

We give an overview of the VT3D model in Section 2, and show its application to areas bare of volatiles (Section 3), volatile-covered areas with local atmospheres (Section 4), and volatile-covered areas with efficient transport of mass and latent heat (Section 5). In each of Section 3-5, we present the continuous equations; recap the analytic results of Young 2012a (hereafter Paper I) for use both in initial conditions and the numerical solution; and present the numerical implementation, in which these equations are linearized, discretized, cast into matrix form, and solved. Compared with previous models of volatile transport (and, to some extent, thermophysical models of airless bodies), the model developed here and in Paper I introduces several new concepts. These include the following:

1. Using analytic approximations for the temperature variation for initial conditions.

2. Approximating a surface in transition between a global, Pluto-like surface and a local, Io-like surface as a splice between areas of local and non-local transport of mass and latent heat (See Fig 2-2).

3. Closing the time-dependent energy equation for volatile-covered areas by requiring mass balance.

4. Linearizing the equations for volatile transport, and casting them in matrix form, simplifying the program structure needed for modeling Pluto, Triton, Eris, and other KBO.

5. Implementing numerically stable methods that allow larger time steps, and take advantage of the fast matrix-based calculations of modern computer languages.

## 2. VT3D (Volatile Transport, 3-Dimensions) overview

This paper and Young 2012a (Paper I) are intended to be the first two in a series of models of increasing complexity. Paper I derived an analytic expression for surface and subsurface temperatures assuming (i) albedos and compositions that vary with latitude and longitude, (ii) a static distribution of volatile and bare surface elements, (iii) a single volatile species, (iv) volatile temperatures that are constant within a volatile slab and across the surface, and (v) substrate properties (density, specific heat, thermal conductivity) that could vary with location but not depth. The analytic results presented in Paper I can be used for physical intuition, diurnal variation, seasonal variation under certain circumstances (such as completely volatile-covered bodies), quick estimations of temperatures, testing of numerical code, and initialization of temperatures for numerical simulations. These are elaborated in Sections 3.2, 4.2, and 5.2.

This paper, Paper II, focuses on the numerical calculations for volatile transport and surface and subsurface temperatures. As with Paper I, this model assumes a single volatile species and volatile temperatures that are constant within a volatile slab. This paper extends Paper I in that: (i) substrate properties are allowed to be variable with depth, (ii) the latent heat of solid-phase transitions are treated (e.g., between $\alpha$ and $\beta$ $N_2$), (iii) volatile-covered areas can sublime to become bare, and cold bare areas can become volatile-covered, and (iv) there is a smooth transition between the case where the surface pressure is globally isobaric (similar to Pluto's current atmosphere), and one where surface temperatures and pressures can vary with location. Multiple species and layers within the volatile slab are planned for Paper III. A more accurate model of the variation involved in the transition case is planned for Paper IV.

As in Paper I, the conceptual framework is built on the physical processes considered by Spencer and Moore (1992), Moore and Spencer (1996), Hansen and Paige (1992) and Hansen and Paige (1996, HP96), as illustrated in Fig 2-1. These include thermal conduction into and within a substrate, a heat flux at the lower boundary, absorbed sunlight, and thermal emission. I assume absorption of solar energy and thermal emission occur at the uppermost surface. Grundy & Stansberry (2000) discuss the "solar gardening" that can occur from a separation between the shallow depth of the source of thermal emission and the deeper penetration of solar heating. This distinction is deferred for a later formulation that treats layering within the volatile slab.

*Insert fig 2-1 here.*

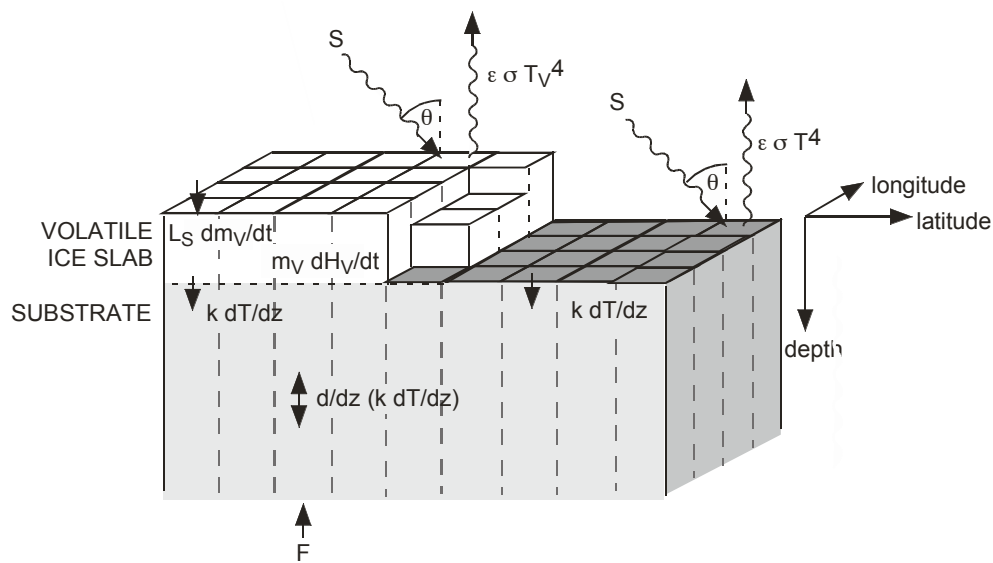

Fig. 2-1. Schematic of the heat balance equation solved by the analytic model (based on Hansen and Paige 1996). Locally, we balance absorbed insolation, $S$, emitted thermal energy $\varepsilon \sigma T^4$, and latent heat of sublimation or condensation, $L_S \, dm_V/dt$, where $m_V$ is the mass per area of the volatile slab and $L_S$ is the latent heat of sublimation. Additionally we balance (i) heat to and from the substrate, $k \, dT/dz$, where $k$ is the thermal conductivity and $dT/dz$ is the vertical gradient of temperature, and (ii) the heat capacity of the isothermal ice layer, $m_V \, dH_V/dt \approx m_V \, c_V \, dT_V/dt$, where $H_V$ is the enthalpy and $c_V$ is the specific heat of the volatile slab (subscript $V$ for volatile). At the lower boundary, there is a heat flow of $F$. All variables except $T_V$ are free to vary with latitude and longitude. Compared with Young (2012a; Paper I), this figure illustrates (i) heating within the substrate for vertically varying $k$, and (ii) enthalpy of the ice slab, $H_V$, to allow the treatment of solid-phase transitions.

The latent heat of sublimation term of the energy equation depends on the mass flux ($dm_V/dt$ in Fig. 2-1). For extremely thin atmospheres, such as on Io or possibly currently on Eris, some atmospheric flow occurs, but is ineffective in changing local surface temperatures (Fig 2-2A). In this case, the volatile slab temperature is controlled by local conditions only. The volatile slab temperature and the local atmospheric pressure are generally higher in areas of high insolation. For thin atmospheres, we assume no atmosphere over the bare areas. This approach allows efficient calculation of surface and subsurface temperatures. Once temperatures are calculated, one can calculate Iapetus-style cold trapping or Io style flow after-the-fact; this is not treated in this paper.

In thicker atmospheres, such as on the current Pluto and Triton, the atmosphere efficiently transports mass and latent heat across the entire globe (Fig 2-2B). As quantified by Trafton and Stern (1983), the pole-to-pole pressure differences are small as long as the sublimation wind is much less than the sound speed. In this case, the mass flux is calculated by ensuring global mass balance, including the mass of the atmosphere over both volatile-covered and bare areas (Trafton & Stern 1983; Young 1992; Hansen and Paige 1992, 1996; Young 2012a).

Accurately modeling the transition between a global and local atmosphere is too complex and computationally expensive to treat here. In future papers, we plan to treat this using vertically integrated hydrodynamic equations, as has been done for Io (Ingersoll et al. 1985, Ingersoll 1989). In this paper, I treat the atmosphere as a *splice* between isolated locations with local atmospheres and ineffective transport of mass and latent heat (Fig 2-2C, Areas I and II), and interacting locations that share a single surface pressure, with effective transport of mass and latent heat (Fig 2-2C, Areas III and IV).

*Insert fig 2-2 here.*

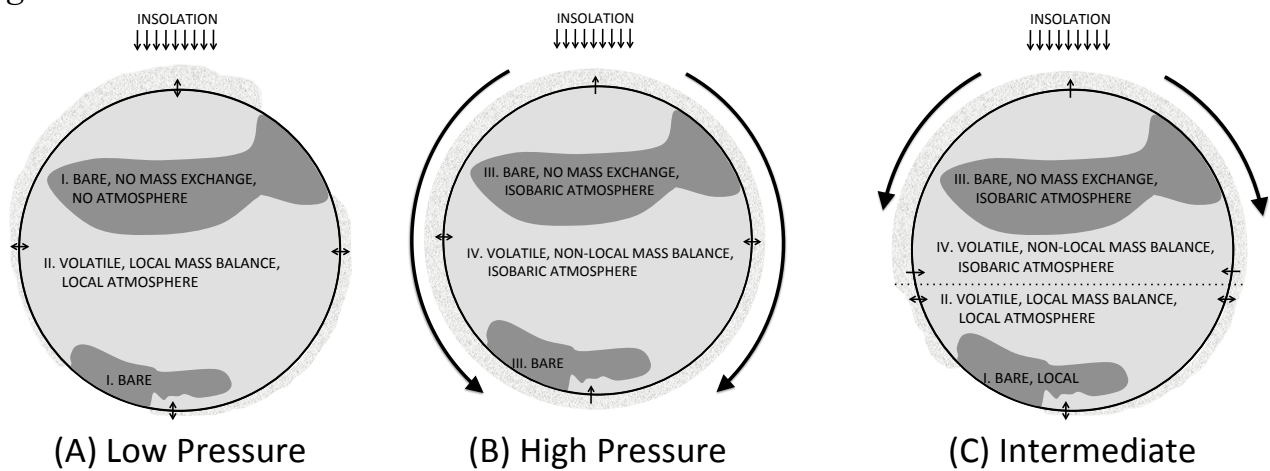

(A) Low Pressure  (B) High Pressure  (C) Intermediate

Fig. 2-2. Schematic of the transport of mass and latent heat over a surface in this model. Dark and light surface areas represent areas devoid of or covered with volatiles, respectively. The mottled outline around the body represents an atmosphere of varying surface pressure. Short arrows represent mass exchange between the surface and atmosphere, and large curved arrows represent net flow within the atmosphere. (A) For low-pressure atmospheres (such as current Eris), the mass and latent heat balance are local, the surface

pressure depends on local heating, and the surface is covered with Areas I (bare) and II (volatile-covered). At each location, the surface and atmosphere exchange volatiles, indicated by the double-headed arrows, but neighboring locations are isolated from each other. (B) For high-pressure atmospheres, such as the current Pluto or Triton, all the volatiles interact with each other. The mass and energy balance is global, the surface pressure is constant over the globe, and the surface is covered with Areas III (bare) and IV (volatile-covered). There is net sublimation at the summer pole (indicated by an arrow pointing from the surface at the top of the figure), net deposition at the winter pole (indicated by an arrow pointing from the atmosphere at the bottom of the figure), and balanced exchange of volatiles at mid latitudes (indicated by double-headed arrows). (C) Intermediate cases are treated with the computationally efficient method of splicing local and isobaric areas.

**Table 1. Area Types**

|  | **Bare** | **Volatile-covered** |
|---|---|---|
| **Isolated** | **Area I** <br> Energy budget does not include latent heat. Atmospheric pressure is negligible. <br> Section 3. | **Area II** <br> Energy budget includes latent heat, without mass or energy transfer from other locations. Atmosphere is in vapor-pressure equilibrium with the surface, with surface pressures that vary with location. <br> Section 4. |
| **Interacting** | **Area III** <br> Energy budget does not include latent heat. Atmospheric pressure is constant over all locations in Area III and IV. <br> Section 3. | **Area IV** <br> Energy budget includes latent heat, with mass or energy transfer from other locations. Atmosphere is in vapor-pressure equilibrium with the surface, with a surface pressures is constant over all locations in Area III and IV. <br> Section 5. |

In the following sections, I develop the VT3D model for both isolated and interacting bare areas (Areas I and III), the local volatile-covered areas (Area II), and the interacting, isobaric volatile-covered areas (Area IV). In each, I present the equations for energy balance, and describe the analytic approximation that can be used as effective initial conditions. The problem is discretized in location, depth, and time, and the energy balance equations are linearized. This leads to two forms of matrix equations. The explicit equations involve multiplication of the current temperature array by a tridiagonal or banded tridiagonal matrix, while the implicit equations (in this case, a Crank-Nicholson timestep scheme; Press et al. 2007) involve multiplication of both the current and next temperature array by tridiagonal or banded tridiagonal matrices. The elements of the matrices are derived, and notes on how to solve the resulting equations are given.

The primary advantage of VT3D is its speed (Table 2). Many of the speed gains are applicable to bare locations, and can speed up calculations involving spatially resolved thermal measurements, measurements involving seasonal, diurnal, and eclipse time scales, and other applications. Therefore, I present the formulation for bare locations first, in Section 3, and only then turn to volatile areas.

**Table 2. Elements of VT3D relating to computational speed**

| Feature | Speed factor | Notes | See Sections |
|---|---|---|---|
| Initialize with analytic approximation | ~4 | Allows spin-up in only a few periods, rather than 10-20. | 3.2 (Areas I, III) 4.2, 5.2 (Areas II, IV) |
| Uneven layer thickness | ~2-1000 | Allowing layers to increase their thickness with depth decreases the number of layers needed. | 3.3 (All Areas) |
| Implicit time steps (Crank-Nicholson) | ~50 | Stable for large timesteps. Requires solution of tridiagonal or banded tridiagonal matrices | 3.3 (Areas I, III) 4.3, 5.3 (Areas II, IV) |
| Lower-Upper (LU) decomposition of substrate sub-matrix | ~2 | Speeds up the solution of tridiagonal or banded tridiagonal matrices | 3.4 (All Areas) |
| Decouple seasonal and diurnal forcing | ~15,000 | Decreases the number of time steps by the ratio of orbital to rotational period | 3.3 (All Areas) |
| Calculate temperatures on multiple locations with a single operation | Depends on the language | Avoids costly "for loops" | 3.4 (Area I, II, III) 5.4 (Area IV) |
| Global tiling | 1.6 | Can decrease the number of required tiles | 3.4 (All Areas) |
| Combine mass and energy equations | 10 | Avoids having to calculate a range of mass fluxes and then test for conservation of mass | n/a (Area I, III) 4.2 (Area II) 5.2 (Area IV) |

I end this section with a few words about the modularity of the techniques described here.

• Anyone interested in this model should read Section 2 (this section) because it is short, and provides an overview.

• If your object has no volatiles, you do not need to read past Section 3.

• If you want to characterize which processes are important in controlling surface temperatures, you can stop at the calculation of the thermal parameters, or Eq. 3.2-11 for volatile-free bodies, plus Eq. 4.2-7 and 4.2-8 for isolated volatile-covered areas, or Eq. 5.2-2a-d for volatile-covered interacting areas.

• If you want to very quickly approximate a temperature field based on the solar forcing, read Section 3.1-3.2, plus Section 4.1-4.2 if you have isolated volatiles, and Section 5.1-5.2 if you have interacting volatile covered areas. The critical equations are Eq. 3.2-12 or 3.2-17, 4.2-9 or 4.2-10, and 5.2-3 or 5.2-4.

• If you are calculating temperatures at one volatile-free location at a time, you can stop at Section 3.3. If you are calculating one isolated volatile-covered location at a time, read through Section 3.3, then skip ahead to Sections 4.1-4.3.

• If you are calculating roughly several hundred timesteps per period (e.g., to gain insights at short timescales or to make smooth plots), then the explicit equations will be stable, and the implicit equations will not save much computation time. In that case, you can skip those equations in Section 3.3, 4.3, and 5.3 that are described as implicit (roughly half of them), and all of sections 3.4b and 5.4b.

## 3. VT3D for bare locations (Areas I and III)

Fig 2-2 and Table 1 recap the definitions of Areas I and III and their interaction with the atmosphere. For Area I (bare, no mass exchange, no atmosphere), the physics in VT3D is identical to the well-known thermophysical model (TPM) used to interpret thermal emission from airless bodies (e.g., Thomas et al. 2000; Spencer et al. 1989; Harris 1988). Heating in the top-most layer is balanced by thermal emission, insolation, and conduction; heating in interior layers is balanced by conduction only; heating in the lower layer is balanced by conduction and a flux condition at the lower boundary.

Area III (bare, no mass exchange, isobaric atmosphere) represents, for example, the "bedrock" $H_2O$ on current-day Triton. There is no volatile slab and no sublimation. The difference between the two bare area types are (i) Area III is a potential deposition site, and (ii) an increase in the volatile temperature for Area IV (volatile-covered, isobaric) also increases the pressure over Area III, so the atmosphere above Area III needs to be included in mass balance equation for Area IV. As long as there is no condensation (which will alter the state from bare to volatile-covered), the energy balance for Area III is the same as for Area I. Therefore, both bare areas, I and III, are treated in this section.

These equations demonstrate several aspects of the numerical power of VT3D. In Section 3.1 and 3.2, I show the analytic expressions for the initial conditions, and show that a simple calculation can approximate the numerical solution. In Section 3.3, I present the explicit and implicit (Crank-Nicholson) numerical solutions for a single bare location, showing that solutions spin up in less than a quarter period. In Section 3.4, I show a compact representation of the linearized, discretized equations. In Section 3.5, I present a worked example of Mimas's diurnal temperatures, with code and output in the supplementary materials.

*3.1 Areas I and III : Continuous expressions for bare areas*

At the lower boundary, there may be positive (or negative) heat flow, $F$, which is balanced by upward (or downward) thermal conduction from a negative (or positive) thermal gradient:

$$k \frac{\partial T}{\partial z}\bigg|_{z \to z_{min}} = -F \qquad (3.1\text{-}1)$$

where $k$ is the thermal conductivity, and $T$ is the temperature. As with Paper I, $z$ is a height coordinate, defined to be zero at the top of the substrate and decreasing downward. Thus, $z = 0$ at the substrate-volatile interface for locations where there is a volatile slab, or at the surface on volatile-free areas.

Within the substrate, I assume there are no heating sources, so net conductive heat flux is balanced by changes in the temperature of the substrate:

$$\underbrace{\rho c \frac{\partial T}{\partial t}}_{\text{Enthalphy of substrate}} = \underbrace{\frac{\partial}{\partial z} k \frac{\partial T}{\partial z}}_{\text{Conduction}} \qquad (3.1\text{-}2)$$

where $\rho$ is the density, $c$ is the specific heat at constant pressure for the substrate, $t$ is time, and $T$ is the temperature.

The energy balance at the surface balances net heating with absorbed sunlight, thermal emission, and thermal conduction. There is no latent heat of sublimation or condensation. The total equation is

$$0 = \underbrace{S}_{\text{Insolation}} - \underbrace{\varepsilon \sigma T^4}_{\text{Emission}} - \underbrace{k \frac{\partial T}{\partial z}\bigg|_{z=0}}_{\text{Conduction}} \qquad (3.1\text{-}3)$$

where $S$ is the absorbed solar energy, and $\varepsilon$ is the emissivity, and $\sigma$ is the Stefan-Boltzmann constant.

The first term of Eq. (3.1-3) describes the solar energy absorbed by the volatile slab, in power per area. For Triton, Pluto, Eris and other large KBOs, the fraction of sunlight absorbed by the atmosphere is small, and we do not need to alter $S$ to account for atmospheric absorption. The absorbed solar energy at a particular location and time of day depends on the solar flux at 1 AU, $S_{1AU}$, the heliocentric distance, $r$, the hemispheric albedo, $A_h$, and the cosine of the solar incidence angle, $\mu_0$ (where $\mu_0$ is 0 when the sun is below the horizon).

$$S = \frac{S_{1AU}}{r^2}(1 - A_h)\mu_0 = S_{SS}\mu_0 \qquad (3.1\text{-}4)$$

where $S_{SS}$ is the absorbed insolation at the sub-solar point. $\mu_0$ depends on latitude, $\lambda$, sub-solar latitude, $\lambda_0$, and the hour angle, $h$ (where $h$ is the difference between the location's longitude and the subsolar longitude, defined to increase with time at any given location).

$$\mu_0 = \max(0, \sin\lambda \sin\lambda_0 + \cos\lambda \cos\lambda_0 \cos h) \qquad (3.1\text{-}5)^1$$

The hemispheric albedo, $A_h$, is a local quantity, also known as the directional-hemispherical reflectance, hemispherical reflectance, or plane albedo (Hapke 1993). It is defined as the ratio of the total scattered power to the incident *collimated* power, $(S_{1AU}/r^2)\mu_0$, and depends on the location on the surface and the incidence angle. It is useful to approximate the hemispheric albedo by its average over all incidence angles, or $A_S = 2\int A_h \mu_0 d\mu_0$, where $A_S$ is known as the spherical reflectance, spherical albedo, or the Bond albedo (note, however, that Bond albedo is strictly defined for an entire surface). For typical phase functions in the outer solar system, substituting $A_S$ for $A_h$ tends to slightly underestimate solar heating for direct illumination and overestimate solar heating for large incidence angles. Since there is typically large uncertainty in the values of $A_S$ or $A_h$ due to uncertain phase functions, this distinction is usually ignored. The remainder of the paper uses $A$ for $A_h$, and does not distinguish between $A_h$ and $A_S$.

The second term of Eq. (3.1-3) represents thermal energy emitted by the substrate. For a physical surface, this term might include such effects as self-heating from crater sides (Spencer 1990; Rozitis and Green 2011). In VT3D the emissivity, $\varepsilon$, is treated as a parameter that defines the power per area lost by thermal emission. Since ε can vary with location and time, it can be used to encompass these more subtle physical effects.

---

[1] `mu0 = `**`vt3d_solar_mu`**`(lat, lon, lat0, lon0)`

The final term of Eq. (3.1-3) represents thermal conduction from the substrate. If the substrate just below the interface is warmer than the surface temperature ($dT/dz < 0$), then conduction expressed by this term warms the surface.

*3.2 Areas I and III : Analytic approximation and initialization bare areas*

This section expands on key results of Paper I. The purpose is to introduce variables that will be used later, and to show the equations that will be used to initialize numerical calculations. For more discussion of the derivation, see Paper I.

If the solar insolation, $S$, at latitude $\lambda$ and longitude $\phi$, is a known function of time, $t$, with period $P$, then it can be approximated as a sum of $M+1$ sinusoidal terms

$$S(\lambda,\phi,t) = \mathrm{Re}\left[\sum_{m=0}^{M} \hat{S}_m(\lambda,\phi)e^{im\omega t}\right] \qquad (3.2\text{-}1)^1$$

where $\omega = 2\pi/P$ is the frequency of the diurnal or seasonal forcing, and $\hat{S}_m(\lambda,\phi)$ are the complex sinusoidal coefficients, with the hat indicating complex quantities (note, however, that $S_0$ is real). The coefficients are derived from the insolation in an expression closely related to the Fourier transform:

$$S_0(\lambda,\phi) = \frac{1}{P}\int_0^P S(\lambda,\phi,t)\,dt \qquad (3.2\text{-}2a)^2$$

$$\hat{S}_m(\lambda,\phi) = \frac{2}{P}\int_0^P S(\lambda,\phi,t)e^{-im\omega t}\,dt \qquad (3.2\text{-}2b)^2$$

A common application is diurnal forcing. For areas in permanent darkness, the solution is trivially $\hat{S}_m(\lambda,\phi) = 0$. For others, the diurnally averaged insolation can be expressed analytically (e.g., Levine et al. 1977). One first finds the maximum hour angle of illumination, $h_{max}$, ($h_{max} = \pi$ for areas of constant illumination)

$$\cos h_{max} = \max\left(-1, \min\left(-\tan\lambda\tan\lambda_0, 1\right)\right) \qquad (3.2\text{-}3)$$

---

[1] `sol = `**`vt3d_solwave`**`(sol_terms, phase)`

[2] `f_t = `**`sft`**`(p, f, nfreq)` (Slow Fourier Transform, in `~layoung/math`)

where $\lambda$ is the latitude and $\lambda_0$ is the sub-solar latitude, as before. The average insolation is a real quantity, so written without the hat, and is given by

$$S_0 = \frac{\sin\lambda \sin\lambda_0 h_{max} + \cos\lambda \cos\lambda_0 \sin h_{max}}{\pi} S_{SS} \quad (3.2\text{-}4a)^1$$

where the ratio $S_0/S_{SS}$ is the longitudinal average of $\mu_0$. The decomposition of the solar forcing can also be written analytically. For a location that has hour angle $h_0$ at time $t = 0$, the first term is

$$\hat{S}_1 = \left[\frac{2\sin\lambda \sin\lambda_0 \sin h_{max}}{\pi} + \frac{\cos\lambda \cos\lambda_0 (h_{max} + \sin h_{max} \cos h_{max})}{\pi}\right] S_{SS} e^{ih_0} \quad (3.2\text{-}4b)^2$$

and, for $m > 1$,

$$\hat{S}_m = \left[\frac{2\sin\lambda \sin\lambda_0 \sin mh_{max}}{m\pi} + \frac{2\cos\lambda \cos\lambda_0 (m\cos h_{max} \sin mh_{max} - \sin h_{max} \cos mh_{max})}{\pi(m^2 - 1)}\right] S_{SS} e^{imh_0} \quad (3.2\text{-}4c)^2$$

If the latitude of the surface element or the sub-solar latitude are near equatorial, then the solar terms are dominated by the first two terms, then diminish quickly with higher order; at the equator, the magnitudes of the terms are proportional to 1, $\pi/2$, 2/3, 0, –2/15, etc., (Paper I). Fig. 3-1 shows an example of the decomposition of insolation for a body with a sub-solar longitude of 2.24° and at a latitude of +30° into a constant plus one term (dashed) or seven terms (dot-dashed).

*Insert fig 3-1 here.*

---

[1] mu0 = **vt3d_solar_mu**(lat, lon, lat0, lon0, /lonavg)

[2] sol_terms = **vt3d_sol_terms_diurnal**(dist_sol_au, albedo, lat, h_phase0, lat_sol, n_terms)

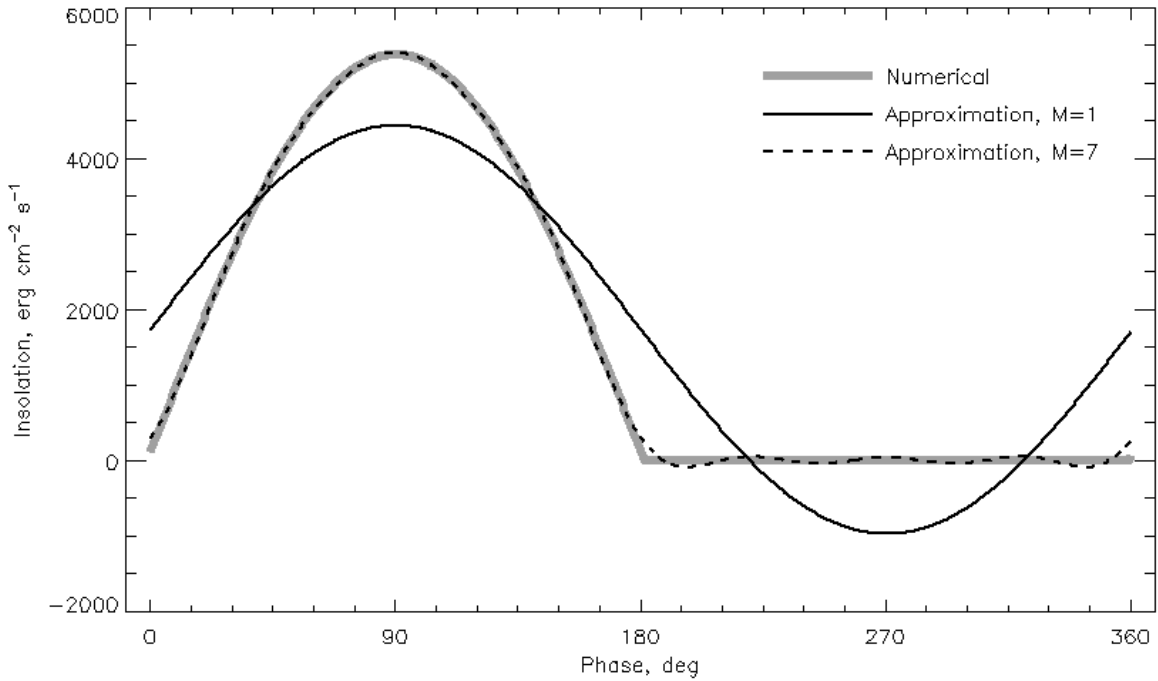

Fig 3-1.[1] Solid gray: numerical calculation of insolation on a bare spot at 9.5 AU with $A = 0.6$, at latitude 30°, a sub-solar latitude of 2.24°, and an hour angle at zero phase of -6 hours (-90°). The off-center maximum heating was chosen to force complex coefficients of the sinusoidal expansion. Solid: sinusoidal approximation with $M=1$, which captures the approximate phase and amplitude of the solar forcing. Dashed: sinusoidal approximation, with $M=7$, which is hard to see except at the "corners" near dawn and dusk because of the accuracy of this approximation.

As discussed in more detail in Paper I, the temperature can be written in terms of sinusoidal terms as well. If the density, specific heat, and thermal conductivity are constant with depth, then the solution to the diffusion equation (Eq. 3.1-2) with flux specifying the lower boundary condition (Eq. 3.1-1) is the sum of damped waves with wavelength $2\pi\sqrt{2mZ}$ and e-folding distance of $\sqrt{2mZ}$ (Fig. 3-2), where

$$\Gamma = \sqrt{k\rho c} \qquad (3.2\text{-}5)[2]$$

---

[1] **vty16_fig3_1**, phase, flux_sol, flux_sol_1, flux_sol_7

[2] therminertia = **vt3d_thermalinertia**(dens, specheat, thermcond)

is the thermal inertia (in cgs units of erg cm$^{-2}$ K$^{-1}$ s$^{-1/2}$, or MKS units of tiu = J m$^{-2}$ K$^{-1}$ s$^{-1/2}$, where tiu, or thermal inertia units, is the SI unit proposed in Putzig 2006), and

$$Z = \frac{k}{\Gamma\sqrt{\omega}} = \sqrt{\frac{k}{\rho c \omega}} \qquad (3.2\text{-}6)^1$$

is the skin depth, as defined by Spencer et al. (1989) and HP96. Other authors use definitions of the skin depths that differ by a constant from Eq. 3.2-6 (e.g., Mellon et al. 2008).

The solution to the conduction equation (3.1-2) can be written as

$$T(\lambda,\phi,z,t) = -\frac{F(\lambda,\phi)}{\sqrt{\omega}}\frac{\zeta}{\Gamma} + T_0(\lambda,\phi) + \text{Re}\left[\sum_{m=1}^{M}\hat{T}_m(\lambda,\phi)e^{im\omega t + \sqrt{im}\zeta}\right] \qquad (3.2\text{-}7)^2$$

where $\zeta = z/Z$ is identical to the unitless scaled depth introduced by Spencer et al. (1989). Temperatures for cases where the thermal-physical properties are variable with depth are treated elsewhere (Fivez & Thoen 1996; Grossel & Depasse 1998; Karam 2000).

*Insert fig 3-2 here.*

---

[1] `z_skin = `**`vt3d_skindepth`**`(dens, specheat, thermcond, freq)`

[2] The general routine, for any number of locations, phases or depths, is
   `temp = `**`vt3d_thermwave`**`(temp_terms, phase, dtemp_dzeta, z_skin, zeta)`
For one or multiple location, phase = 0, and multiple depths, you can use
   `temp = `**`vt3d_thermwave_1loc_p0_nz`**`(temp_terms, dtemp_dzeta, zeta)`
   `temp = `**`vt3d_thermwave_nloc_p0_nz`**`(temp_terms, dtemp_dzeta, zeta)`

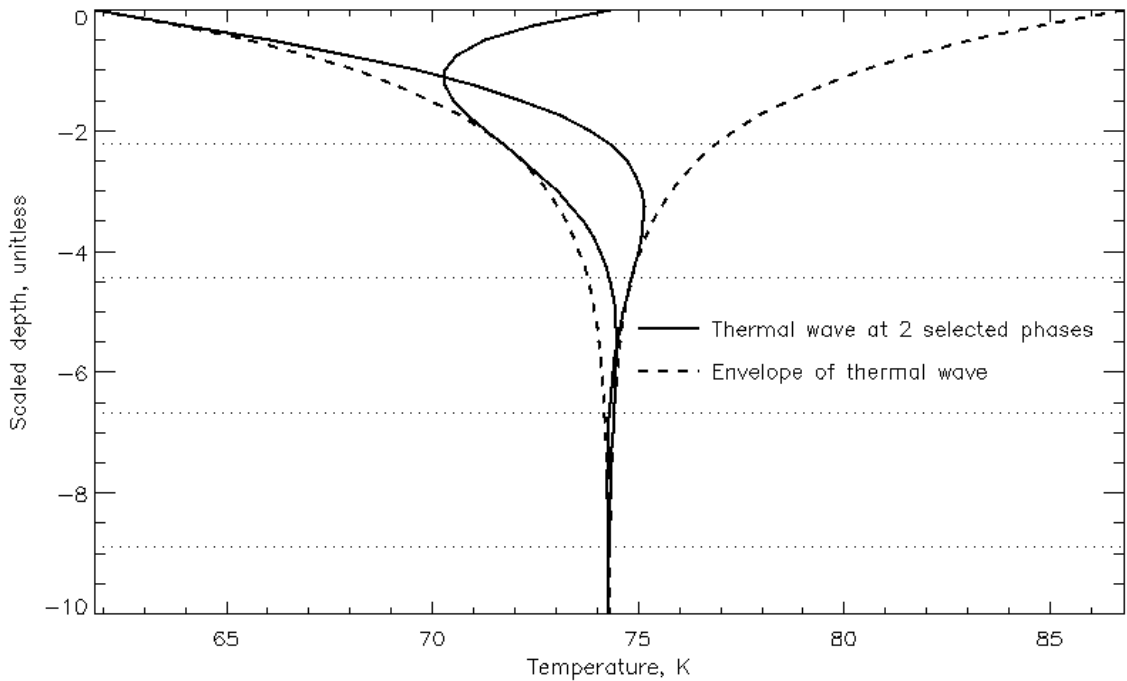

Fig 3-2.[1] Example thermal wave with $\hat{T}_0$ = 74.3 K and $\hat{T}_1$ = 12.5 K at two phases (solid lines). The envelope of the damped waves is shown with dashed lines, and every quarter wavelength is shown with horizontal dotted lines.

The goal is to use Eq. (3.2-7) to create initial conditions for numerical calculations in the three dimensions of latitude, longitude, and depth, given the coefficients for the temperature. There are three ways to do this. The simplest is to expand Eq. (3.1-3) to get the Fourier terms of the temperature directly, as described in Paper I and recapped here. The second is to follow this step by an adjustment of the average temperature, to ensure time-averaged energy balance. The third is to expand into Fourier terms of ($T^4$).

The average temperature, $T_0$, is found by substituting the sinusoidal forms of *S* and *T* into Eq. (3.1-3) and taking the first-order, time-averaged component, resulting in Eq. (3.2-8). This simply states that the mean temperature balances the mean solar insolation and the flux at the lower boundary condition.

$$0 = \underbrace{S_0(\lambda,\phi)}_{\text{Insolation}} - \underbrace{\varepsilon\sigma\left[T_0(\lambda,\phi)\right]^4}_{\text{Emission}} + \underbrace{F(\lambda,\phi)}_{\text{Lower boundary flux}} \qquad (3.2\text{-}8)^2$$

---

[1] `vty16_fig3_2`

[2] `temp_0 = `**`vt3d_temp_term0_bare`**`(sol_0, flux_int, emis)`

The temperature coefficients, $\hat{T}_m$, are found for each $m$ by also substituting $S$ and $T$ into Eq. (3.1-3), taking the appropriate derivatives ($d/dt \to im\omega$, $d/dz \to \sqrt{im}Z^{-1}$), and taking only those terms proportional to exp($im\omega t$). The results are most simply expressed by defining the following variables, which represent the derivative of energy flux or heating with respect to temperature (in cgs units of erg cm$^{-2}$ s$^{-1}$ K$^{-1}$) for the fundamental frequency:

$$\Phi_S = \sqrt{\omega}\Gamma \qquad (3.2\text{-}9a)[1]$$

$$\Phi_E(T) = 4\varepsilon\sigma T^3 \qquad (3.2\text{-}9b)[2]$$

As described in Paper I, a system where $\Phi_S$ is zero has temperatures that track the solar forcing, while positive $\Phi_S$ serves to dampen the amplitude of the temperature variation and introduce a lag. The temperature variation ($\hat{T}_m$) as a function of the solar variation for bare or volatile-covered area is found from:

$$\left[ \underbrace{\sqrt{im}\Phi_S}_{\text{Conduction}} + \underbrace{\Phi_E(T_0)}_{\text{Emission}} \right] \hat{T}_m = \underbrace{\hat{S}_m}_{\text{Insolation}} \qquad (3.2\text{-}10)[3]$$

The temperature is then calculated from $T_0$ and $\hat{T}_m$ from Eq. (3.2-7).

Eq. (3.2-8) overestimates mean temperatures, with the discrepancy being worse with larger peak-to-peak temperature variations, because the time average of $T^4$ is larger than $T_0^4$. Once an estimate of the peak-to-peak variation is found, the value of $T_0$ can be adjusted downward so that the time-average thermal emission equals the sum of the insolation and internal heat flux, iterating over Eq. (3.2-9b) and (3.2-10) until the mean thermal emission converges on the mean absorbed insolation.

As described in Paper I, the time lag and smaller temperature variation can be described by a dimensionless parameter, $\Theta_S$.

$$\Theta_S(T) = \frac{\Phi_S}{\Phi_E(T_0)/4} = \frac{\sqrt{\omega}\Gamma}{\varepsilon\sigma T_0^3} \qquad (3.2\text{-}11)$$

---

[1] `phi_s = `**`vt3d_dfluxdtemp_substrate`**`(freq, therminertia)`

[2] `phi_e = `**`vt3d_dfluxdtemp_emit`**`(emis, temp)`

[3] `temp_terms = `**`vt3d_temp_terms_bare`**`(sol_terms,flux_int,emis,freq,`
`              therminertia)`
`  temp_terms= `**`vt3d_temp_terms_bare_iter`**`(sol_terms,flux_int,emis,freq,`
`              therminertia,thermcond)`

The ratio $\Theta_S = 4\Phi_S/\Phi_E$ is essentially the thermal parameter of Spencer et al. (1989), but defined at the time-averaged local temperature, rather than at the subsolar temperature. As in Paper I, $\Theta_S$ can be used to quantify the shift and decrease in amplitude of the response to solar forcing (Eq. 3.2-12). See Paper I for more discussion on the interpretation of Eq. (3.2-12) in terms of real quantities.

$$\hat{T}_m = \frac{\hat{S}_m}{\Phi_E(T_0)} \frac{4}{4 + \sqrt{im}\Theta_S} \qquad (3.2\text{-}12)$$

Figure 3-3 shows an example of the initial temperature at the surface for a purely sinusoidal wave ($M = 1$) and for $M = 7$ given the same conditions as in Fig 3-2. The combination of a 22.6 hour period and $\Gamma = 16$ tiu (appropriate for Mimas Region 1, Howett etl al 2011) gives $\Theta_S = 6.0$.

*Insert fig 3-3 here.*

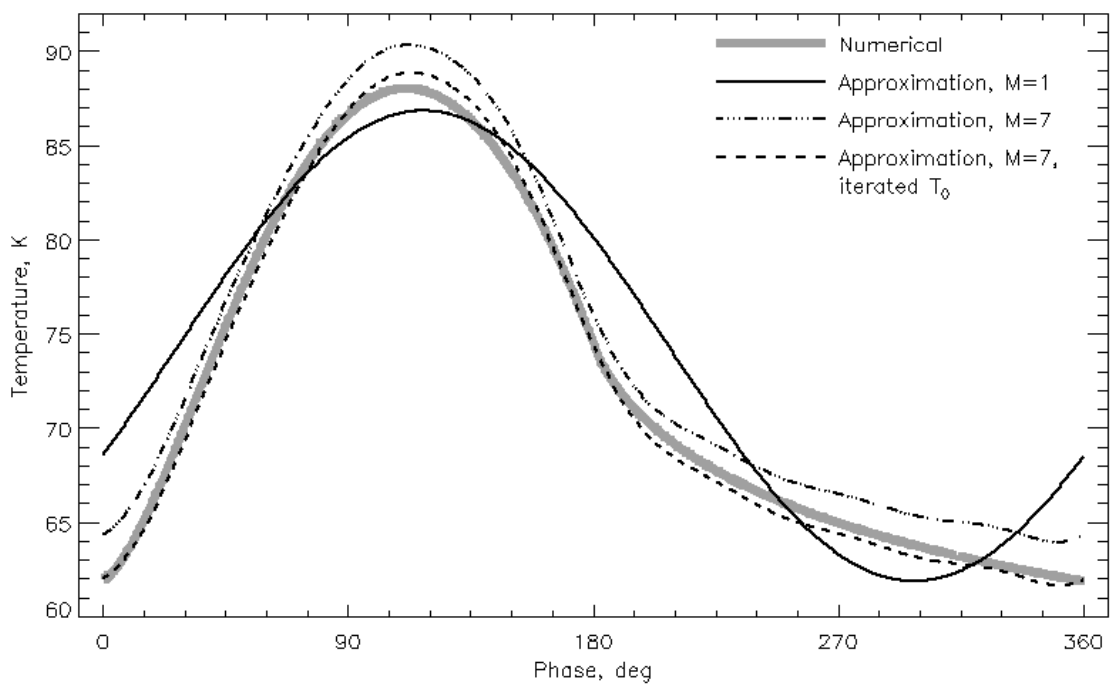

Fig 3-3.[1] Example surface temperatures for the insolation shown in Fig 3-1, with unit emissivity. The result of the numerical integration is shown as solid, thick gray. Initial approximations to the temperature are shown for $M = 1$, without an adjustment of $T_0$ to balance energy fluxes (solid), $M = 7$, without

---

[1] **vty16_fig3_3**

an adjustment of $T_0$ (triple-dot-dash), and $M = 7$ with an adjusted value of $T_0$ (dashed), for a period of 22.6 hours and a thermal inertia of 16 tiu (for a thermal parameter $\Theta_S = 6.0$).

In some situations of large variation in the solar forcing and small values of $\Theta_S$, the linearization of $T^4$ is poor, and it is better to expand in the emitted flux instead.

$$F^E(\lambda,\phi,t) \equiv \varepsilon\sigma T^4 = \text{Re}\left[\sum_{m=0}^{M} \hat{F}^E_m(\lambda,\phi)e^{im\omega t}\right] \tag{3.2-13}$$

The mean term is found from Eq. (3.2-8): $F^E_0 = S_0 + F$. Before, we expanded the emitted flux in terms of temperature, but now we expand temperature in terms of emitted flux:

$$\frac{\hat{T}_m}{T_0} \approx \frac{\hat{F}^E_m}{4 F^E_0} \tag{3.2-14}$$

The conduction is a now a small correction to the thermal emission, so the error in the linearization is confined to the second-order term. Substituting Eq. 3.2-14 into the original equation for energy balance, Eq. (3.1-3), gives:

$$0 = \underbrace{\hat{S}_m}_{\text{Insolation}} - \underbrace{\hat{F}^E_m}_{\text{Emission}} - \underbrace{\frac{k\sqrt{im}}{Z}\hat{T}_m}_{\text{Conduction}} \tag{3.2-15}$$

which, with some manipulation, gives an expression similar to Eq. (3.2-10):

$$\underbrace{\hat{F}^E_m}_{\text{Emission}} + \underbrace{\sqrt{im}\frac{\Theta_S}{4}\hat{F}^E_m}_{\text{Conduction}} = \underbrace{\hat{S}_m}_{\text{Insolation}} \tag{3.2-16}$$

and one similar to Eq. (3.2-12):

$$\hat{F}^E_m = \hat{S}_m \frac{4}{4 + \sqrt{im}\Theta_S} \tag{3.2-17}^1$$

From $F^E$, calculate the surface temperature and its Fourier terms from $T = (F^E/\varepsilon\sigma)^{1/4}$. In some cases, more Fourier terms ($M = 30$ to $100$) need to be used than when calculating the temperature terms directly, to avoid ringing at sharp transitions in the solar forcing.

---

[1] `flux_terms = `**`vt3d_eflux_terms_bare`**`(sol_terms, flux_int, emis, freq, therminertia)`

The analysis in this section can be used for more complex insolation patterns as well. Any insolation pattern, no matter how complex, can be decomposed with a Fast Fourier Transform (FFT) algorithm. If the forcing happens on two different frequencies, such as seasonal and diurnal, then the sums (in e.g., 3.2-7) can be performed over a discrete set of *m*, not necessarily contiguous. For the specific case of combined seasonal and diurnal variation, we can often decouple the two timescales (Young 2012a). First, calculate the seasonal thermal wave as a function of time, using longitudinally averaged insolation. If the seasonal and diurnal skin depths are sufficiently different, then the diurnal wave is superimposed on the uppermost portion of the seasonal one, and the seasonal wave can be treated as a linear contribution to the diurnal wave. This is mathematically identical to an internal heat flux term, *F*, already introduced. In other words, the seasonal thermal heat flow to and from deeper layers affects the diurnal temperatures by affecting the energy flux at the lower boundary. This works because the orbital periods in the outer solar system are orders of magnitude longer than the rotational periods. Pluto, for example, has an orbital period of 248 years and a rotational period of only 6.4 days. The seasonal scale height is larger than the diurnal one by $(248 \text{ years}/6.4 \text{ days})^{1/2}$, or a factor of 119. A typical depth for the lower boundary is 6 diurnal skin depths. This is only 0.05 times the seasonal skin depth, or a tenth of a tick in Fig 3-2, clearly in the linear regime of the seasonal wave.

*3.3 Numerical solution for a single bare location (Area I or III)*

For some applications, the results of the analytic calculations may be adequate. For others, higher accuracy is needed. Even for these applications, the analytic solution provides an initial condition that improves convergence.

The continuous equations of Section 3.1 are converted to a form suitable for computation. This is done by discretizing the variables into *L* locations on the surface (indexed by *l*), *J* + 1 layers within the substrate (indexed by *j*), and choosing time step schemes that take the state from time *n* to time *n* + 1 (i.e., no leap-frogging time step schemes). The general approach is to treat the time step as a finite-difference diffusion problem, with flux conditions at both the upper and lower boundaries (Press et al., 2007; Haltiner and Williams 1984).

Figure 3-4 represents the discretization of the numerical model. The substrate is divided into *J+1* layers, indexed with *j* = 0 for the top-most layer to *j* = *J* for the lowest layer, and defined by a depth $z_j$ and a thickness $\Delta_j$. Depths (*z*) are less than or equal to zero, and become more negative with increasing index. Thicknesses of the layers ($\Delta_j$) are positive. Thickness can vary with index *j* to speed computation (Table 2). All layers except the top layer extend from $z_j - \Delta_j/2$ to $z_j + \Delta_j/2$, with temperature $T_{l,j,n}$ defined at the center of the layer. The top layer extends from $z = 0$ to $z = -\Delta_0$, with the temperature $T_{l,0,n}$ defined at the top of the layer. If the layering is the same across the globe, then $z_j$ and $\Delta_j$ are functions only of $j$. Density, specific heat, and thermal conductivity are constant within a layer, but can vary with both depth and location, with values $\rho_{l,j}$, $c_{l,j}$, and $k_{l,j}$. Substrate temperature varies with location, depth, and time. Temperatures are continuous, and are linear between $z_j$ and the layer boundaries. Conducted fluxes ($k\, dT/dz$) are continuous at layer boundaries.

The use of layers that are free to vary their thickness[1] with depth improves efficiency, since the computational time is proportional to the number of layers, requiring only a little additional computation at the beginning of a calculation. A common layering approach uses a geometrically increasing thickness, where the thickness of each layer is some factor larger than the layer above (typically a factor of 1.1 to 1.5, e.g., Hansen and Paige 1996; Keiffer 2013). When modeling a diurnal wave, this allows modest computational savings, since geometrically thickening layers can span down to six skin depths with 2-3 times fewer layers than for layers of equal thickness. Unevenly spaced layers is even more important for practical modeling of the diurnal and seasonal wave simultaneously. Because the skin depth is proportional to $\omega^{-1/2}$, the ratio of diurnal and seasonal skin depths equals the square root of the ratio of their periods, if thermophysical properties are constant with depth. This is important even for Mars, where the orbital period is roughly 669 times the rotation period, so the seasonal skin depth is roughly 25 times the diurnal skin depth (if thermophysical properties are constant with depth). In the outer solar system, the orbital periods can be quite long, so that the equivalent ratio of seasonal to diurnal skin depths is 88 for Enceledus, 118 for Pluto, and 700 for Eris. If the thermal conductivity is greater at depth, these ratios can be even larger. Here the savings for geometrically thickening layers is dramatic, allowing calculation to 100, 1000, or even 10,000 diurnal skin depths with computational savings of ~20, ~100, or ~1000 respectively. For example, layers that begin with a thickness of 0.25 diurnal skin depths can reach 10,000 diurnal skin depths with only 41 layers for a thickening factor of 1.5, or with 87 layers for a thickening factor of 1.2.

*Insert fig 3-4 here.*

---

[1] `z = `**`vt3d_zgrid`**`(skindepth,z_delta,n_z)`

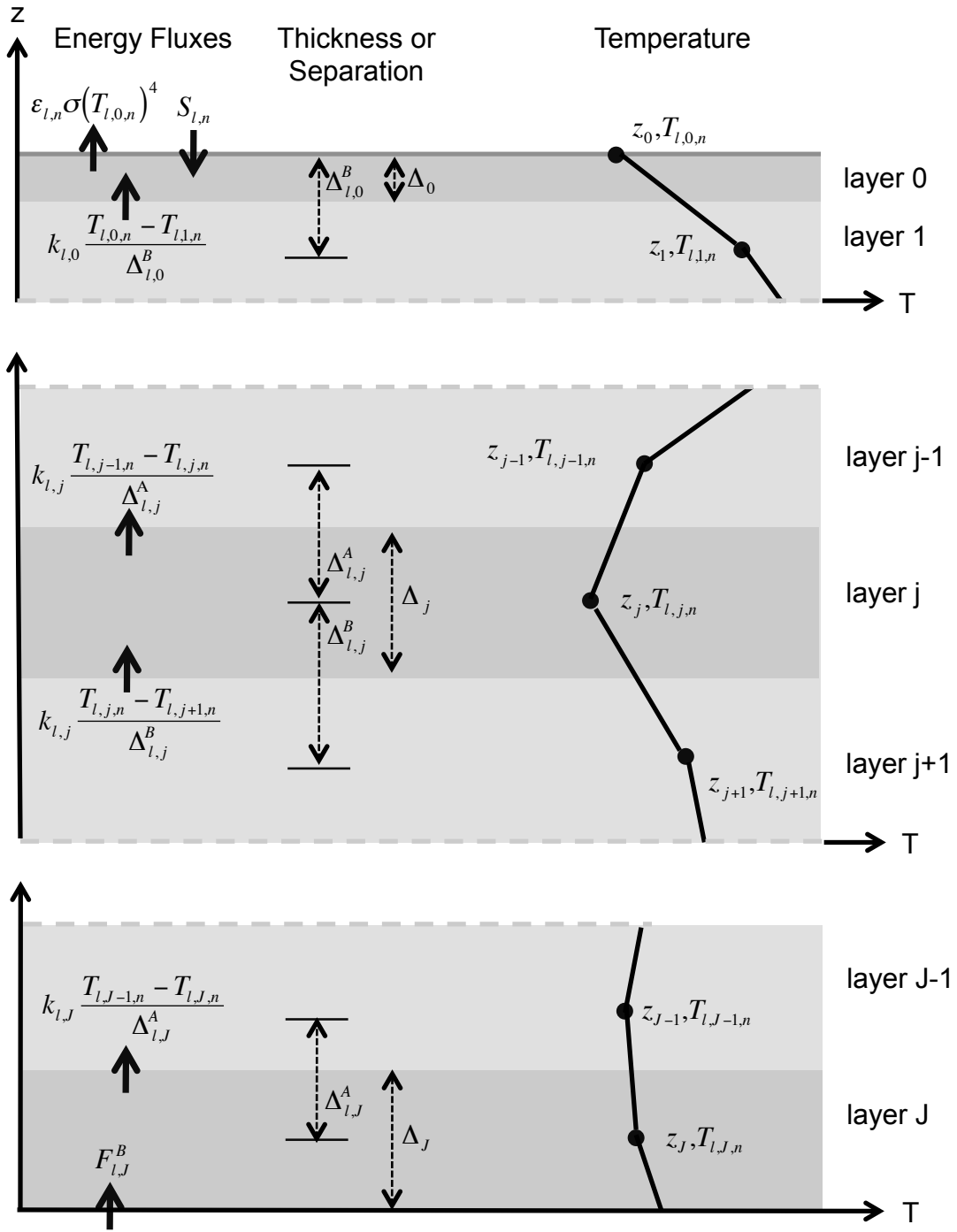

Fig. 3-4. Schematic of the layering scheme and energy fluxes. Top: layer $j = 0$ for bare areas (Areas I and III). Middle: interior layers, valid for $j = 1$ to $J–1$. Bottom: $j = J$. Thick arrows near the left side of the plot represent energy fluxes due to emission, insolation, or conduction. Thin arrows near the middle

of the plot indicate layer depths and distances between layers. Thick lines with large dots near the right side of the plot schematically represent temperatures. Temperatures are tabulated at the top of layer 0, and at the middle of all other layers, and are assumed to be linear with $z$ within the layer. Conductive flux ($k\, dT/dz$) is continuous across layer boundaries.

The goal is to cast the equations as matrix operations to take advantage of the fast array operations that are available in many modern computer languages. The continuous equations of Section 3.1 can be cast as explicit equations (Fig 3-5), where the new temperature depends explicitly only on the previous temperature (Press et al. 2007; Haltiner and Williams 1984). The explicit expressions for diffusion equations are only accurate to first order in the time step, $\Delta t$, and require small time steps for stability. For explicit equations, the timesteps must satisfy $(\Delta t/P) \leq (\Delta z/Z)^2/4\pi$, or slightly more than 200 steps per period for a vertical sampling of 4 layers per skin depth.

The explicit linearized problem can be described with a $(J+1) \times (J+1)$ tridiagonal matrix (Fig. 3-5). The new temperatures depend on the current temperatures in the layer above (with matrix element $\alpha$, mnemonically "*a* for Above"), the current temperature in that layer (with matrix element $\eta$, mnemonically "*h* for Here"), and the current temperatures in the layer below (with matrix element $\beta$, mnemonically "*b* for Below").

*Insert fig 3-5 here.*

$$\begin{pmatrix} T_{l,0,n+1} \\ T_{l,1,n+1} \\ T_{l,2,n+1} \\ T_{l,3,n+1} \\ \vdots \\ T_{l,J-1,n+1} \\ T_{l,J,n+1} \end{pmatrix} = \begin{pmatrix} \eta_{l,0,n} & \beta_{l,0,n} & & & & & \\ \alpha_{l,1} & \eta_{l,1} & \beta_{l,1} & & & & \\ & \alpha_{l,2} & \eta_{l,2} & \beta_{l,2} & & & \\ & & \alpha_{l,3} & \eta_{l,3} & \beta_{l,3} & & \\ & & & \ddots & \ddots & \ddots & \\ & & & & \alpha_{l,J-1} & \eta_{l,J-1} & \beta_{l,J-1} \\ & & & & & \alpha_{l,J} & \eta_{l,J} \end{pmatrix} \times \begin{pmatrix} T_{l,0,n} \\ T_{l,1,n} \\ T_{l,2,n} \\ T_{l,3,n} \\ \vdots \\ T_{l,J-1,n} \\ T_{l,J,n} \end{pmatrix} + \begin{pmatrix} \gamma_{l,0,n} \\ \\ \\ \\ \\ \\ \gamma_{l,J} \end{pmatrix}$$

Fig 3-5. Schematic of an explicit time-step from time *n* to time *n*+1 for a location *l* in Area I (bare, isolated), II (volatile-covered, isolated) or III (bare, interacting). Dark gray elements (the temperatures and the elements of the upper row) change with each time step. Light gray elements are independent of time. White elements are zero.

Accuracy and stability can be improved by using implicit (Crank-Nicholson) methods, which solve equations involving both the current and the next temperatures (Fig. 3-7), at the cost of computational complexity (Press et al. 2007; Haltiner and Williams 1984). The Crank-Nicholson scheme results in an equation that is accurate to second order in the time step, and satisfies von Neumann stability criteria for all sizes of time step. The implicit (Crank-Nicholson) problem uses two $(J+1) \times (J+1)$ tridiagonal matrices, with primed elements on the right-hand side of the equation and double-primed elements on the left. The goal of this section is to derive the matrix elements, which are summarized in Tables 3 to 5.

*Insert fig 3-6 here.*

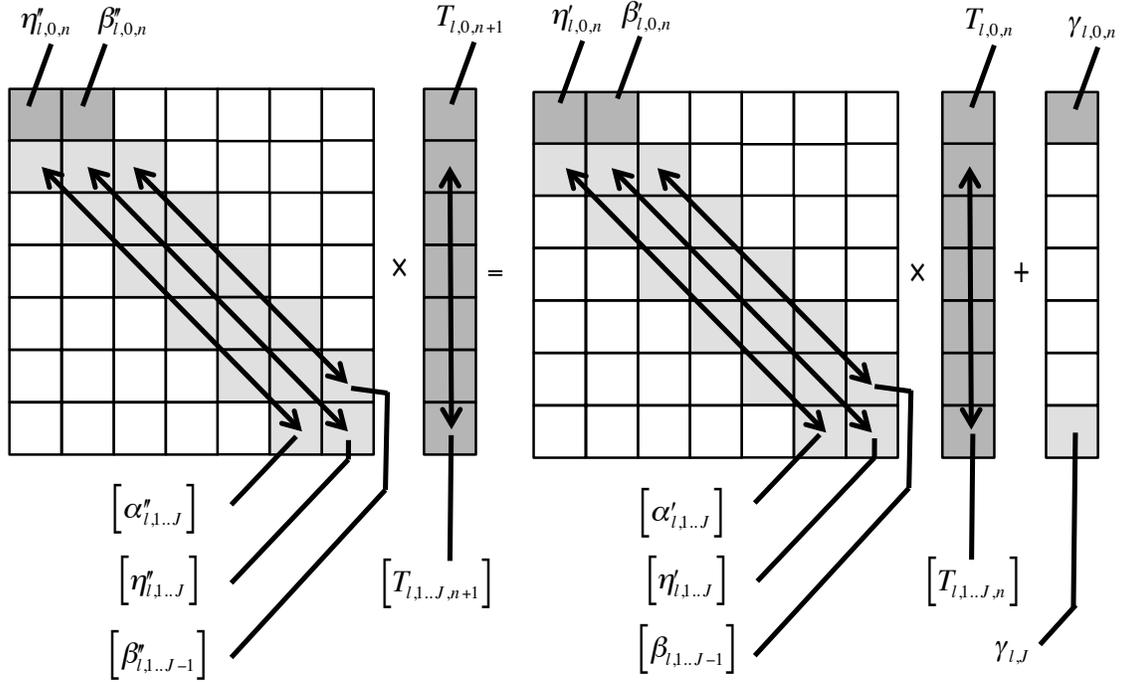

Fig 3-6. Schematic of an implicit time-step from time $n$ to time $n+1$ for a location $l$ in Area I (bare, local), II (volatile-covered, local) or III (bare, isobaric), using the Crank-Nicholson scheme. Dark gray elements (the temperatures and the elements of the upper row) change with each time step. Light gray elements are independent of time. White elements are zero. The variables in brackets refer to the vectors of length $J$ or $J–1$ indicated by the double-arrowed lines.

**Table 3. matrix elements for $j = 0$, Area I and III (bare)**

| Matrix equation | Matrix elements |
|---|---|
| Explicit $$T_{l,0,n+1} = \eta_{l,0,n} T_{l,0,n} + \beta_{l,0,n} T_{l,1,n} + \gamma_{l,0,n}$$ | $$\beta_{l,0,n} = \frac{\Phi_{l,0}^{K,B}}{\Phi_{l,n}^{T}}$$ $$\eta_{l,0,n} = 1 - \beta_{l,0,n}$$ $$\gamma_{l,0,n} = \frac{\overline{S_{l,n'}} - \varepsilon_{l,n}\sigma(T_{l,0,n})^4}{\Phi_{l,n}^{T}}$$ |
| Implicit (Crank-Nicholson) $$\eta''_{l,0,n} T_{l,0,n+1} + \beta''_{l,0,n} T_{l,1,n+1} = \eta'_{l,0,n} T_{l,0,n} + \beta'_{l,0,n} T_{l,1,n} + \gamma_{l,0,n}$$ | $$\beta'_{l,0,n} = \frac{\beta_{l,0,n}}{2}; \quad \beta''_{l,0} = -\frac{\beta_{l,0,n}}{2}$$ $$\eta'_{l,0,n} = 1 - \beta'_{l,0,n}; \quad \eta''_{l,0,n} = 1 + \beta''_{l,0,n}$$ |

$\Phi_{l,0}^{K,B}$ is given by Eq. 3.3-10. $\Phi_{l,n}^{T}$ is given by 3.3-18.

**Table 4. Matrix elements for $j = 1 .. J-1$, all Areas**[1]

| Matrix equation | Matrix elements |
|---|---|
| Explicit<br>$T_{l,j,n+1} = \alpha_{l,j} T_{l,j-1,n} + \eta_{l,j} T_{l,j,n} + \beta_{l,j} T_{l,j+1,n}$ | $\alpha_{l,j} = \dfrac{\tau}{\delta^A_{l,j} \delta_{l,j}}$ <br> $\beta_{l,j} = \dfrac{\tau}{\delta^B_{l,j} \delta_{l,j}}$ <br> $\eta_{l,j} = 1 - \alpha_{j,l} - \beta_{l,j}$ |
| Implicit (Crank-Nicholson)<br>$\alpha''_{l,j} T_{l,j-1,n+1} + \eta''_{l,j} T_{l,j,n+1} + \beta''_{l,j} T_{l,j+1,n+1} =$<br>$\alpha'_{l,j} T_{l,j-1,n} + \eta'_{l,j} T_{l,j,n} + \beta'_{l,j} T_{l,j+1,n}$ | $\alpha'_{l,j} = \dfrac{\alpha_{l,j}}{2}; \quad \alpha''_{l,j} = -\dfrac{\alpha_{l,j}}{2}$ <br> $\beta'_{l,j} = \dfrac{\beta_{l,j}}{2}; \quad \beta''_{l,j} = -\dfrac{\beta_{l,j}}{2}$ <br> $\eta'_{l,j} = 1 - \alpha'_{j,l} - \beta'_{l,j}; \quad \eta''_{l,j} = 1 - \alpha'_{j,l} - \beta'_{l,j}$ |

**Table 5. Matrix elements for $j = J$, all Areas**[1]

| Matrix equation | Matrix elements |
|---|---|
| Explicit<br>$T_{l,J,n+1} = \alpha_{l,J} T_{l,J-1,n} + \eta_{l,J} T_{l,J,n} + \gamma_{l,J}$ | $\alpha_{l,J} = \dfrac{\tau}{\delta^A_{l,J} \delta_{l,J}}$ <br> $\eta_{l,J} = 1 - \alpha_{J,l}$ <br> $\gamma_{l,J} = \dfrac{F^B_{l,J}}{\Phi^H_{l,j}}$ |
| Implicit (Crank-Nicholson)<br>$\alpha''_{l,J} T_{l,J-1,n} + \eta''_{l,J} T_{l,J,n} =$<br>$\alpha'_{l,J} T_{l,J-1,n} + \eta'_{l,J} T_{l,J,n} + \gamma_{l,J}$ | $\alpha'_{l,J,n} = \dfrac{\alpha_{l,j,n}}{2}; \quad \alpha''_{l,J,n} = -\dfrac{\alpha_{l,j,n}}{2}$ <br> $\eta'_{l,J} = 1 - \alpha'_{l,J}; \quad \eta''_{l,J} = 1 - \alpha'_{l,J}$ |

$\Phi^H_{i,j}$ is given by Eq. 3.3-5.

To find the energy balance in layer 0, integrate the conduction equation (Eq. 3.1-2) over the top layer, from $z = -\Delta_0$ to $z = 0$. Add this to the energy balance equation (Eq. 3.1-3) to get:

$$\underbrace{\overline{\rho_{l,0} c_{l,0} \int_{-\Delta_0}^{0} \frac{\partial T}{\partial t} dz}}_{\text{Enthalpy, layer 0}} = \underbrace{\overline{S_{l,n'}}}_{\text{Insolation}} - \underbrace{\overline{\varepsilon_{l,n'} \sigma T^4_{l,0,n'}}}_{\text{Emission}} - \underbrace{\overline{k \left.\dfrac{dT}{dz}\right|_{z=-\Delta_0}}}_{\text{Conduction}} \qquad (3.3\text{-}1)$$

where the overbar indicates the time-averaged value over the time step $t_n$ to $t_{n+1}$. The subscript for time in the insolation and emission terms is $n'$ to indicate that it varies over the

---

[1] ```
alpha_i = vt3d_alpha_interior(tau, del, del_a)
beta_i  = vt3d_beta_interior(tau, del, del_b)
```



time interval from $n$ to $n+1$. The change in enthalpy over layer 0 can be approximated as a function of the temperature sampled at the top of the layer:

$$\rho_{l,0} c_{l,0} \overline{\int_{-\Delta_0}^{0} \frac{\partial T}{\partial t} dz} \approx \Phi_{l,0}^{H} \left( T_{l,0,n+1} - T_{l,0,n} \right) \qquad (3.3\text{-}2)$$

where $\Phi_{l,j}^{H}$ has units of erg cm$^{-2}$ s$^{-1}$ K$^{-1}$, with the superscript $H$ representing heat or enthalpy. Eq. 3.3-2 samples the temperature of layer 0 at the top of the layer. If the temperature is integrated over layer 0 instead, then the slope of the temperature through layer 0 needs to be included; this depends on $T_{l,1,n}$, and is a second-order effect that I ignore here.

Defining a unitless measure of the time step, $\tau$, (radians per timestep):

$$\tau = (t_{n+1} - t_n)\omega \qquad (3.3\text{-}3)$$

and a unitless measure of the thickness of layer $j$ expressed as a fraction of the skin depth (c.f., Spencer et al. 1989)

$$\delta_{l,j} = \frac{\Delta_j}{Z_{l,j}} \qquad (3.3\text{-}4)$$

gives

$$\Phi_{l,j}^{H} = \frac{\rho_{l,j} c_{l,j}}{t_{n+1} - t_n} \Delta_0 = \frac{\delta_{l,j} \Phi_S(\Gamma_{l,j})}{\tau} \qquad (3.3\text{-}5)$$

where $\Phi_S$ is defined in Eq. (3.2-9a) and only depends on the physical properties of the problem. In contrast, $\Phi_{i,j}^{H}$ additionally includes non-dimensional factors that depend on the numerical choices of $\tau$ and $\delta_{l,j}$. In general, I represent the fluxes-per-temperature that depend only on the physical properties with a single subscript for the physical process (e.g., $S$ or $H$), and the ones that are discretized and depend on $\tau$ and $\delta_{l,j}$ with the superscript for the process and a subscript for the indices of location and time.

The average solar insolation between $t_n$ and $t_{n+1}$, $\overline{S_{l,n'}}$, depends on the geometry (heliocentric distance, and subsolar latitude and longitude) and the albedo. If the insolation is evaluated at the start of the timestep ($\overline{S_{l,n'}} \approx S_{l,n}$), then the results will be skewed in time by half a timestep, which is acceptable when timesteps are small (e.g., Spencer et al. 1989), but not at the larger timesteps allowed by the Crank-Nicholson method. A simple correction is to average the insolation at the start and end of the timestep



$$\overline{S_{l,n'}} \approx (S_{l,n} + S_{l,n+1})/2 \tag{3.3-6}$$

The average thermal emission at the midpoint of the time interval is found by evaluating the first-order Taylor expansion of $T^4$ at the average temperature for the time interval, $(T_{l,0,n+1} - T_{l,0,n})/2$, assuming that the emissivity is constant over the time interval.

$$\overline{\varepsilon_{l,n'} \sigma T^4_{l,0,n'}} = \varepsilon_{l,n} \sigma \left(T_{l,0,n}\right)^4 + \Phi^E_{l,n} \left(T_{l,0,n+1} - T_{l,0,n}\right) \tag{3.3-7}$$

where $\Phi^E_{l,n}$ has units cgs of erg cm$^{-2}$ s$^{-1}$ K$^{-1}$, with the superscript *E* representing *emission*:

$$\Phi^E_{l,n} = 2\varepsilon_{l,n}\sigma\left(T_{l,0,n}\right)^3 = \frac{\Phi_E(T_{l,0,n})}{2} \tag{3.3-8}$$

where $\Phi_E$ is defined in Eq. (3.2-9b). As with the enthalpy term, $\Phi_E$ only depends on the physical properties of the problem, and $\Phi^E_{i,j}$ is the value used in the descretized calculation. Unlike the enthalpy term, $\Phi_E$ and $\Phi^E_{l,n}$ changes with each time step.

The next term in Eq. (3.3-1) is the thermal conduction. For explicit equations, the derivative is evaluated at the start of the time interval:

$$\overline{\left(k\frac{dT}{dz}\bigg|_{z=-\Delta_0}\right)} \approx \Phi^{K,B}_{l,0}\left(T_{l,0,n} - T_{l,1,n}\right) \tag{3.3-9}$$

where $\Phi^{K,B}_{l,j}$ has cgs units of erg cm$^{-2}$ s$^{-1}$ K$^{-1}$. The superscript *K* represents thermal conduction, and the superscript *B* represents conduction from the layer *below*. The expression for $\Phi^{K,B}_{l,j}$ is

$$\Phi^{K,B}_{l,j} = \frac{k_{l,j}}{\Delta^B_{l,j}} = \frac{\Phi_S(\Gamma_{l,j})}{\delta^B_{l,j}} \tag{3.3-10}$$

where $\Delta^B_{l,j}$ is essentially the distance to the middle of the layer *below*, modified to ensure continuity of fluxes at layer boundaries:

$$\Delta^B_{l,j} = \frac{\Delta_j}{2} + \frac{k_{l,j}}{k_{l,j+1}} \frac{\Delta_{j+1}}{2}, j = 1...J-1 \tag{3.3-11}[1]$$

and the unitless distances used for calculating thermal gradients from the layer below is

---

[1] **vt3d_zdelta,** z, z_delta, z_delta_top, z_delta_bot



$$\delta_{l,j}^{B} = \frac{\Delta_{l,j}^{B}}{Z_{l,j}} \qquad (3.3\text{-}12)$$

Even if $z_j$ and $\Delta_j$ are constant from one location to the next, the dependence on $k$ means that $\Delta_{l,j}^{A}$ and $\Delta_{l,j}^{B}$ may vary with location. Again, $\Phi_S$ only depends on the physical properties of the problem, and $\Phi_{i,j}^{K,B}$ additionally includes non-dimensional factors that depend on the numerical implementation.

The more accurate and more stable Crank-Nicholson scheme (Press et al. 2007) replaces the derivative in Eq (3.3-9) with the average of the derivatives calculated at the start and end of the time step (at time $t_n$ and time $t_{n+1}$):

$$\overline{\left(k\frac{dT}{dz}\bigg|_{z=-\Delta_0}\right)} \approx \frac{1}{2}\Phi_{l,0}^{K,B}\left(T_{l,0,n} - T_{l,1,n}\right) + \frac{1}{2}\Phi_{l,0}^{K,B}\left(T_{l,0,n+1} - T_{l,1,n+1}\right) \qquad (3.3\text{-}13)$$

The explicit discretized equation for energy balance of layer 0 becomes

$$\underbrace{\Phi_{l,0}^{H}\left(T_{l,0,n+1} - T_{l,0,n}\right)}_{\text{Enthalpy, layer 0}} = \underbrace{\overline{S_{l,n'}}}_{\text{Insolation}} \underbrace{- \varepsilon_{l,n}\sigma\left(T_{l,0,n}\right)^4 - \Phi_{l,n}^{E}\left(T_{l,0,n+1} - T_{l,0,n}\right)}_{\text{Emission}} - \underbrace{\Phi_{l,0}^{K,B}\left(T_{l,0,n} - T_{l,1,n}\right)}_{\text{Conduction}} \qquad (3.3\text{-}14)$$

while the implicit equation is

$$\begin{aligned}\underbrace{\Phi_{l,0}^{H}\left(T_{l,0,n+1} - T_{l,0,n}\right)}_{\text{Enthalpy, layer 0}} &= \underbrace{\overline{S_{l,n'}}}_{\text{Insolation}} \underbrace{- \varepsilon_{l,n}\sigma\left(T_{l,0,n}\right)^4 - \Phi_{l,n}^{E}\left(T_{l,0,n+1} - T_{l,0,n}\right)}_{\text{Emission}} \\ &\underbrace{- \frac{1}{2}\Phi_{l,0}^{K,B}\left(T_{l,0,n} - T_{l,1,n}\right) - \frac{1}{2}\Phi_{l,0}^{K,B}\left(T_{l,0,n+1} - T_{l,1,n+1}\right)}_{\text{Conduction}}\end{aligned} \qquad (3.3\text{-}15)$$

Collecting terms for the explicit equation (only $T_{l,0,n+1}$ on the left-hand side) results in:

$$\left(\Phi_{l,0}^{H} + \Phi_{l,n}^{E}\right)T_{l,0,n+1} = \left(\Phi_{l,0}^{H} + \Phi_{l,n}^{E} - \Phi_{l,0}^{K,B}\right)T_{l,0,n} + \left(\Phi_{l,0}^{K,B}\right)T_{l,1,n} + \overline{S_{l,n'}} - \varepsilon_{l,n}\sigma\left(T_{l,0,n}\right)^4 \qquad (3.3\text{-}16a)$$

and for the implicit equation ($T_{l,0,n+1}$ and $T_{l,1,n+1}$ on the left-hand side) results in:

$$\begin{aligned}\left(\Phi_{l,0}^{H} + \Phi_{l,n}^{E} + \frac{\Phi_{l,0}^{K,B}}{2}\right)T_{l,0,n+1} - \left(\frac{\Phi_{l,0}^{K,B}}{2}\right)T_{l,1,n+1} &= \left(\Phi_{l,0}^{H} + \Phi_{l,n}^{E} - \frac{\Phi_{l,0}^{K,B}}{2}\right)T_{l,0,n} + \left(\frac{\Phi_{l,0}^{K,B}}{2}\right)T_{l,1,n} \\ &+ \overline{S_{l,n'}} - \varepsilon_{l,n}\sigma\left(T_{l,0,n}\right)^4\end{aligned} \qquad (3.3\text{-}16b)$$



The goal is to now turn Eq. (3.3-16a) and (3.3-16b) into equations that express the matrix multiplication shown in Fig. 3-5 and Fig. 3-6, respectively. For the top layer, the matrix equations are

$$T_{l,0,n+1} = \eta_{l,0,n} T_{l,0,n} + \beta_{l,0,n} T_{l,1,n} + \gamma_{l,0,n} \tag{3.3-17a}$$

for explicit and

$$\eta''_{l,0,n} T_{l,0,n+1} + \beta''_{l,0,n} T_{l,1,n+1} = \eta'_{l,0,n} T_{l,0,n} + \beta'_{l,0,n} T_{l,1,n} + \gamma_{l,0,n} \tag{3.3-17b}$$

for implicit time step schemes. Divide Eq. (3.3-15) by the total flux per temperature

$$\Phi^T_{l,n} = \Phi^H_{l,0} + \Phi^E_{l,n}, \tag{3.3-18}$$

with units erg cm$^{-2}$ s$^{-1}$ K$^{-1}$, where the superscript *T* represents *total*, to get the matrix elements for $j = 0$, Areas I and III (Table 3). The forcing is a function of time, and is subscripted *n*. Because the derivative of the thermal emission depends on time, the matrix elements $\beta_{l,0,n}$ and $\eta_{l,0,n}$ also depend on time.

For interior layers, the integral form of the diffusion equations (Eq. 3.1-2), averaged over time step *n* is

$$\underbrace{\overline{\int_{z_j-\Delta_j/2}^{z_j+\Delta_j/2} \rho c \frac{\partial T}{\partial t} dz}}_{\text{Enthalpy of layer } j} = \underbrace{\overline{\left(k \frac{dT}{dz}\bigg|_{z_j+\Delta_j/2}\right)} - \overline{\left(k \frac{dT}{dz}\bigg|_{z_j-\Delta_j/2}\right)}}_{\text{Conduction}}, \quad j = 1..J-1 \tag{3.3-19}$$

where the overbar indicates the time-averaged value over the time step $t_n$ to $t_{n+1}$.

In the lowest layer, as in the interior layers, the net change in enthalpy of the layer is balanced by the difference between the flux entering from below and leaving from above (Fig. 3-4). For layer *J*, unlike for layers $j = 1... J$-1, the flux from below is specified as a lower boundary condition. The energy balance equation for the lowest layer is:

$$\underbrace{\overline{\int_{z_j-\Delta_j/2}^{z_j+\Delta_j/2} \rho c \frac{\partial T}{\partial t} dz}}_{\text{Enthalpy of layer } J} = \underbrace{\overline{\left(k \frac{dT}{dz}\bigg|_{z_j+\Delta_j/2}\right)}}_{\text{Conduction}} + \underbrace{F_l}_{\text{Lower flux}}, \quad j = J \tag{3.3-20}$$

where $F_l$ is the heat flux at the lower boundary for location *l*.

The change in enthalpy over layer *j* ($j = 1.. J$) can be approximated as a function of the temperature sampled at the middle of the layer:



$$\overline{\rho_{l,j} c_{l,j} \int_{z_j - \Delta_j/2}^{z_j + \Delta_j/2} \frac{dT}{dz} dz} \approx \Phi_{l,j}^{H} \left( T_{l,j,n+1} - T_{l,j,n} \right) \tag{3.3-21}$$

The expressions for conduction into the layer above for from the layer $j$ are similar to those into layer 0 from layer 1 (Eq. 3.3-9 and 3.3-13). For the explicit scheme, it is:

$$\overline{\left( k \frac{dT}{dz} \bigg|_{z_j - \Delta_j/2} \right)} \approx \Phi_{l,j}^{K,B} \left( T_{l,j,n} - T_{l,j+1,n} \right) \tag{3.3-22a}$$

$$\overline{\left( k \frac{dT}{dz} \bigg|_{z_j + \Delta_j/2} \right)} \approx \Phi_{l,j}^{K,A} \left( T_{l,j-1,n} - T_{l,j,n} \right) \tag{3.3-22b}$$

and for the Crank-Nicholson implicit scheme it is:

$$\overline{\left( k \frac{dT}{dz} \bigg|_{z_j - \Delta_j/2} \right)} \approx \frac{\Phi_{l,j}^{K,B}}{2} \left( T_{l,j,n} - T_{l,j+1,n} \right) + \frac{\Phi_{l,j}^{K,B}}{2} \left( T_{l,j,n+1} - T_{l,j+1,n+1} \right) \tag{3.3-23a}$$

$$\overline{\left( k \frac{dT}{dz} \bigg|_{z_j + \Delta_j/2} \right)} \approx \frac{\Phi_{l,j}^{K,A}}{2} \left( T_{l,j-1,n} - T_{l,j,n} \right) + \frac{\Phi_{l,j}^{K,A}}{2} \left( T_{l,j-1,n+1} - T_{l,j,n+1} \right) \tag{3.3-23b}$$

where $\Phi_{l,j}^{K,B}$ is specified by Eq. (3.3-10) and

$$\Phi_{l,j}^{K,A} = \frac{k_{l,j}}{\Delta_{l,j}^{A}} = \frac{\Phi_S(\Gamma_{l,j})}{\delta_{l,j}^{A}} \tag{3.3-24}$$

$$\Delta_{l,j}^{A} = \frac{\Delta_j}{2} + \frac{k_{l,j}}{k_{l,j-1}} \Delta_{j-1}, \quad j = 1 \tag{3.3-25a}$$

$$\Delta_{l,j}^{A} = \frac{\Delta_j}{2} + \frac{k_{l,j}}{k_{l,j-1}} \frac{\Delta_{j-1}}{2}, \quad j = 2 \ldots J \tag{3.3-25b}[1]$$

and

$$\delta_{l,j}^{A} = \frac{\Delta_{l,j}^{A}}{Z_{l,j}} \tag{3.3-26}$$

Substituting Eq. (3.3-21) and (3.3-22) into (3.3-19), the explicit equation for $j = 1 \ldots J–1$ is

---

[1] **vt3d_zdelta,** z, z_delta, z_delta_top, z_delta_bot



$$\underbrace{\Phi^H_{l,j}\left(T_{l,j,n+1} - T_{l,j,n}\right)}_{\text{Enthalpy of layer } j} = \underbrace{\Phi^{K,A}_{l,j}\left(T_{l,j-1,n} - T_{l,j,n}\right) - \Phi^{K,B}_{l,j}\left(T_{l,j,n} - T_{l,j+1,n}\right)}_{\text{Conduction}} \tag{3.3-27}$$

and substituting Eq. (3.3-21) and (3.3-22) into (3.3-20), the explicit equation for the lowest layer, $j = J$, is

$$\underbrace{\Phi^H_{l,J}\left(T_{l,J,n+1} - T_{l,J,n}\right)}_{\text{Enthalpy of layer } J} = \underbrace{\Phi^{K,A}_{l,J}\left(T_{l,J-1,n} - T_{l,J,n}\right)}_{\text{Conduction}} + \underbrace{F_{l,J}}_{\text{Lower flux}} \tag{3.3-28}$$

Similarly, substituting Eq. (3.3-21) and (3.3-23) into (3.3-19), the implicit equation for $j = 1 \ldots J-1$ is

$$\underbrace{\Phi^H_{l,j}\left(T_{l,j,n+1} - T_{l,j,n}\right)}_{\text{Enthalpy of layer } j} = \underbrace{\left[\frac{\Phi^{K,A}_{l,j}}{2}\left(T_{l,j-1,n} - T_{l,j,n}\right) + \frac{\Phi^{K,A}_{l,j}}{2}\left(T_{l,j-1,n+1} - T_{l,j,n+1}\right)\right] - \left[\frac{1}{2}\Phi^{K,B}_{l,0}\left(T_{l,0,n} - T_{l,1,n}\right) + \frac{1}{2}\Phi^{K,B}_{l,0}\left(T_{l,0,n+1} - T_{l,1,n+1}\right)\right]}_{\text{Conduction}} \tag{3.3-29}$$

and substituting Eq. (3.3-21) and (3.3-23) into (3.3-20), the implicit equation for the lowest layer, $j = J$, is

$$\underbrace{\Phi^H_{l,J}\left(T_{l,J,n+1} - T_{l,J,n}\right)}_{\text{Enthalpy of layer } J} = \underbrace{\frac{\Phi^{K,A}_{l,j}}{2}\left(T_{l,j-1,n} - T_{l,j,n}\right) + \frac{\Phi^{K,A}_{l,j}}{2}\left(T_{l,j-1,n+1} - T_{l,j,n+1}\right)}_{\text{Conduction}} + \underbrace{F_{l,J}}_{\text{Lower flux}} \tag{3.3-30}$$

Collect terms and divide by $\Phi^H_{l,j}$, to get the matrix elements (Tables 4 and 5). For the interior layers, the matrix elements $\alpha_{1,j}$, $\beta_{1,j}$ and $\eta_{1,j}$ are independent of time.

Fig. 3-7 compares the sinusoidal, explicit, and implicit calculations (at large and small timesteps) for a bare spot at 9.5 AU with $A = 0.6$, $\varepsilon = 1$, and $\Gamma = 16000$ erg cm$^{-2}$ s$^{-1/2}$ K$^{-1}$ = 16 tiu, at latitude of 30°, a sub-solar latitude of 2.24°, $P = 22.6$ hours, and an hour angle at zero phase of -6 hours (-90°). Calculations were performed on a vertical grid with $\delta_{l,j} = \delta^A_{l,j} = \delta^B_{l,j} = 1/4$, except for the upper layer, where $\delta_{l,0} = 1/8$. At small time steps, the explicit and implicit calculations agree, and only the implicit calculations are plotted. The implicit temperatures calculated at large time steps are similar to calculations with small time steps. As expected from stability analysis (Haltiner & Williams 1984), the explicit temperatures with large time steps show large, unstable fluctuations after only two time steps, and are not plotted.

*Insert fig 3-7 here*



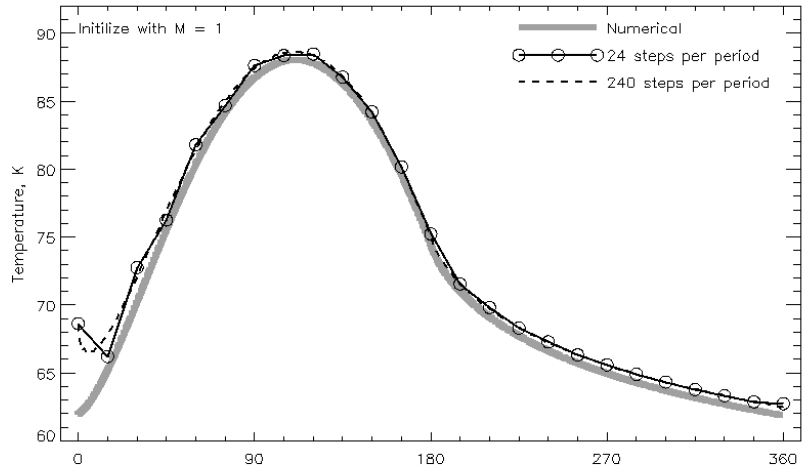

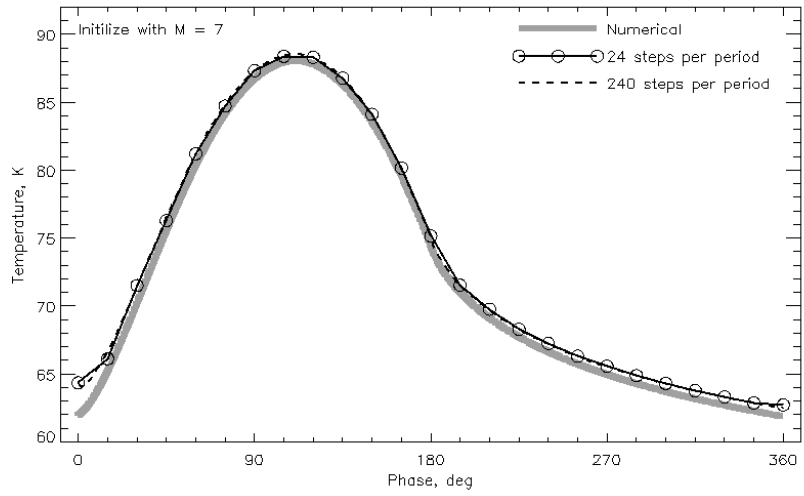

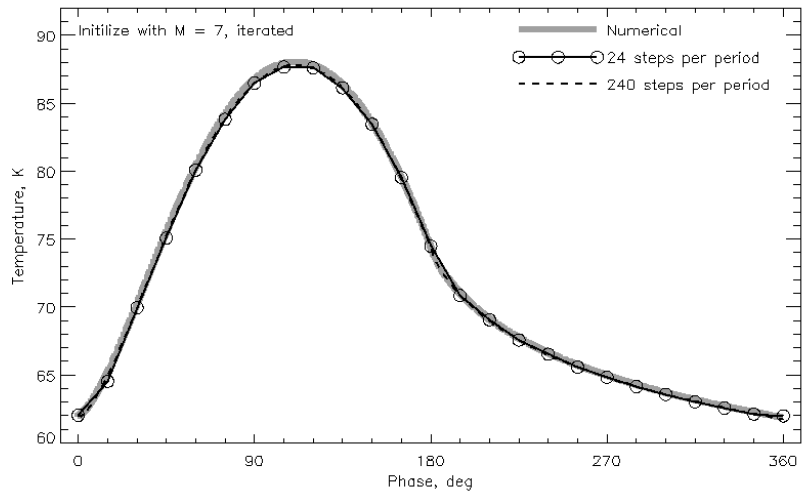



Fig 3-7.[1] Surface temperatures as a function of phase for the first period of numerical calculation, showing the benefit of a good initial condition. Top: Initial condition with $M = 1$ (one sinusoidal term), without an adjustment of $T_0$ to balance energy fluxes. Middle: Initial condition with $M = 7$ (7 sinusoidal terms), without an adjustment of $T_0$. Bottom: Initial condition with $M = 7$, with an adjustment of $T_0$ (see note after Eq. 3.2-10). All plots show the converged calculation calculated at 5000 time step per period for 20 periods (thick gray line), explicit calculations at 240 time steps per period for a single period (dashed line), and Crank-Nicholson implicit calculations at 24 time steps per period for a single period (open circles, solid line). The thin lines overlay the gray in places, and are difficult to see, indicating the quality of the calculation.

Fig. 3-7 also compares the effect of the choices of three of the initial conditions shown in Fig 3-3. For the simplest, the single frequency sine-wave ($M = 1$) with no adjustment to the mean temperature, the numerical answer agrees well with the converged answer within 60° rotational phase (Fig 3-8, top); the other initial conditions agree with the converged answer even more quickly (within one time step, for the $M=7$ case with adjusted mean temperature). The calculated temperatures for both $M = 1$ and $M=7$ are too warm at the end of one period if the mean temperature for the initial condition was not adjusted (Fig 3-7, top and middle), but reaches the proper temperature with adjustment (Fig 3-7, bottom). All three cases shown have a similar convergence rate. Most of the gain is in the first period, with subsequent periods improving the solution by 12-20% per period.

*3.4 Matrix operations for single or multiple bare locations (Areas I and III)*

*3.4a. Overview and explicit timesteps*

In this section, I present notes on how to solve the matrix equations in Figs. 3-5 and 3-6 in a way that takes advantage of the fact that for many problems substrate properties are often constant with time and location. I show how the implicit and explicit equations can be computed as a single matrix operation for those locations which share common substrate properties. This speeds calculation because it avoids "for-loop" constructions, with a speed savings that depends on the computer language involved. This section also shows how to

---

[1] `vty16_fig3_7a`
  `vty16_fig3_7b`
  `vty16_fig3_7c`



precompute the matrices associated with the substrate: both the elements for explicit calculations (the light gray elements in Fig 3-5, and the light-gray single-primed elements on the right-hand side of Fig 3-6), and the Lower-Upper (LU) decomposition of the matrix needed for implicit calculation (the light gray double-primed elements on the left-hand side of the equation in Fig 3-6.). Since LU decomposition is the first of the two steps needed in solving a tridiagonal matrix (Press et al. 1997), precomputing the LU decompositon of the substrate portion of the tridiagonal matrix cuts computation time roughly in half.

The key to these efficiencies is to separate the calculations for the uppermost layer ($j = 0$) from the lower layers ($j = 1$ to $J$). In addition to helping with the bare calculations, some of the notions introduced here will be required for implicit calculations of the interacting surfaces.

We separate the temperatures at a given location into a scalar describing the temperature of layer 0, $T_{l,0,n}$, and a row vector of length $J$ describing the temperatures of interior layers, $\mathbf{T}_{l,1..J,n}$:

$$\mathbf{T}_{l,1..J,n} = \left[T_{l,1,n}, \cdots, T_{l,J,n}\right]^T \qquad (3.4\text{-}1)$$

With this separation, the timestep for Areas I and III can be written for the explicit timestep (Fig 3-5) as

$$\begin{bmatrix} T_{l,0,n+1} \\ \mathbf{T}_{l,1..J,n+1} \end{bmatrix} = \begin{bmatrix} \eta_{l,0,n} & \mathbf{b}_{l,n} \\ \mathbf{a}_l & \mathbf{S}_l \end{bmatrix} \times \begin{bmatrix} T_{l,0,n} \\ \mathbf{T}_{l,1..J,n} \end{bmatrix} + \begin{bmatrix} \gamma_{l,0,n} \\ \mathbf{g}_l \end{bmatrix} \qquad (3.4\text{-}2)^1$$

which is displayed more graphically in Fig 3-8. $T_{l,0,n}$ and $T_{l,0,n+1}$ are the scalar initial and final temperatures in the top layer. $\mathbf{T}_{l,1..J,n}$ and $\mathbf{T}_{l,1..J,n+1}$ are $J$-element column vectors with the initial and final temperatures in layers 1 to $J$. This notation separates the time-varying matrix elements ($\eta_{l,0,n}$, $\mathbf{b}_{l,n}$, $\gamma_{l,0,n}$) from the ones that are constant with time ($\mathbf{a}_l$, $\mathbf{S}_l$, $\mathbf{g}_l$, ). $\mathbf{b}_{l,n}$ is a $J$-element row vector with one non-zero element

$$\mathbf{b}_{l,n} = \left[\beta_{l,0,n}, 0, \cdots, 0\right] \qquad (3.4\text{-}3)$$

$\mathbf{a}_l$, $\mathbf{S}_l$ and $\mathbf{g}_l$, are constant for each timestep, and so are not subscripted with $n$. $\mathbf{a}_l$ is a $J$-element column vector with one non-zero element.

$$\mathbf{a}_l = \left[\alpha_{l,1}, 0, \cdots, 0\right]^T \qquad (3.4\text{-}4)$$

---

[1] **vt3d_step_expl_1loc**, alpha_i, beta_0, beta_i, gamma_0, gamma_J, temp_0, temp_i



$\mathbf{S}_l$ is a $J \times J$ tridiagonal matrix whose $J$-1 lower elements are $[\alpha_{l,2}, \cdots \alpha_{l,J}]$; $J$ diagonal elements are $[\eta_{l,1}, \cdots \eta_{l,J}]$; and $J$-1 upper elements are $[\beta_{l,1}, \cdots \beta_{l,J-1}]$. $\mathbf{g}_l$ is a $J$-element column vector with one non-zero element:

$$\mathbf{g}_l = [0, \cdots, 0, \gamma_{l,J}]^T \qquad (3.4\text{-}5)$$

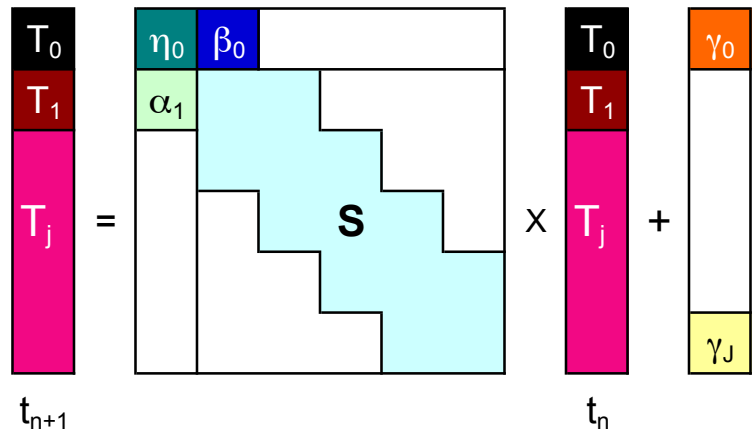

Fig 3-8. Graphical schematic of an explicit time-step from time *n* to time *n+1* for a single non-interacting location *l*, such as Area I (bare, isolated), Area II (volatile-covered, isolated) or III (bare, interacting); Eq. (3.4-2). Compare with Fig 3-5. To simplify the graphic, the time and location subscripts are dropped (e.g., $\eta_0$ for $\eta_{l,0,n}$). The temperature array is divided into the uppermost layer, $T_0$, the next lower layer, $T_1$, and remaining layers for $j = 2..J$, $T_j$. The elements of the substrate matrix $S$ consist of the three arrays $\alpha_{2..J}$, $\eta_{1..J}$, and $\beta_{1..J-1}$. Darker elements with white lettering correspond to the dark gray elements in Fig. 3-5, and change with each time step. Lighter elements with black lettering correspond to the light gray elements in Fig. 3-5, and are independent of time. White elements are zero.

Computation of Eq. (3.4-2) is displayed graphically in Fig 3-9. The uppermost temperature, $T_{l,0,n}$, is calculated by simple scalar arithmetic. The interior temperatures are calculated by matrix multiplication using a matrix that is likely to be time-independent, with additional terms added for $T_1$ and $T_J$.



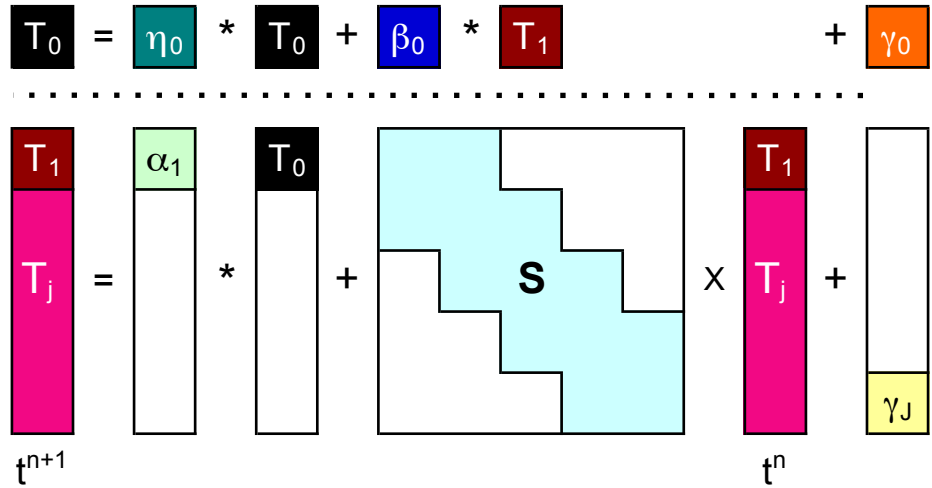

Fig 3-9. Graphical schematic of the implementation of an explicit time-step from time *n* to time *n+1* for a single non-interacting location); Eq. (3.4-2). Elements are labeled as in Fig 3-8. "*" indicates scalar multiplication (above the dotted line) or element-by-element multiplication of two arrays (below the dotted line). "X" indicates matrix multiplication.

In many applications, the substrate properties and internal heat flux are assumed to be constant over much of the body. In that case, in Eq. 3.4-2, the substrate arrays, $\mathbf{a}_l$ and $\mathbf{g}_l$, and the substrate matrices, $\mathbf{S}_l$, are independent of location $l$. In this case, it is particularly efficient to calculate an array of new temperatures in terms of old ones. If $\{L\} = \{l, m, \cdots\}$ (where $m$ is simply a second location index, not to be confused with the order of Fourier decomposition in Section 3.2) represents the set of locations which share a common $\mathbf{a}_{\{L\}}$, $\mathbf{S}_{\{L\}}$, and $\mathbf{g}_{\{L\}}$ (so that $\mathbf{a}_{\{L\}} = \mathbf{a}_l = \mathbf{a}_m = [\alpha_{\{L\},1}, 0, ..., 0]^T$ is a $J$ element column vector, $\alpha_{\{L\},1}$ is a scalar, $\mathbf{S}_{\{L\}} = \mathbf{S}_l = \mathbf{S}_m$ is a $J \times J$ element tridiagonal matrix, $\mathbf{g}_{\{L\}} = \mathbf{g}_l = \mathbf{g}_m = [\gamma_{\{L\},1}, 0, ..., 0]^T$ is a $J$ element column vector, and $\gamma_{\{L\},1}$ is a scalar), then we can write the temperatures in layer 0 as a row array of length $L$

$$\mathbf{T}_{\{L\},0,n} = \left[ T_{l,0,n}, T_{m,0,n}, \cdots \right] \quad (3.4\text{-}6)$$

and the temperatures in the interior layers 1 .. $J$ as a $J \times L$ matrix with $J$ rows and $L$ columns formed by the concatenation of $L$ temperature arrays of length $J$:

$$\mathbf{T}_{\{L\},1..J,n+1} = \left[ \mathbf{T}_{l,1..J,n+1}, \mathbf{T}_{m,1..J,n+1}, \cdots \right] \quad (3.4\text{-}7)$$

It is admittedly awkward that $\mathbf{T}_{\{L\},0,n}$ is an array, while $\alpha_{\{L\},1}$ is a scalar. I hope that context and Appendix A can help.



The surface temperatures are listed as a single 1-D array covering all the locations, rather than as a rectangular matrix of longitude and latitude. This is to simplify the matrix expressions of multiple locations. In addition, this allows for other divisions of the surface rather than simply a rectangular division, which tends to have needlessly small surface elements near the poles. Tiling schemes that maintain similar areas per tile need π/2 fewer tiles than equirectangular tiling schemes.

The new temperatures can be calculated in a way that takes advantage of array arithmetic:

$$\mathbf{T}_{\{L\},0,n+1} = [\eta_{l,0,n}, \eta_{m,0,n}, \cdots] \cdot \mathbf{T}_{\{L\},0,n} + [\beta_{l,0,n}, \beta_{m,0,n}, \cdots] \cdot \mathbf{T}_{\{L\},1,n} + [\gamma_{l,0,n}, \gamma_{m,0,n}, \cdots] \quad (3.4\text{-}8a)^1$$

$$\mathbf{T}_{\{L\},1..J,n+1} = \mathbf{S}_{\{L\}} \cdot \mathbf{T}_{\{L\},1..J,n} + \begin{Bmatrix} \alpha_{\{L\},1} T_{\{L\},0,n} \\ 0 \\ \vdots \\ 0 \\ \gamma_{\{L\},J} \end{Bmatrix} \quad (3.4\text{-}8b)^1$$

The computation represented by Eq. (3.4-8) is represented graphically in Fig 3-10, where Eq. (3.4-8a) is represented by the portion above the dotted line, and Eq. (3.4-8b) is represented by the portion below the dotted line.

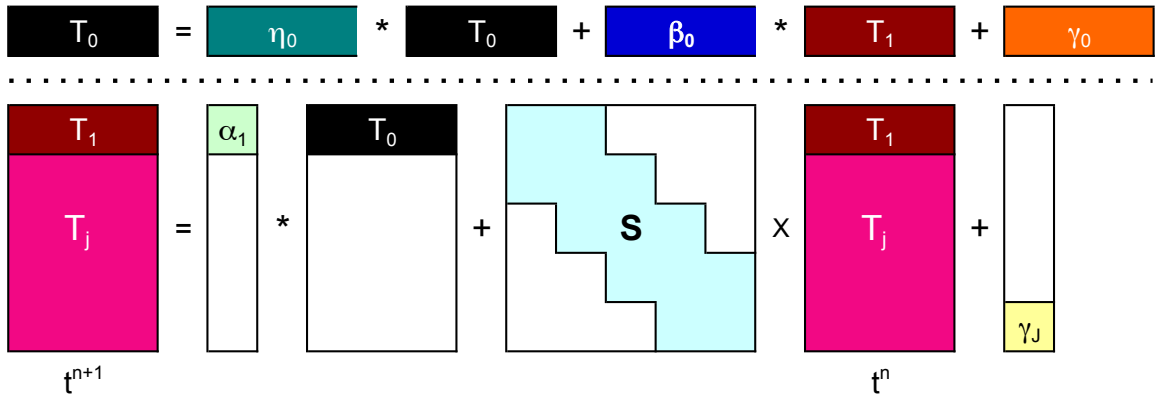

Fig 3-10. Graphical schematic of the implementation of an explicit time-step from time *n* to time *n+1* for multiple non-interacting locations (Eq. 3.4-8).

---

[1] **vt3d_step_expl_nloc**, alpha_i, beta_0, beta_i, gamma_0, gamma_J, temp_0, temp_i



Elements are labeled as in Fig 3-8. . "*" indicates element-by-element multiplication of two arrays (above the dotted line) or the multiplication of each row by a scalar (below the dotted line). "X" indicates matrix multiplication.

### 3.4b. Implicit timesteps

With the division of temperatures into layer 0 and layer 1 .. *J* in Eq. (3.4-1), the implicit timestep for a single location for Areas I and III (Fig 3-6) can be written as

$$\begin{bmatrix} \tilde{T}_{l,0,n} \\ \tilde{\mathbf{T}}_{l,1..J,n} \end{bmatrix} = \begin{bmatrix} \eta'_{l,0,n} & \mathbf{b}'_{l,n} \\ \mathbf{a}'_l & \mathbf{S}'_l \end{bmatrix} \begin{bmatrix} T_{l,0,n} \\ \mathbf{T}_{l,1..J,n} \end{bmatrix} + \begin{bmatrix} \gamma_{l,0,n} \\ \mathbf{g}_l \end{bmatrix} \qquad (3.4\text{-}9a)^1$$

$$\begin{bmatrix} \eta''_{l,0,n} & \mathbf{b}''_{l,n} \\ \mathbf{a}''_l & \mathbf{S}''_l \end{bmatrix} \begin{bmatrix} T_{l,0,n+1} \\ \mathbf{T}_{l,1..J,n+1} \end{bmatrix} = \begin{bmatrix} \tilde{T}_{l,0,n} \\ \tilde{\mathbf{T}}_{l,1..J,n} \end{bmatrix} \qquad (3.4\text{-}9b)^1$$

$T_{l,0,n}$, $\tilde{T}_{l,0,n}$ and $T_{l,0,n+1}$ are the (scalar) initial, intermediate, and final temperatures in the top layer. $\mathbf{T}_{l,1..J,n}$, $\tilde{\mathbf{T}}_{l,1..J,n}$, and $\mathbf{T}_{l,1..J,n+1}$ are the *J*-element column vectors with the initial, intermediate, and final temperatures in layers 1 to *J*. As with the explicit equation, this notation separates the time-varying matrix elements ($\eta'_{l,0,n}$, $\eta''_{l,0,n}$, $\mathbf{b}'_{l,n}$, $\mathbf{b}''_{l,n}$, $\gamma_{l,0,n}$) from the ones that are constant with time ($\mathbf{a}'_l$, $\mathbf{a}''_l$, $\mathbf{S}'_l$, $\mathbf{S}''_l$, $\mathbf{g}_l$). Similarly to Eq. (3.4-3), $\mathbf{b}'_{l,n}$ and $\mathbf{b}''_{l,n}$ are *J*-element row vectors with one non-zero element

$$\mathbf{b}'_{l,n} = [\beta'_{l,0,n}, 0, \cdots 0] \qquad (3.4\text{-}10a)$$

$$\mathbf{b}''_{l,n} = [\beta''_{l,0,n}, 0, \cdots 0] \qquad (3.4\text{-}10b)$$

---

[1] **vt3d_step_cn_1loc**, alpha_i, beta_0, beta_i, gamma_0, gamma_J,
    temp_0, temp_i, which calls:
  **vt3d_step_impl_1loc**, alpha_i, beta_0, beta_i, gamma_0, gamma_J,
    temp_0, temp_i

The lower-upper (LU) decomposition of the S'' matrix can be computed before the time steps with:

  spp_tridc = **vt3d_cn_tridc**(alpha_i, beta_i, spp_indx)

in which case use

  **vt3d_step_cn_trisol_1loc**, alpha_i, beta_0, beta_i, gamma_0, gamma_J,
    spp_tridc, spp_indx, temp_0, temp_i, which calls:
  **vt3d_step_impl_trisol_1loc**, alpha_1, beta_0, gamma_0, gamma_J,
    spp_tridc, spp_indx, temp_0, temp_i



$\mathbf{a}'_l$ and $\mathbf{a}''_l$ are *J*-element column vectors with one non-zero element (compare Eq. 3.4-4). Since they are constant with time, they are not subscripted with *n*.

$$\mathbf{a}'_l = [\alpha'_{l,1}, 0, \cdots 0]^T \tag{3.4-11a}$$

$$\mathbf{a}''_l = [\alpha''_{l,1}, 0, \cdots 0]^T \tag{3.4-11b}$$

$\mathbf{S}'_l$ and $\mathbf{S}''_l$ are $J \times J$ tridiagonal matrices, also constant with time, whose *J*-1 lower elements are $[\alpha'_{l,2}, \cdots \alpha'_{l,J}]$ and $[\alpha''_{l,2}, \cdots \alpha''_{l,J}]$; *J* diagonal elements are $[\eta'_{l,1}, \cdots, \eta'_{l,J}]$ and $[\eta''_{l,1}, \cdots, \eta''_{l,J}]$; and *J*-1 upper elements are $[\beta'_{l,1}, \cdots, \beta'_{l,J-1}]$ and $[\beta''_{l,1}, \cdots, \beta''_{l,J-1}]$ respectively. $\mathbf{g}_l$ is a *J*-element column vector with one non-zero element, defined the same as Eq. (3.4-5). The first half of the implicit calculation, Eq. (3.4-9a), is computed in the same way as the explicit calculation, Eq. (3.4-2). The second half of the implicit calculation, Eq. (3.4-9b) is a tridiagonal linear problem, shown graphically in Fig 3-11, and can be solved in time (*J*+1).

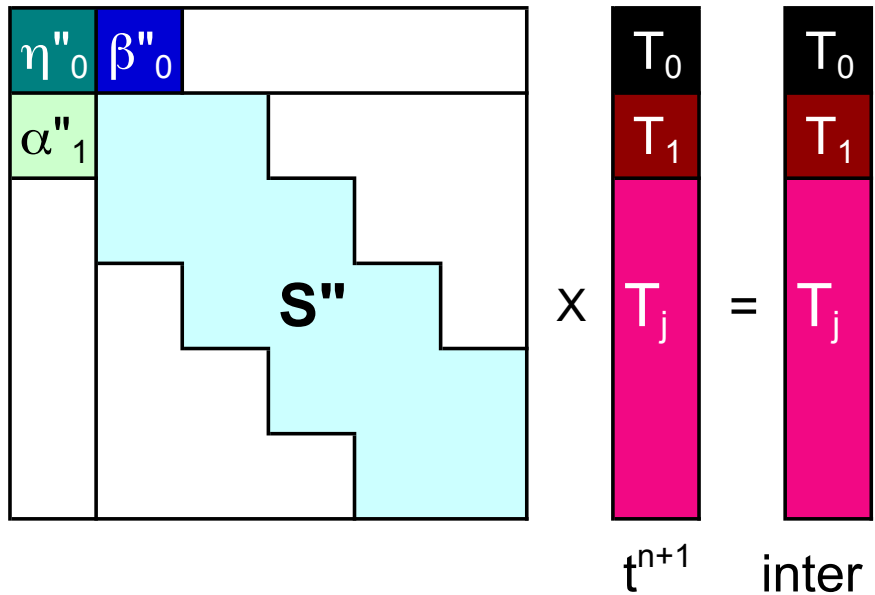

Fig 3-11. Graphical schematic of an implicit time-step from an intermediate temperature to the temperature at time *n+1* for a single non-interacting location, with local energy balance, such as Area I (bare, isolated), Area II (volatile-covered, isolated) or III (bare, interacting); Eq. (3.4-9b). Compare with Fig 3-6. The temperature array is divided into the uppermost layer, $T_0$, the next lower layer, $T_1$, and remaining layers for $j = 2..J$, $T_J$. The elements of the substrate matrix $S''$ consist of the three arrays $\alpha''_{2..J}$, $\eta''_{1..J}$, and $\beta''_{1..J-1}$. Darker elements with white lettering correspond to the dark gray elements in



Fig. 3-6, and change with each time step. Lighter elements with black lettering correspond to the light gray elements in Fig. 3-7, and are independent of time. White elements are zero.

We treat Eq. (3.4-9b) as a banded tridiagonal matrix to take advantage of the fact that the terms $\mathbf{a}''_l$ and $\mathbf{S}''_l$ are constant with time. This is a special case of inversion by partitioning, whose solution is presented in Press et al. (2007; section 2.7.4). A similar problem was addressed by Xing-Bo (2009). This allows us to precompute the lower-upper (LU) decomposition of $\mathbf{S}''_l$. The solution to Eq. (3.4-9b) can be written by defining two column vectors $\mathbf{y}_l$ and $\mathbf{z}_{l,n}$ of length $J$, and two scalars $c_{l,n}$ and $d_{l,n}$:

$$\mathbf{y}_l = \mathbf{S}''^{-1}_l \times \mathbf{a}''_l \tag{3.4-12a}$$

$$\mathbf{z}_{l,n} = \mathbf{S}''^{-1}_l \times \tilde{\mathbf{T}}_{l,1..J,n} \tag{3.4-12b}$$

$$c_{l,n} = \mathbf{b}''_{l,n} \cdot \mathbf{y}_l = \beta''_{l,0,n} y_{l,0} \tag{3.4-12c}$$

$$d_{l,n} = \mathbf{b}''_{l,n} \cdot \mathbf{z}_{l,n} = \beta''_{l,0,n} z_{l,0,n} \tag{3.4-12d}$$

with which the temperatures at the next time step for location $l$ are

$$T_{l,0,n+1} = \frac{\tilde{T}_{l,0,n} - d_{l,n}}{\eta''_{l,0} - c_{l,n}} \tag{3.4-13a}$$

$$\mathbf{T}_{l,1..J,n+1} = \mathbf{z}_{l,n} - T_{l,0,n+1} \mathbf{y}_l \tag{3.4-13b}$$

This solution can be confirmed by direct substitution of Eqs. (3.4-13a,b) into Eq. (3.4-9b). The solution is shown graphically in Fig 3-12. Note that the only the time-independent substrate matrix needs to be inverted, and this can be done at the start of the computation, rather than for each time step. Furthermore, the array $y$ is also independent of time.



$$\begin{pmatrix} y_0 \\ y_j \end{pmatrix} = S''^{-1} \times \begin{pmatrix} \alpha''_1 \\ \end{pmatrix}$$

$$\begin{pmatrix} z_0 \\ z_j \end{pmatrix} = S''^{-1} \times \begin{pmatrix} T_1 \\ T_j \end{pmatrix}_{inter}$$

$$T_0 \big|_{t^{n+1}} = \frac{T_0 - \beta''_0 * z_0}{\eta''_0 - \beta''_0 * y_0}$$

$$\begin{pmatrix} T_1 \\ T_j \end{pmatrix}_{t^{n+1}} = \begin{pmatrix} z_0 \\ z_j \end{pmatrix} - T_0\big|_{t^{n+1}} * \begin{pmatrix} y_0 \\ y_j \end{pmatrix}$$



Fig 3-12. Graphical schematic of the solution to the banded tridiagonal matrix for a single location. "*" indicates scalar multiplication (above the lowest dotted line) or element-by-element multiplication of an array by a scalar (below the dotted line). "X" indicates matrix multiplication.

For those locations with the same substrate properties (so that $\mathbf{S}''_{\{L\}} = \mathbf{S}''_l = \mathbf{S}''_m \cdots$ and $\mathbf{a}''_{\{L\}} = \mathbf{a}''_l = \mathbf{a}''_m \cdots$), the solution can be calculated for several locations simultaneously, as with the explicit scheme. Define the intermediate temperatures in layer 0 as a row vector of length $L$

$$\tilde{\mathbf{T}}_{\{L\},0,n} = \left[\tilde{T}_{l,0,n}, \tilde{T}_{m,j,n}, \cdots \right] \quad (3.4\text{-}14)$$

and the intermediate temperatures in the interior layers 1 .. $J$ as a $J \times L$ matrix:

$$\tilde{\mathbf{T}}_{\{L\},1..J,n} = \left[\tilde{\mathbf{T}}_{l,1..J,n}, \tilde{\mathbf{T}}_{m,1..J,n}, \cdots \right] \quad (3.4\text{-}15)$$

Define $J$ column vector $\mathbf{y}_{\{L\}}$ (the same for all locations in $\{L\}$ and independent of time, so that $\mathbf{y}_{\{L\}} = \mathbf{y}_l = \mathbf{y}_m$), a $J \times L$ matrix $\mathbf{Z}_{\{L\},n}$ and row vectors $\mathbf{c}_n$, $\mathbf{d}_n$, and $\mathbf{h}''_{0,n}$ of length $L$:

$$\mathbf{y}_{\{L\}} = \mathbf{y}_l = \mathbf{y}_m \cdots = \mathbf{S}''^{-1}_{\{L\}} \times \mathbf{a}''_{\{L\}} \quad (3.4\text{-}16\text{a})$$

$$\mathbf{Z}_{\{L\},n} = \left[\mathbf{z}_{l,n}, \mathbf{z}_{m,n}, \cdots \right] = \mathbf{S}''^{-1}_{\{L\}} \tilde{\mathbf{T}}_{\{L\},1..J,n} \quad (3.4\text{-}16\text{b})$$

$$\mathbf{c}_n = \left[c_{l,n}, c_{m,n}, \cdots \right] \quad (3.4\text{-}16\text{c})$$

$$\mathbf{d}_n = \left[d_{l,n}, d_{m,n}, \cdots \right] \quad (3.4\text{-}16\text{d})$$

$$\mathbf{h}''_{0,n} = \left[\eta''_{l,0,n}, \eta''_{m,0,n}, \cdots \right] \quad (3.4\text{-}16\text{e})$$

The new temperatures are then



$$\mathbf{T}_{\{L\},0,n+1} = \frac{\tilde{\mathbf{T}}_{\{L\},0,n} - \mathbf{d}_n}{\mathbf{h}''_{0,n} - \mathbf{c}_n} \quad (3.4\text{-}17a)\,[1]$$

$$\mathbf{T}_{\{L\},1..J,n+1} = \mathbf{Z}_{\{L\},n} - \mathbf{y}_{\{L\}} \times \mathbf{T}_{\{L\},0,n+1} \quad (3.4\text{-}17b)$$

where $\mathbf{y}_{\{L\}} \times \mathbf{T}_{\{L\},0,n+1}$ is an outer product of a *J*-length column vector and an *L*-length row vector, yielding a $J \times L$ matrix obtained by

$$\mathbf{y}_{\{L\}} \times \mathbf{T}_{\{L\},0,n+1} = \left[ T_{l,0,n+1}\mathbf{y}_{\{L\}}, T_{m,0,n+1}\mathbf{y}_{\{L\}}, \cdots \right] \quad (3.4\text{-}18)$$

The graphical schematic is shown in Fig 3-13.

---

[1] **vt3d_step_cn_nloc**, alpha_i, beta_0, beta_i, gamma_0, gamma_J,
    temp_0, temp_i, which calls:

  **vt3d_step_impl_nloc**, alpha_i, beta_0, beta_i, gamma_0, gamma_J,
    temp_0, temp_i

The lower-upper (LU) decomposition of the S'' matrix can be computed before the time steps with:

  spp_tridc = **vt3d_cn_tridc**(alpha_i, beta_i, spp_indx)

in which case use

  **vt3d_step_cn_trisol_nloc**, alpha_i, beta_0, beta_i, gamma_0, gamma_J,
      spp_tridc, spp_indx, temp_0, temp_i, which calls:

  **vt3d_step_impl_trisol_nloc**, alpha_1, beta_0, gamma_0, gamma_J,
      spp_tridc, spp_indx, temp_0, temp_i



Fig 3-13. Graphical schematic of the solution to the banded tridiagonal matrix for mutliple locations. "*" indicates element-by-element multiplication of two arrays (above the lowest dotted line). "X" indicates matrix multiplication



(equivalent to the outer product of two arrays for the multiplication below the lowest dotted line).

## *3.5 Example: Mimas*

As a worked example, Fig 3-14 shows the surface temperatures on Mimas, following Howett et al. (2011). Most of Mimas (Region 1) has a thermal inertia of 9 tiu and a bond albedo of 0.6, while the anomaly (Region 2) has a high thermal inertia of 66 tiu and a bond albedo of 0.59. The three snapshots are for sub-solar west longitudes of 167, 87, and 43°, and the dashed lines indicate the visible surface (30° from the limb) for sub-spacecraft west longitudes of 147, 180, and 83°. The color bar and scale are chosen to allow direct comparison with Howett et al. (2011), their Figure 2. The temperatures were calculated on a grid of 45 latitude by 90 longitude bins, each of which have their own thermal inertia. The time step was once per degree of sub-solar longitude.



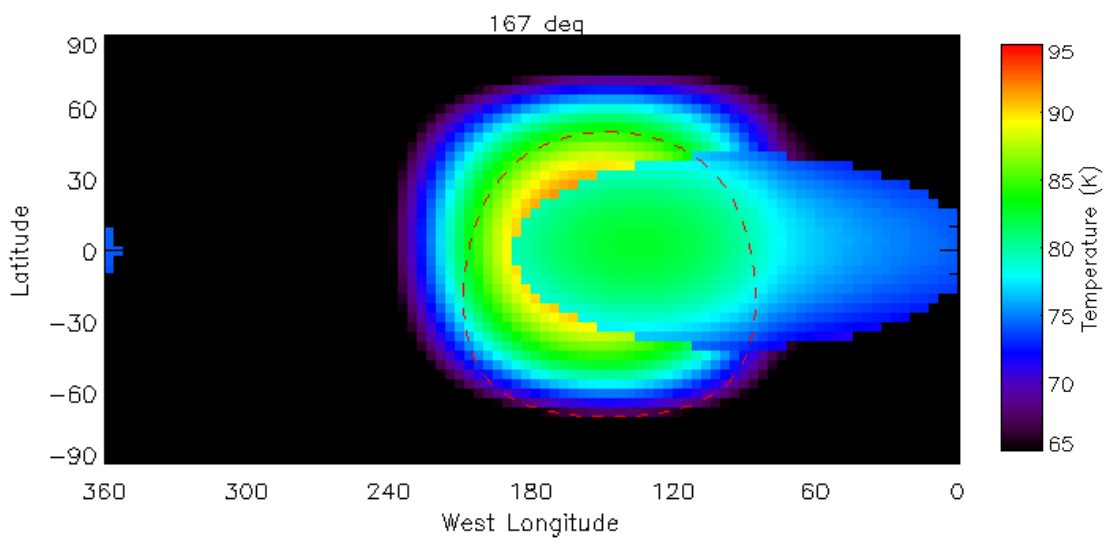
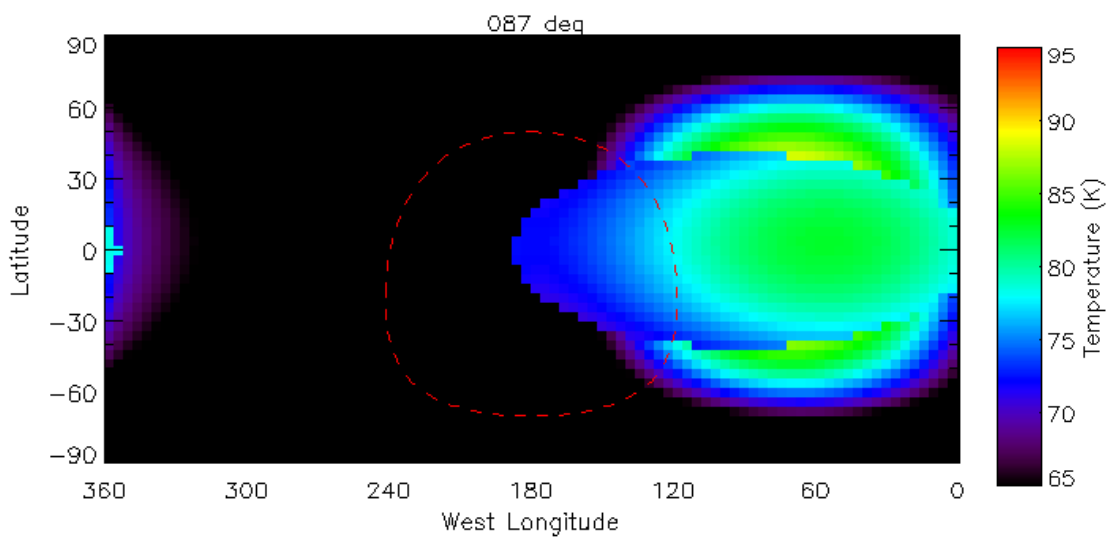
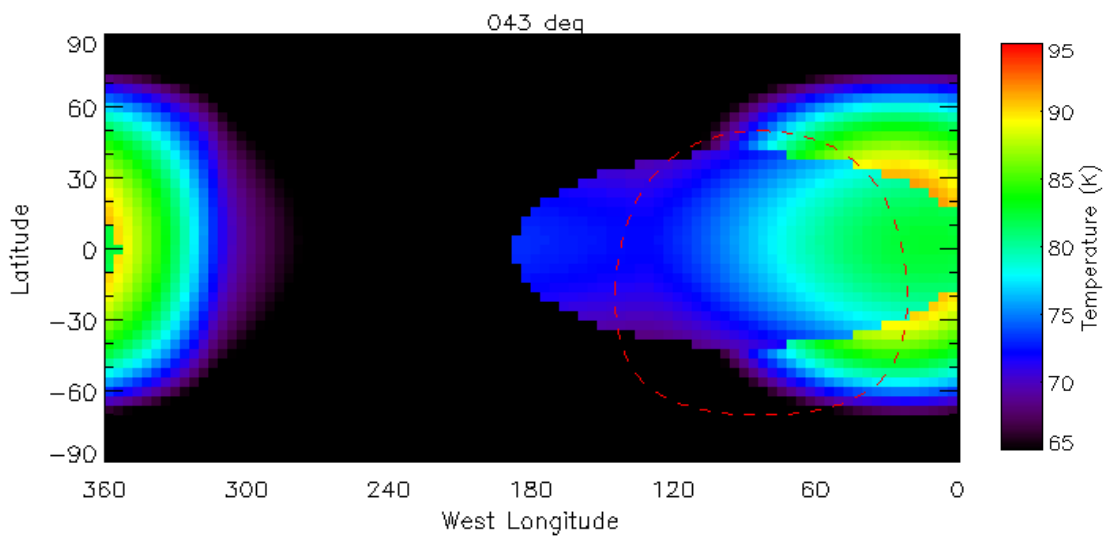



Fig 3-14.[1] Model of Mimas's global temperature at selected subsolar longitudes, following the thermophysical properties derived from Howett et al. (2011). The dashed line represents the visible area of Mimas that correspond to three particular observations of Mimas from Cassini (See Howett et al. 2011, their Fig 2).

## 4. VT3D for local volatile-covered locations (Area II)

In this section, I consider locations that have volatiles on their surfaces, but for which the energy balance is essentially local. For worlds where the surface pressure is too low to effectively transport volatiles over the surface, the transport of energy, through latent heat of sublimation and deposition, does not effectively influence on the surface temperatures. This is the case on Io, and almost certainly the case on the large volatile-covered Kuiper-belt objects when far from perihelion. These are the isolated, volatile-covered areas (Area II) in Fig 2-2.

Within the substrate, the physics of thermal conduction and the lower boundary condition for the volatile covered locations (Area II) is identical as for the bare locations (Areas I and III, Section 3), and will not be repeated here. At the surface, on the other hand, the energy equation contains two new terms, one related to the energy needed to heat the volatile slab, and another related to latent heat exchange between the surface and the local gas column via deposition and sublimation. The continuous form is discussed in Section 4.1 and analytic expression for an initial condition is discussed in Section 4.2. Because the energy equations are strictly local, the form of the numerical implementation is very similar to that in Section 3. Only the form of the matrix elements $\eta_0$ and $\beta_0$ change, as discussed in Section 4.3.

*4.1 Analytic expressions for isolated volatile-covered locations (Area II)*

The energy equation at the surface balances net heating or crystalline phase changes with absorbed sunlight, thermal emission, thermal conduction, and latent heat of sublimation/condensation. The total energy equation is

$$\underbrace{m_V \frac{\partial H_V}{\partial t}}_{\text{Enthalphy of volatile slab}} = \underbrace{S}_{\text{Insolation}} - \underbrace{\varepsilon \sigma T^4}_{\text{Emission}} - \underbrace{k \frac{\partial T}{\partial z}\bigg|_{z=0}}_{\text{Conduction}} + \underbrace{L_S \frac{\partial m_V}{\partial t}}_{\text{Latent heat}} \qquad (4.1\text{-}1)$$

---

[1] `vty16_fig3_14`



where $m_V$ is the mass per area of the volatile slab, $\partial H_V/\partial t$ is the time derivative of the enthalpy of the volatile slab in energy per mass (equal to $c_V \partial T/\partial t$ if there is no phase change, see Eq. 4.1-2, where $c_V$ is the specific heat of the volatile slab. Note $c_V$ is subscripted $V$ for volatile, not $V$ for constant volume), and $L_S$ is the latent heat of sublimation. $L_S$ is subscripted with $S$ to distinguish it from the latent heat of crystalline phase change ($L_C$) and or the number of discrete locations on the surface ($L$, Section 3.3).

At the surface, a volatile slab is assumed to be isothermal within its vertical extent (See Fig 4.1), with a temperature equal to that at the top of the substrate. As described in Paper I, the isothermal slab was assumed in Hansen and Paige (1992) and Hansen and Paige (1996). This has been justified (David Paige, personal communication) by assuming that if the slab porous, it is in contact with the local atmosphere and the gas can isothermalize the solid; conversely, if the slab is not porous (e.g., from annealing, Eluszkiewicz et al., 1998) then its conductivity will be high, helping to isothermalize a thin enough slab. For very thick deposits, such as the suspected $N_2$ reservoir seen on Pluto, one approach is to keep track of mass per area of the volatiles available for sublimation as a separate quantity from the mass per area that is isothermalized (Young et al., 2016). Layering within the volatile slab will be treated in a later paper.

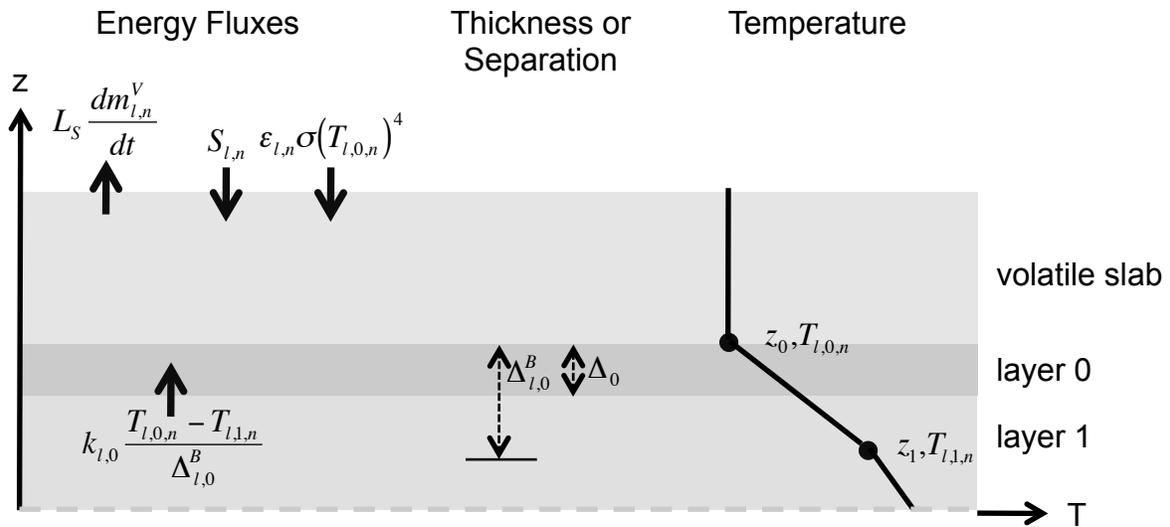

Fig. 4-1. Schematic of the layering scheme and energy fluxes for layer $j = 0$, Area II (volatile slab, local).

The insolation, emission, and conduction terms are identical to those in Eq. (3.1-3). The first term of Eq. (4.1-1) describes the change in the enthalpy per area of the volatile slab, which depends on the volatile-ice temperature and crystalline ice phase. Away from the



temperature of a crystalline phase transition, the derivative of $H_V$ with respect to $T$ at constant pressure equals $c_V$, the specific heat of the volatile slab, Eq. (4.1-2a). Adding energy to the slab raises its temperature. At the temperature of a crystalline ice phase transition, the latent heat equals the difference in $H_V$ between two phases ($L_C$); adding energy to the slab converts ice from the low-temperature to the high-temperature phase without changing the temperature. This gives:

$$\frac{\partial H_V}{\partial t} = c_V \frac{\partial T}{\partial t}, \quad T \neq T_C \tag{4.1-2a}$$

$$\frac{\partial H_V}{\partial t} = L_C \frac{\partial X}{\partial t}, \quad T = T_C \tag{4.1-2b}$$

where $T_C$ is the temperature of a crystalline phase transition, $L_C$ is the latent heat of crystalline phase change, and $X$ is the mass fraction of the high-temperature phase. If $c_V$ is treated as a constant, then we can write $H_V = c_V T + L_C X$, which is proportional to the "pseudo temperature" used by John Spencer (personal communication).

Tracking the enthalpy of the slab, rather than its temperature, was introduced because $N_2$ has a reversible transition between the α and β phase at 35.6 K (e.g., Scott 1976), a relevant temperature for Pluto, Triton, and elsewhere in the outer solar system. Some volatile ices have no solid-state phase transitions at relevant temperatures, which simplifies matters. Others have multiple transitions, or non-reversible transitions. In all cases, the enthalpy is the general quantity that can account for phases as well as temperatures, and Eq. 4.1-2b represents the "special case" of enthalpy change at a phase transition temperature.

Area II satisfies local energy and mass balance. Assuming negligible horizontal transport of mass, any mass lost by the atmosphere either condenses or escapes.

$$\frac{dm_V}{dt} + \frac{dm_A}{dt} + E = 0 \tag{4.1-3}$$

where $m_A$ is the mass per area of the atmosphere, and $E$ is the escape rate in mass per area per time. Negative values of $E$ can be used to account for injection into the atmosphere from non-sublimation sources such as geysers (see Paper I). If the atmosphere is in vapor-pressure equilibrium with the surface, then the mass of the atmosphere is a function only of the surface pressure and effective gravity (defined by $g = p_S/m_A$, which is smaller than the surface gravity for extended atmospheres by a factor of $1 - 2H/R$, where $H$ is the scale height and $R$ is the surface radius, see Paper I):



$$\frac{dm_V}{dt} = -\frac{1}{g}\frac{dp_S(T)}{dT}\frac{dT}{dt} - E \qquad (4.1\text{-}4)$$

where $p_S(T)$ is the equilibrium vapor pressure at temperature $T$. The pressure derivative in Eq. (4.1-4) can be evaluated using the Clausius-Clapeyron relation,

$$\frac{dp_S(T)}{dT} = \frac{L_S m_{molec} p_S}{k_B T^2} \qquad (4.1\text{-}5)$$

where $m_{molec}$ is the mass of one molecule and $k_B$ is Bolzmann's constant. Substituting Eqs. (4.1-2a), (4.1-2b), (4.1-3), and (4.1-4) into Eq. (4.1-1) and collecting like terms yields:

$$\left(\underbrace{m_V c_V}_{\text{Enthalpy of volatile slab}} + \underbrace{\frac{L_S}{g}\frac{dp_S(T)}{dT}}_{\substack{\text{Latent heat}\\\text{of sublimation}}}\right)\frac{\partial T}{\partial t} = \underbrace{S}_{\text{Insolation}} - \underbrace{\varepsilon\sigma T^4}_{\text{Emission}} - \underbrace{k\frac{dT}{dz}\bigg|_{z=0}}_{\text{Conduction}} - \underbrace{L_S E}_{\text{Latent heat of escape}} \;,\; T \ne T_C \qquad (4.1\text{-}6a)$$

$$\underbrace{m_V L_C \frac{dX}{dt}}_{\text{Enthalpy of volatile slab}} = \underbrace{S}_{\text{Insolation}} - \underbrace{\varepsilon\sigma T^4}_{\text{Emission}} - \underbrace{k\frac{dT}{dz}\bigg|_{z=0}}_{\text{Conduction}} - \underbrace{L_S E}_{\text{Latent heat of escape}} \;,\; T = T_C \qquad (4.1\text{-}6b)$$

Eq (4.1-6a) is strikingly similar to the equivalent equation for the bare areas (3.1-3), differing only by the inclusion of the enthalpy and latent heat terms on the left-hand side, and the latent heat of the escaping atmosphere on the right side. The enthalpy and latent heat of sublimation introduce terms proportional to the frequency, ω, in the analytic equations (Section 4.2). They also introduce two additional terms to the total expression for the change in energy flux per temperature for the upper-most layer ($\Phi_{l,n}^T$) in the numeric solutions (Section 4.3), but the form of the matrix equations is unchanged. When there is a phase change, (4.1-6b), the analytic and numeric forms are both simpler, as the temperature does not change with time.

*4.2 Analytic approximation and initialization for isolated volatile-covered areas (Area II)*

As in Section 3.2, an analytic form of the continuous equations (Eq. 4.1-6a, b) can be found by decomposing the solar insolation and temperature into a sum of sinusoidal terms of frequency ω (Eqs. 3.2-1, 3.2-7). Additionally, we specify that the temperature of the volatile slab equals the substrate temperature at the substrate-slab interface



$$T_V(\lambda,\phi,t) = T(\lambda,\phi,z=0,t) \tag{4.2-1}$$

The escape rate is decomposed into a sum of sinusoidal terms in an analogous manner to the solar forcing

$$E(\lambda,\phi,t) = \text{Re}\left[\sum_{m=0}^{M} \hat{E}_m(\lambda,\phi)e^{im\omega t}\right] \tag{4.2-2}$$

where $\omega = 2\pi/P$ is the frequency of the diurnal or seasonal forcing, and $\hat{E}_m$ is the complex sinusoidal coefficient (the complexity is indicated by the hat).

As in Section 3.2, the average temperature is found by substituting the sinusoidal forms of *S* and *T* into Eqs. (4.1-6a, b) and taking the first-order, time-averaged component.

$$0 = \underbrace{\hat{S}_0(\lambda,\phi)}_{\text{Insolationn}} - \underbrace{\varepsilon\sigma\left(\hat{T}_0(\lambda,\phi)\right)^4}_{\text{Emission}} + \underbrace{F(\lambda,\phi)}_{\text{Flux at lower boundary}} - \underbrace{L_S E_0(\lambda,\phi)}_{\text{Latent heat of escaping gas}} \tag{4.2-3}[1]$$

As in Section 3.2, the temperature coefficients, $\hat{T}_m$, are found by substituting the periodic functions into Eq. (4.1-6), and taking only those terms proportional to exp(*imωt*). For simplicity, if $T \neq T_C$, then we assume that the temperature does not cross a crystalline phase boundary in the expansion. In addition to the expressions for the temperature dependence of conducted and emitted energy flux, $\Phi_S$ and $\Phi_E$ (Eq. 3.2-9a, b), I define two new variables:

$$\Phi_V(T) = \omega m_V c_V \tag{4.2-4a}[2]$$

$$\Phi_A(T) = \omega \frac{L_S}{g} \frac{dp_S}{dT_V} \tag{4.2-4b}[3]$$

$\Phi_V$ is simply related the to specific heat per area of the volatile slab, being the energy per degree per area. $\Phi_A$ is related to the energy needed for the atmosphere to vary its column mass (atmospheric "breathing"). If the surface temperature rises, the equilibrium pressure rises too. The column mass of the equilibrium atmosphere increases due to sublimation from the surface. This takes energy, through the latent heat of sublimation. The result is that the specific heat of the volatile slab and the atmospheric "breathing" delay and decrease the thermal response (Paper I). The resulting expansion of 4.1-6a is:

---

[1] `temp_0 = `**`vt3d_temp_term0_local`**`(sol_0, flux_int, emis, latheat, mflux_esc)`

[2] `phi_v = `**`vt3d_dfluxdtemp_slab`**`(freq, mass_0, specheat)`

[3] `phi_a = `**`vt3d_dfluxdtemp_atm`**`(freq, temp_v, frac_varea, gravacc, name_species)`



$$\left[\underbrace{\sqrt{im}\Phi_S(T_0)}_{\text{Conduction}}+\underbrace{\Phi_E(T_0)}_{\text{Emission}}+\underbrace{im\Phi_V(T_0)}_{\text{Enthalpy of volatile slab}}+\underbrace{im\Phi_A(T_0)}_{\text{Latent heat}}\right]\hat{T}_m$$
$$=\underbrace{\hat{S}_m}_{\text{Insolation}}-\underbrace{L_s\hat{E}_m}_{\text{Latent heat of escaping gas}}\quad,\ T\neq T_C \quad (4.2\text{-}5)^1$$

If the equilibrium temperature is at a crystalline phase boundary, then the corresponding equation for the change in the slab's state is

$$\underbrace{i\omega m m_V L_C \hat{X}_m}_{\text{Enthalpy of volatile slab}}=\underbrace{\hat{S}_m}_{\text{Insolation}}-\underbrace{L_s\hat{E}_m}_{\text{Latent heat of escaping gas}}\quad,\ T=T_C \quad (4.2\text{-}6)$$

As described in Paper I, we can define non-dimensional thermal parameters, analogous to the thermal parameter of Spencer et al. (1989), to quantify the importance of heating of the volatile slab and atmospheric breathing. The substrate thermal parameter, $\Theta_S$, is defined in Eq. 3.2-11. Two new parameters are:

$$\Theta_V(T_0)=\frac{\Phi_V(T_0)}{\Phi_E(T_0)/4} \quad (4.2\text{-}7)$$

$$\Theta_A(T_0)=\frac{\Phi_A(T_0)}{\Phi_E(T_0)/4} \quad (4.2\text{-}8)$$

Substituting into Eq. (4.2-5) shows how the amplitude and phase of the thermal response depends on the thermal inertia, the specific heat and depth of the volatile slab, and the extent of the atmospheric "breathing."

$$\hat{T}_m=\frac{\hat{S}_m-L_S\hat{E}_m}{\Phi_E(T_0)}\frac{4}{4+\sqrt{im}\Theta_S+im\Theta_V+im\Theta_A} \quad (4.2\text{-}9)$$

As for the bare areas (Areas I and III), the expansion can be written in terms of the emitted thermal flux in the case of large temperature variations, giving

---

[1] `temp_terms = `**`vt3d_temp_terms_local`**`(sol_terms,flux_int,emis,freq,`
`    therminertia, is_volatile, mass_volatile, specheat_volatile,`
`    gravacc, name_species)`

`temp_terms = `**`vt3d_temp_terms_local_iter`**`(sol_terms,flux_int,emis,freq,`
`    therminertia, thermcond, is_volatile, mass_volatile,`
`    specheat_volatile, gravacc, name_species)`



$$\hat{F}^E_m = \left(\hat{S}_m - L_S \hat{E}_m\right) \frac{4}{4 + \sqrt{im\Theta_S + im\Theta_V + im\Theta_A}} \quad (4.2\text{-}10)^1$$

*4.3 Numerical solution for isolated volatile-covered areas (Area II)*

The discretization for the interior layers ($j = 1..J–1$) and the lowest layer ($j = J$) is the same for the isolated, volatile-covered locations (Area II) as it is for the bare locations (Areas I and III). The discretization for the volatile slab and the upper two layers are shown in Fig 4-1. Although the physics is different in the presence of a volatile, the numerics are nearly identical for all calculations on a local level, whether volatiles are present or not.

First consider usual case where the volatile slab is not at a crystalline phase transition temperature. As with Areas I and III, to find the energy balance in layer 0, integrate the conduction equation (Eq. 3.1-2) over the top layer, from $z = -\Delta_0$ to $z = 0$. Add this to the energy balance equation (Eq. 4.1-6a) to get Eq. (4.3-1), the volatile-covered equivalent to Eq. (3.3-1):

$$\underbrace{\overline{\rho_{l,0} c_{l,0} \int_{-\Delta_0}^{0} \frac{\partial T}{\partial t} dz}}_{\text{Enthalpy, layer 0}} + \underbrace{\overline{m^V_{l,n} c^V_l \frac{\partial T}{\partial t}}}_{\text{Enthalpy, volatile slab}} + \underbrace{\overline{\frac{L_S}{g} \frac{dp_s(T)}{dT} \frac{dT}{dt}}}_{\text{Latent heat, volatile slab}} = \\ \underbrace{\overline{S_{l,n'}}}_{\text{Insolation}} - \underbrace{\overline{(\varepsilon_{l,0,n} \sigma T^4)}}_{\text{Emission}} - \underbrace{\overline{\left(k \frac{dT}{dz}\bigg|_{z=-\Delta_0}\right)}}_{\text{Conduction}} - \underbrace{\overline{L_S E_{l,n}}}_{\text{Latent heat, escape}} \quad (4.3\text{-}1)$$

where the overbar indicates the time-averaged value over the time step $t_n$ to $t_{n+1}$.

The enthalpy of layer 0, insolation, emission, and conduction are the same as for Areas I and III (Section 3.3).

The second term in Eq (4.3-1) reflects the change in the enthalpy of the volatile slab with temperature. The volatile slab mass, $m^V_{l,n}$, can change over the time interval. However, this change is going to be small unless the slab is about to completely sublime, in which case this term contributes little. Ignoring the change in volatile slab mass during the time interval, this term becomes:

---

[1] `flux_terms = `**`vt3d_eflux_terms_local`**`(sol_terms, flux_int, emis, freq, therminertia, is_volatile, mass_volatile, specheat_volatile, gravacc, name_species, term_terms=temp_terms)`



$$\overline{m_{l,n}^V c_l^V \frac{\partial T}{\partial t}} \approx \Phi_{l,n}^V \left( T_{l,0,n+1} - T_{l,0,n+1} \right) \tag{4.3-2}$$

where $c_l^V$ is the specific heat of the volatile slab at location $l$, $\Phi_{l,n}^V$ has units of erg cm$^{-2}$ s$^{-1}$ K$^{-1}$, and the superscript $V$ stands for *volatile slab*

$$\Phi_{l,n}^V = \frac{\Phi_V(m_V)}{\tau} = \frac{m_{l,n}^V c_l^V \omega}{\tau} \tag{4.3-3}$$

The third term in Eq (4.3-1) is related to the amount of latent heat required sublime the atmospheric mass needed to maintain vapor-pressure equilibrium with a higher surface temperature. Linearizing the change in surface pressure with respect to time gives

$$\frac{L_S}{g} \overline{\frac{dp_s(T)}{dT} \frac{dT}{dt}} = \Phi_{l,n}^A \left( T_{l,0,n+1} - T_{l,0,n+1} \right) \tag{4.3-4}$$

where $\Phi_{l,n}^A$ has units of erg cm$^{-2}$ s$^{-1}$ K$^{-1}$, and the superscript $A$ stands for *atmosphere*.

$$\Phi_{l,n}^A = \frac{\Phi_A(T_{l,0,n})}{\tau} = \frac{L_S}{g} \left( \frac{dp_s(T)}{dT} \bigg|_{T_{l,0,n}} \right) \frac{\omega}{\tau} \tag{4.3-5}$$

The temperature dependence of pressure is highly non-linear. If this is a dominant source of error, then one either chooses a small $\tau$, or iterates from an initial guess at a temperature $T_{l,0,n+1}^{approx}$ to an improved temperature $T_{l,0,n+1}$. In the latter case, by Taylor expansion of $p$ around $T_{l,0,n+1}^{approx}$,

$$\frac{L_S}{g} \overline{\frac{dp_s(T)}{dT} \frac{dT}{dt}} = \frac{L_S}{g \Delta t} \left[ p\left(T_{l,0,n+1}^{approx}\right) + \left(T_{l,0,n+1} - T_{l,0,n+1}^{approx}\right) \frac{dp}{dT} \bigg|_{T_{l,0,n+1}^{approx}} - p\left(T_{l,0,n}\right) \right] \tag{4.3-6}$$

This can be cast in a form parallel to that of Eq. (4.3-4) by

$$\frac{L_S}{g} \overline{\frac{dp_s(T)}{dT} \frac{dT}{dt}} = \Phi_{l,n}^A \left( T_{l,0,n+1} - T_{l,0,n} \right) + F_{l,n}^A \tag{4.3-7}$$

where the derivative in $\Phi_{l,n}^A$ is evaluated at the current guess at a temperature $T_{l,0,n+1}^{approx}$. The term $F_{l,n}^A$ has units of erg cm$^{-2}$ s$^{-1}$, and combines mathematically with the solar forcing.

$$F_{l,n}^A = \frac{L_S \omega}{g \tau} \left[ p\left(T_{l,0,n+1}^{approx}\right) - p\left(T_{l,0,n}\right) \right] - \left( T_{l,0,n+1}^{approx} - T_{l,0,n} \right) \Phi_{l,n}^A \tag{4.3-8}$$



By writing Eq. (4.4-7) in terms of the change in temperature relative to the previous time step (i.e., $T_{l,0,n+1} - T_{l,0,n}$), rather than in terms of the smaller change in temperature relative to the current guess (i.e., $T_{l,0,n+1} - T_{l,0,n+1}^{approx}$), Eq. (4.4-7) can be simply combined with the other terms in the discretized energy equation. On the first iteration, $T_{l,0,n+1}^{approx} = T_{l,0,n}$, and Eq. (4.3-7) reduces to Eq (4.3-4), so Eq. (4.3-7) can be used with very little added computational complexity.

The escape rate, $E$, if present, can be calculated at the start or mid time, similarly to the insolation.

Substituting the expressions for the explicit equations gives an equation similar to Eq. 3.3-14:

$$\underbrace{\Phi_{l,0}^H \left(T_{l,0,n+1} - T_{l,0,n}\right)}_{\text{Enthalpy, layer 0}} + \underbrace{\Phi_{l,n}^V \left(T_{l,0,n+1} - T_{l,0,n}\right)}_{\text{Enthalpy, volatile slab}} + \underbrace{\Phi_{l,n}^A \left(T_{l,0,n+1} - T_{l,0,n}\right) + F_{l,n}^A}_{\text{Latent heat, volatile slab}} = \\ \underbrace{\overline{S_{l,n'}}}_{\text{Insolation}} - \underbrace{\varepsilon_{l,n} \sigma \left(T_{l,0,n}\right)^4 - \Phi_{l,n}^E \left(T_{l,0,n+1} - T_{l,0,n}\right)}_{\text{Emission}} - \underbrace{\Phi_{l,0}^{K,B} \left(T_{l,0,n} - T_{l,1,n}\right)}_{\text{Conduction}} - \underbrace{L_S \overline{E_{l,n'}}}_{\text{Latent heat, escape}} \tag{4.3-9}$$

Collecting terms for the explicit equation gives

$$\begin{aligned} \left(\Phi_{l,0}^H + \Phi_{l,n}^E + \Phi_{l,n}^V + \Phi_{l,n}^A\right) T_{l,0,n+1} = \\ \left(\Phi_{l,0}^H + \Phi_{l,n}^E + \Phi_{l,n}^V + \Phi_{l,n}^A - \Phi_{l,0}^{K,B}\right) T_{l,0,n} + \left(\Phi_{l,0}^{K,B}\right) T_{l,1,n} \\ + \left(\overline{S_{l,n'}} - \varepsilon_{l,n} \sigma \left(T_{l,0,n}\right)^4 - F_{l,n}^A - L_S \overline{E_{l,n'}}\right) \end{aligned} \tag{4.3-10}$$

As in Section 3.3, divide by $\Phi_{l,n}^T$, with units erg cm$^{-2}$ s$^{-1}$ K$^{-1}$, where the superscript $T$ represents *total*, and the total "flux-per-temperature" now includes terms for enthalpy of the slab and interaction with the atmosphere

$$\Phi_{l,n}^T = \Phi_{l,0}^H + \Phi_{l,n}^E + \Phi_{l,n}^V + \Phi_{l,n}^A \tag{4.3-11}$$

The explicit equations for Area II can be written in a form that is identical to the explicit equation for the bare areas, Areas I and II (See Fig 3-6), with the resulting matrix elements given in the first row of Table 6.



**Table 6. Matrix elements for $j = 0$, Area II, $T \neq T_C$**

| Matrix equation | Matrix elements |
|---|---|
| Explicit<br>$T_{l,0,n+1} = \eta_{l,0,n} T_{l,0,n} + \beta_{l,0,n} T_{l,1,n} + \gamma_{l,0,n}$ | $\beta_{l,0,n} = \dfrac{\Phi_{l,0}^{K,B}}{\Phi_{l,n}^T}$<br>$\eta_{l,0,n} = 1 - \beta_{l,0,n}$<br>$\gamma_{l,0,n} = \dfrac{\overline{S_{l,n'}} - \varepsilon_{l,n} \sigma \left(T_{l,0,n}\right)^4 - F_{l,n}^A - L_S \overline{E_{l,n'}}}{\Phi_{l,n}^T}$ |
| Implicit (Crank-Nicholson)<br>$\eta''_{l,0} T_{l,0,n+1} + \beta''_{l,0} T_{l,1,n+1} = \eta'_{l,0} T_{l,0,n} + \beta'_{l,0} T_{l,1,n} + \gamma_{l,0,n}$ | $\beta'_{l,0,n} = \dfrac{\beta_{l,0,n}}{2}; \quad \beta''_{l,0} = -\dfrac{\beta_{l,0,n}}{2}$<br>$\eta'_{l,0} = 1 - \beta'_{l,0,n}; \quad \eta''_{l,j} = 1 - \beta''_{l,0,n}$ |

$\Phi_{l,0}^{K,B}$ is given by Eq. 3.3-10. $\Phi_{l,n}^T$ is given by 4.3-11.

The implicit form of the energy balance equation for Area II away from a crystalline transition temperature is found by substituting the Crank-Nicholson expression for the conduction term into Eq. 4.3-1. The energy balance for the implicit equation is

$$\underbrace{\Phi_{l,0}^H \left(T_{l,0,n+1} - T_{l,0,n}\right)}_{\text{Enthalpy, layer 0}} + \underbrace{\Phi_{l,n}^V \left(T_{l,0,n+1} - T_{l,0,n}\right)}_{\text{Enthalpy, volatile slab}} + \underbrace{\Phi_{l,n}^A \left(T_{l,0,n+1} - T_{l,0,n}\right) + F_{l,n}^A}_{\text{Latent heat, volatile slab}} =$$

$$\underbrace{\overline{S_{l,n'}}}_{\text{Insolation}} - \underbrace{\varepsilon_{l,n} \sigma \left(T_{l,0,n}\right)^4 - \Phi_{l,n}^E \left(T_{l,0,n+1} - T_{l,0,n}\right)}_{\text{Emission}} \quad (4.3\text{-}12)$$

$$\underbrace{- \dfrac{\Phi_{l,0}^{K,B}}{2} \left(T_{l,0,n} - T_{l,1,n}\right) - \dfrac{\Phi_{l,0}^{K,B}}{2} \left(T_{l,0,n+1} - T_{l,1,n+1}\right)}_{\text{Conduction}} - \underbrace{L_S \overline{E_{l,n'}}}_{\text{Latent heat, escape}}$$

Collecting terms for the implicit equation gives the volatile-covered equivalent to 3.3-16b:

$$\left(\Phi_{l,0}^H + \Phi_{l,n}^E + \Phi_{l,n}^V + \Phi_{l,n}^A + \dfrac{\Phi_{l,0}^{K,B}}{2}\right) T_{l,0,n+1} - \left(\dfrac{\Phi_{l,0}^{K,B}}{2}\right) T_{l,1,n+1} =$$

$$\left(\Phi_{l,0}^H + \Phi_{l,n}^E + \Phi_{l,n}^V + \Phi_{l,n}^A - \dfrac{\Phi_{l,0}^{K,B}}{2}\right) T_{l,0,n+1} + \left(\dfrac{\Phi_{l,0}^{K,B}}{2}\right) T_{l,1,n} \quad (4.3\text{-}13)$$

$$\left(\overline{S_{l,n'}} - \varepsilon_{l,n} \sigma \left(T_{l,0,n}\right)^4 - F_{l,n}^A - L_S \overline{E_{l,n'}}\right)$$

Again, divide by $\Phi_{l,n}^T$, with the resulting matrix elements given the second row in Table 5. The matrix elements for $j = 1$ to $J$ are identical as for the bare areas, Areas I and III (Section



3.3; Tables 3 and 4). The methods for solving the matrix equations are identical as for the bare areas, Areas I and III (Section 3.4).

If the volatile slab is at a crystalline transition temperature, then $T_{l,0,n+1} = T_{l,0,n}$, and the matrix elements are particularly simple for both the explicit and implicit form (Table 7).

**Table 7. Matrix elements for $j = 0$, Area II, $T=T_C$**

| Matrix equation | Matrix elements |
|---|---|
| Explicit $T_{l,0,n+1} = \eta_{l,J} T_{l,0,n} + \beta_{l,0} T_{l,1,n} + \gamma_{l,0,n}$ | $\beta_{l,0,n} = 0$ <br> $\eta_{l,0,n} = 1$ <br> $\gamma_{l,0,n} = 0$ |
| Implicit (Crank-Nicholson) $\eta''_{l,0} T_{l,0,n+1} + \beta''_{l,0} T_{l,1,n+1} = \eta'_{l,0} T_{l,0,n} + \beta'_{l,0} T_{l,1,n} + \gamma_{l,0,n}$ | $\beta'_{l,0,n} = 0;\ \beta''_{l,0} = 0$ <br> $\eta'_{l,0} = 1;\ \eta''_{l,j} = 1$ |

Once the new temperature is found, the change in the mass flux ($m^V_{l,n+1} - m^V_{l,n}$) is found by using local energy balance, Eq. (4.3-14). This is simply the discretized form of Eq (4.1-1). This applies whether the temperature is at a crystalline transition temperature or not, and whether the time step is calculated explicitly or implicitly.

$$\underbrace{m^V_{l,n} c_V \frac{T_{l,0,n+1} - T_{l,0,n}}{\Delta t}}_{\text{Enthalpy of volatile slab}} =$$

$$\underbrace{\overline{S_{l,n'}}}_{\text{Insolation}} - \underbrace{\left[\varepsilon_{l,n} \sigma (T_{l,0,n})^4 + 4\varepsilon_{l,n} \sigma (T_{l,0,n+1} - T_{l,0,n})^3\right]}_{\text{Emission}} - \underbrace{k \frac{T_{l,0,n} - T_{l,1,n}}{\Delta z}}_{\text{Conduction}} + \underbrace{L_S \frac{m^V_{l,n+1} - m^V_{l,n}}{\Delta t}}_{\text{Latent heat}} \quad (4.3\text{-}14)$$

*4.4 Matrix operations for single or multiple isolated volatile-covered locations (Area II)*

As with Areas I and III, computation can be sped up considerably by taking advantage of matrix operations to calculate the temperature evolution on multiple locations with a single operation. The form of the matrices for isolated volatile-covered locations (Area II) is the same as for bare locations (Areas I and III). Therefore, once the matrix elements are found, the calculations can proceed identically to Section 3.4.



## *4.5 Example: KBOs with bare areas or locally-supported atmospheres*

As an example, consider a point on the equator of a generic KBO with $A = 0.7$, $\varepsilon = 0.9$, no internal heat flux or mass loss, 50 cm s$^{-2}$ surface gravity, and an equatorial sub-solar latitude, at a range of heliocentric distances ($r$) from 30 to 80 AU (Fig 4.2). The thermal parameter for the substrate ($\Theta_S$) ranges from ~4-17 for 5 tiu (similar to those found by Lellouch et al. 2013), and ~1600 to ~7000 for 2100 tiu (pure, compact water ice). The thermal parameter for heating one g cm$^{-2}$ of a volatile slab ($\Theta_V$) is 7 times larger than $\Theta_S$ for the 5-tiu case, or 27 to117, so it is not insignificant. Both $\Theta_S$ and $\Theta_V$ increase with heliocentric distance, since their numerators stay constant and their denominators (proportional to $T^3$) decrease. Thus, the same object can be a slow rotator at perihelion and a fast rotator at aphelion. The atmospheric thermal parameter ($\Theta_A$), which has equilibrium pressure in the numerator, varies by 5-7 orders of magnitude over the range of $r$ from $5.2 \times 10^3$ to $8.6 \times 10^{-2}$ for $N_2$ and 1.0 to $1.5 \times 10^{-7}$ for $CH_4$.

For simplicity, the remainder of Fig 4.2 only contrasts a bare substrate with thermal inertia of 5 tiu, with a surface that is either $N_2$-covered or $CH_4$-covered. The effect of the decreasing temperature and increasing $\Theta_S$ with $r$ is clear in the progression for the substrate temperatures in the second panel, which plots the temperature for a bare substrate as a black solid line. The $N_2$ atmospheric "breathing" (green dashed line) has little effect at 70 AU, modifies the temperatures at 60 AU; by 40 and 30 AU, it nearly flattens out the temperature variation. The atmospheric breathing shifts the maximum by 90° phase, while the thermal conduction into the substrate shifts it by 45°; this is most evident when $\Theta_A$ is comparable to $\Theta_S$, such as for $N_2$ near 60 AU or $CH_4$ (red triple-dot-dashed line) near 30 AU. This has the effect of decreasing the peak temperature, and increasing the temperature at both the dawn and dusk limbs.

The third panel of Fig. 4.2 shows the increase in the mean and amplitude of the temperature for a bare substrate (gray fill) with decreasing heliocentric distance. For $N_2$-covered areas (green slanted fill), the temperatures are similar to the bare temperatures beyond ~70 AU. Closer than that, first the maximum temperature decreases while the dusk temperature rises, then the minimum and dawn temperature rise in tandem, until finally the maximum, dusk, dawn, and minimum temperatures all converge inward of 40 AU. For $CH_4$-covered areas (red vertical fill), the temperatures match the bare temperatures beyond ~40 AU; inward of 40 AU, as with the $N_2$, the maximum temperature decreases while the dusk temperature rises, with the slight rise in the minimum and dawn temperatures. The corresponding minimum, dawn, dusk, and maximum pressures are shown in the final panel.



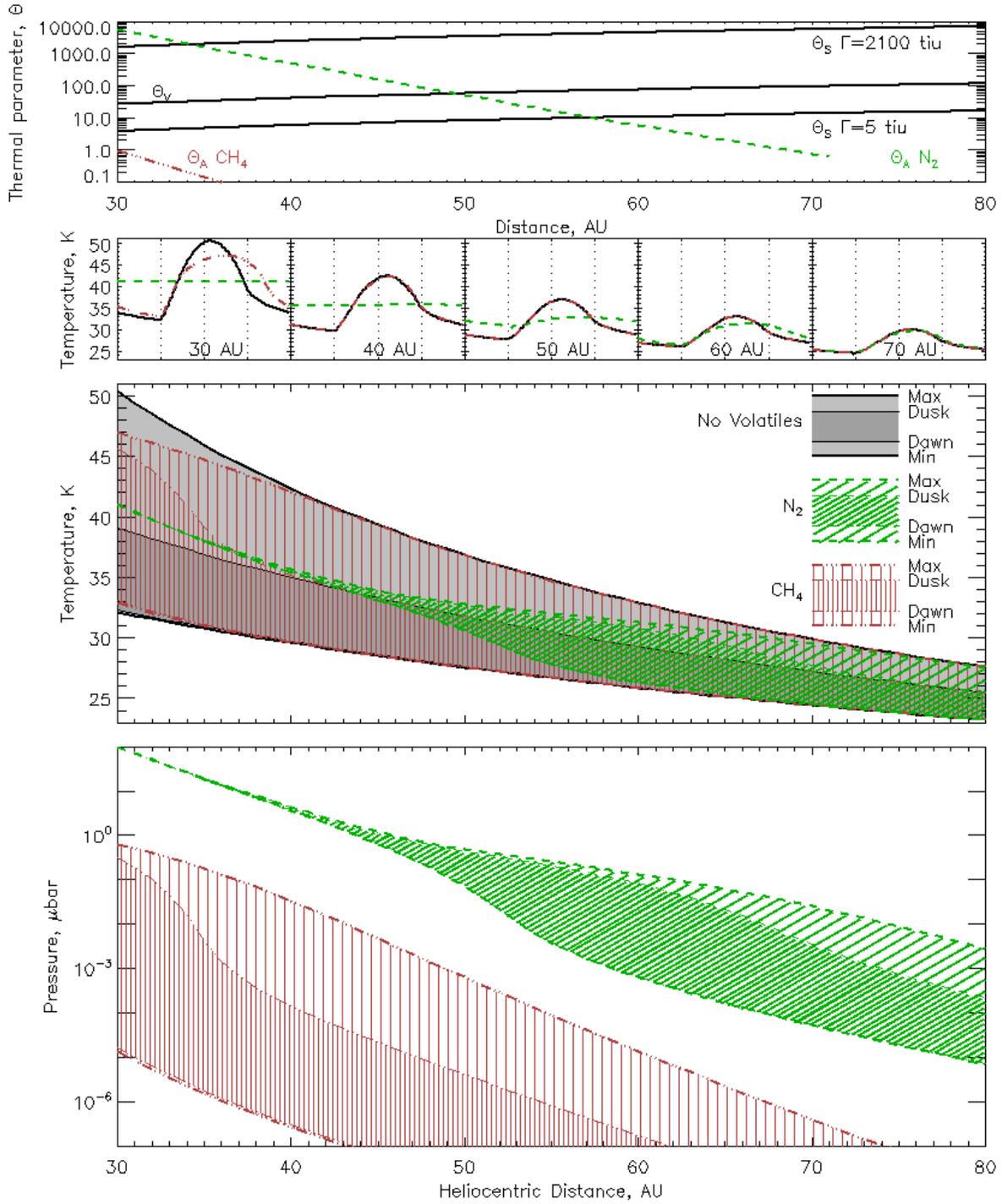



Fig 4-2.[1] Example temperatures and pressures for an equatorial location on a KBO with $A = 0.7$, $\varepsilon = 0.9$, and non-interacting areas with $N_2$, $CH_4$, or bare of volatiles, with equatorial illumination. Top: Thermal parameters for two values of thermal inertia, for 1 g of slab at 1.3e7 erg g$^{-1}$ K$^{-1}$, and for atmospheric "breathing" by $N_2$ (green dashed) and $CH_4$ (red triple-dot-dashed). Second: Temperature for $\Gamma=5$ tiu over a single day for bare (solid, black), $N_2$-covered (green dashed, indistinguishable from bare at 70 AU), and $CH_4$-covered (red triple-dot-dashed, indistinguishable from bare at 40 AU and farther) at selected distances. Third: Minimum, dawn, dusk, and maximum temperatures over a range of distances for bare (gray), $N_2$-covered (green), and $CH_4$-covered (red) areas. Fourth: Minimum, dawn, dusk, and maximum pressures over a range of distances for $N_2$-covered (green) and $CH_4$-covered (red) areas.

## 5. VT3D for interacting volatile-covered areas (Area IV)

Currently, Pluto and Triton are expected to have similar surface pressures over the entire globe, independent of local insolation (Trafton & Stern 1983, Trafton 1984, Spencer et al. 1987). $N_2$ sublimes from areas of high insolation, with latent heat loss balancing the excess insolation. Sublimation winds carry this mass to areas of low insolation, where $N_2$ is deposited, adding latent heat as well as solid $N_2$ (Fig 2-2B). As long as the atmosphere is dense enough, transport of mass and latent heat will keep the volatile ice temperatures nearly constant over the globe. Through vapor-pressure equilibrium, the surface pressures will also be nearly constant. If the atmosphere is thin enough so that the sublimation winds are a significant fraction of the sound speed, then the surface pressures will vary over the globe. This case can be handled efficiently by treating the surface as a "splice" between the interaction regions or isobaric regions, which share the same surface pressure, and local regions, for which the surface pressure varies with location (Fig 2-2C).

---

[1] **vty16_fig4_2** which calls
**vt3d_1loc_diurnal_local,** constants, input, grid, program, output and
**vty16_fig4_2_plot,** dist_sol_au, tod, theta_sva, temp_term_0, temp, p, $
                name_therminertia, n_per_period



In this section, I consider areas that have volatiles on their surfaces and which interact to share the same volatile ice temperature and surface pressure. This includes the entire globe for dense atmospheres, or the interacting portions of the splice for intermediate atmospheres (See Fig 2-2B, 2-2C). I will discuss the continuous equations in Section 5.1, analytic equations in Section 5.2, the discrete equations in Section 5.3, and efficient solutions to the matrix equations in Section 5.4. In Section 5.5, I present a worked example of Pluto's seasonal activity, with code and output in the supplementary materials.

*5.1   Continuous expressions for interacting volatile-covered locations (Area IV)*

For interacting volatile-covered locations, Area IV, energy is transported between locations through mass transport of volatiles through the atmosphere and the latent heat of sublimation. What ties the multiple locations together is (1) a common volatile-ice temperature, $T_V$, and (2) conservation of mass over the interacting regions. This latter includes the atmosphere over all areas that share a single surface pressure, whether bare (Area III) or volatile-covered (Area IV), because raising the surface pressure increases the atmospheric mass over all locations that share a common surface pressure. That is, if the surface pressure of the atmosphere increases in the region of effective transport, the mass of the atmosphere will increase above both the volatile-covered areas (Area IV) and the bare areas (Area III). The expression for mass balance in the area of effective transport is found by integrating Eq. 4.1-4 over both Area III and Area IV:

$$\int_{\Omega_{III}+\Omega_{IV}} \frac{dm_V}{dt} d\Omega = -\int_{\Omega_{III}+\Omega_{IV}} \frac{1}{g}\frac{dp_S(T)}{dT}\frac{dT}{dt} d\Omega - \int_{\Omega_{III}+\Omega_{IV}} E d\Omega \qquad (5.1\text{-}1)$$

where $\Omega_{III}$ and $\Omega_{IV}$ represent the solid angle of areas III and IV. Both the surface pressure and the temperature of the volatile slab are constant over Areas III and IV; the terms involving gravity, pressure, and temperature can be factored out of the middle integral. Futhermore, the mass flux for Area III is zero, so that the first integral can be evaluated over just Area IV. With these changes, the mass balance equation becomes

$$\int_{\Omega_{IV}} \frac{dm_V}{dt} d\Omega = -\frac{1}{g}\frac{dp_S(T_V)}{dT_V}\frac{dT_V}{dt}\left(\Omega_{III}+\Omega_{IV}\right) - \int_{\Omega_{III}+\Omega_{IV}} E d\Omega \qquad (5.1\text{-}2)$$

The areal average of the mass flux over Area IV is:

$$\left\langle \frac{dm_V}{dt} \right\rangle \equiv \frac{1}{\Omega_{IV}} \int_{\Omega_{IV}} \frac{dm_V}{dt} d\Omega \qquad (5.1\text{-}3a)$$



where brackets represent an areal average over Area IV. The atmosphere escapes from above both bare and volatile-covered areas, so the areal average of $E$ is taken over Areas III and IV:

$$\langle E \rangle' \equiv \frac{1}{\Omega_{III} + \Omega_{IV}} \int_{\Omega_{III}+\Omega_{IV}} E\, d\Omega \qquad (5.1\text{-}3b)$$

where primed brackets represent an areal average over Area III and Area IV.

$f_V$ is the fraction of the interacting areas (III and IV) covered with volatiles. In Paper I, which only treated a global atmosphere, this was fraction of the surface covered by volatiles. Here, with the possibility of a spliced atmosphere, the expression is written more generally.

$$f_V \equiv \frac{\Omega_{IV}}{\Omega_{III} + \Omega_{IV}} \qquad (5.1\text{-}4)$$

With these definitions, the equation for mass balance over the areas of isobaric surface pressure becomes

$$\left\langle \frac{dm_V}{dt} \right\rangle = -\frac{1}{f_V g} \frac{dp_S(T_V)}{dT_V} \frac{dT_V}{dt} - \frac{1}{f_V} \langle E \rangle' \qquad (5.1\text{-}5)$$

Eq. 5.1-5 illustrates the significance of the fraction of the surface covered by volatiles, $f_V$. If the volatile ices are confined to a small patch, then that patch has to lose a lot of mass to supply an increase of the entire atmosphere in the isobaric area.

The local energy balance is the same as for localized volatile-covered areas, Eq. (4.1-1). Integrating Eq. 4.1-1 over Area IV, and substituting the equation for conservation of mass over isobaric areas, yields an equation for energy balance over all of Area IV, using the same notation for spatial averages as in Eq. 5.1-3a.

$$\left( \underbrace{\langle m_V c_V \rangle}_{\text{Enthalpy of volatile slab}} + \underbrace{\frac{L_S}{f_V g} \frac{dp_S(T)}{dT}}_{\text{Latent heat}} \right) \frac{\partial T}{\partial t} =$$

$$\underbrace{\langle S \rangle}_{\text{Insolation}} - \underbrace{\langle \varepsilon \rangle \sigma T^4}_{\text{Emission}} - \underbrace{\left\langle k \frac{dT}{dz}\bigg|_{z=0} \right\rangle}_{\text{Conduction}} - \underbrace{\frac{L_S \langle E \rangle'}{f_V}}_{\text{Latent heat of escaping gas}} \qquad ,\ T \neq T_C \qquad (5.1\text{-}6a)$$

$$\underbrace{\langle m_V \rangle L_C \frac{dX}{dt}}_{\text{Enthalpy of volatile slab}} = \underbrace{\langle S \rangle}_{\text{Insolation}} - \underbrace{\langle \varepsilon \rangle \sigma T^4}_{\text{Emission}} - \underbrace{\left\langle k \frac{dT}{dz}\bigg|_{z=0} \right\rangle}_{\text{Conduction}} - \underbrace{\frac{L_S \langle E \rangle'}{f_V}}_{\text{Latent heat of escaping gas}} \qquad ,\ T = T_C \qquad (5.1\text{-}6b)$$



While the temperature of isolated volatile-covered areas depend only on local conditions (Eq 4.1-6a,b), the volatile ice temperature in the interacting areas depends on the *spatial average* of energy sources and sinks.

*5.2 Analytic approximation and initialization for interacting volatile-covered locations (Area IV)*

The analytic form of the continuous equations (Eq. 5.1-6) is very similar to that for the isolated volatile-covered areas, Area II, described in Section 4. As in Section 4, the solar forcing, the atmospheric escape, and the thermal wave are (1) decomposed into sinusoidal terms (3.2-1 for absorbed insolation, 3.2-7 for temperature, and 4.2-2 for escape), (2) substituted into Eq 5.1-6, and (3) isolated term-by-term. The $m=0$ term gives the expression for the time-averaged temperature:

$$0 = \underbrace{\langle S_0 \rangle}_{\text{Insolation}} - \underbrace{\langle \varepsilon \rangle \sigma T_V^4}_{\text{Emission}} + \underbrace{\langle F \rangle}_{\text{Flux at lower boundary}} - \underbrace{L_S \langle E_0 \rangle' / f_V}_{\text{Latent heat of escaping gas}} \qquad (5.2\text{-}1)$$

To find the variation in the temperature (the terms with $m = 1$ and higher), substitute the expressions for solar forcing, temperature, and escape into Eq. 5.1-6, expand the thermal emission term to first order in $T_m$, and take only those terms proportional to $\exp(im\omega t)$. This expression is simpler with the spatially averaged versions of the "flux-per-temperature" expressions:

$$\langle \Phi_S \rangle = \sqrt{\omega} \langle \Gamma \rangle \qquad (5.2\text{-}2a)$$

$$\langle \Phi_E(T) \rangle = 4 \langle \varepsilon \rangle \sigma T^3 \qquad (5.2\text{-}2b)$$

$$\langle \Phi_V \rangle = \omega \langle m_V c_V \rangle \qquad (5.2\text{-}2c)$$

$$\langle \Phi_A(T) \rangle = \omega \frac{L_S}{f_V g} \frac{dp_S}{dT_V} \qquad (5.2\text{-}2d)$$

If the substrate under all of the volatile ices has the same thermophysical properties, then the first two terms reduce to their local equivalents: Eq. 3.2-9a, b. Likewise, if the specific heat of the volatile ices are the same over Area IV, then the third equation (Eq. 5.2-2c) differs from its local equivalent (4.2-4a) simply by replacing the local volatile slab mass with the areal average. If there is no bare ground in the isobaric area (that is, if no locations are Area III), then $f_V = 1$, and the last expression (Eq. 5.2-2d) is identical to its local equivalent (4.2-4b). However, if only part of the isobaric area is volatile-covered, then $\langle \Phi_A(T) \rangle > \Phi_A(T)$. A



change in volatile temperature increases the atmosphere above both bare and volatile-covered locations in the isobaric areas, so more mass is exchanged between the surface and atmosphere, and more latent heat of sublimation is required. This means that the latent heat term is more effective at suppressing the temperature variation when there is a smaller fraction of surface volatiles. For temperatures away from a crystalline phase, with these substitutions, the spatially averaged energy equation is:

$$\left[\underbrace{\sqrt{im}\langle\Phi_S\rangle}_{\text{Conduction}}+\underbrace{\langle\Phi_E(T_V)\rangle}_{\text{Emission}}+\underbrace{im\langle\Phi_V\rangle}_{\text{Enthalpy of volatile slab}}+\underbrace{im\langle\Phi_A(T_V)\rangle}_{\text{Latent heat}}\right]\hat{T}_m = \underbrace{\langle\hat{S}_m\rangle}_{\text{Insolation}} - \underbrace{L_s\langle\hat{E}_m\rangle/f_V}_{\text{Latent heat of escaping gas}} \quad, T \neq T_C \quad (5.2\text{-}3)$$

If the equilibrium temperature is at a crystalline phase transition, then the corresponding equation for the change in the slab's state is

$$\underbrace{i\omega m\langle m_V\rangle L_C \hat{X}_m}_{\text{Enthalpy of volatile slab}} = \underbrace{\langle\hat{S}_m\rangle}_{\text{Insolation}} - \underbrace{\frac{L_s\langle\hat{E}_m\rangle}{f_V}}_{\text{Latent heat of escaping gas}} \quad, T = T_C \quad (5.2\text{-}4)$$

*5.3 Numerical solution for interacting volatile-covered locations (Area IV)*

Fig 5-1 shows the interaction between different locations in Area IV. There is no horizontal heat flow within the substrate. However, the volatile slabs exchange energy through latent heat of sublimation and condensation, and share a single temperature, $T_V$. The temperature of the volatile ice slab therefore depends on the insolation over the entire volatile-covered interacting region, and the conduction from each of the substrate layers (layer 0) that immediately underlie the volatile ice slab. The temperatures of each of the topmost substrate layers depend, in turn, on the single volatile slab temperature, through thermal conduction.



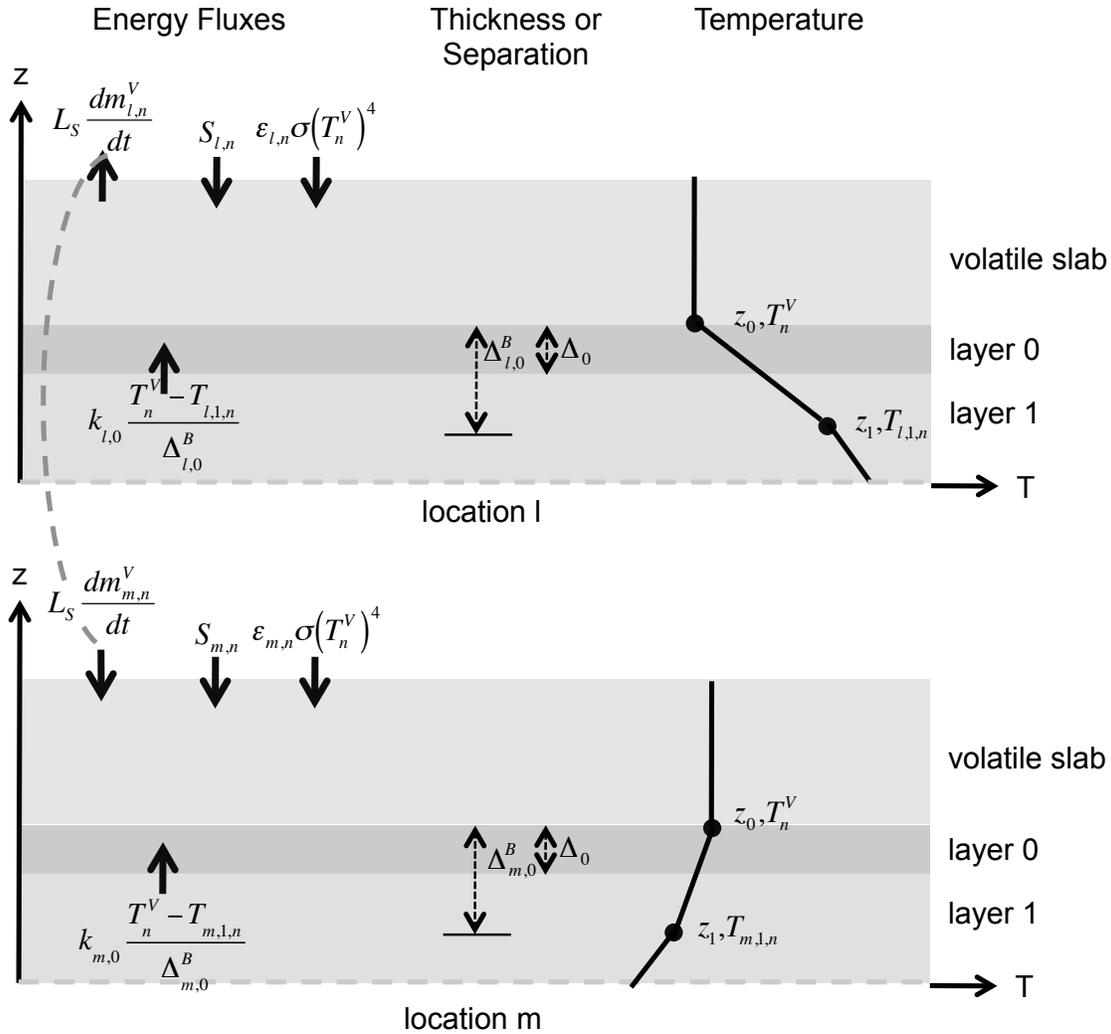

Fig. 5-1. Schematic of the layering scheme and energy fluxes for layer $j = 0$, areas IV (interacting volatiles). Two interacting regions are shown, for locations $l$ and $m$ (not to be confused with Fourier term $m$). Typically there are multiple interacting regions. The dashed gray line indicated that the areas are connected through the latent heat term.

Because there is no horizontal heat flow within the substrate, the discretization for layers $j = 2 .. J$ is the same as the other areas, so that much of the matrix form for the explicit equations is tridiagonal (Fig. 5-2). However, because volatile slabs of the areas interact (Fig 5-1), the explicit discretized equation for the new $T_V$ has non-zero coefficients accounting for the conduction upward from each of the $j = 1$ layers (the upper row of the matrix). Similarly, the explicit discretized equation for each new $T_1$ has non-zero coefficients accounting for the conduction downward from each of the $j = 0$ slabs, all assumed to be at $T_V$. The resulting



matrix, with non-zero elements on the left-most column and top-most row, is a *banded tridiagonal* matrix.

$$
\begin{pmatrix} T^V_{n+1} \\ T_{l,1,n+1} \\ \vdots \\ T_{l,J,n+1} \\ T_{m,1,n+1} \\ \vdots \\ T_{m,J,n+1} \end{pmatrix} = \begin{pmatrix} \eta^V_n & \beta_{l,0,n}w_l & & & \beta_{m,0,n}w_m & & \\ \alpha_{l,1} & \eta_{l,1} & \beta_{l,1} & & & & \\ & \ddots & \ddots & \ddots & & & \\ & & \alpha_{l,J} & \eta_{l,J} & & & \\ \alpha_{m,1} & & & & \eta_{m,1} & \beta_{m,1} & \\ & & & & \ddots & \ddots & \ddots \\ & & & & & \alpha_{m,J} & \eta_{m,J} \end{pmatrix} \times \begin{pmatrix} T^V_n \\ T_{l,1,n} \\ \vdots \\ T_{l,J,n} \\ T_{m,1,n} \\ \vdots \\ T_{m,J,n} \end{pmatrix} + \begin{pmatrix} \gamma^V_n \\ \\ \vdots \\ \gamma_{l,J} \\ \\ \\ \gamma_{m,J} \end{pmatrix}
$$

Fig 5-2. Schematic of an explicit time-step from time *n* to time *n*+1 for two locations, *l* and *m*, in Area IV (volatile-covered, interacting). Dark gray elements (the temperatures and the elements of the upper row) change with each time step. Light gray elements are independent of time. White elements are zero. $w_l$ and $w_m$ are the areal weights for locations *l* and *m*.

The implicit (Crank-Nicholson) form of the matrix equations has a similar form, with a banded tridiagonal matrix on both the left and right hand sides of the equation.



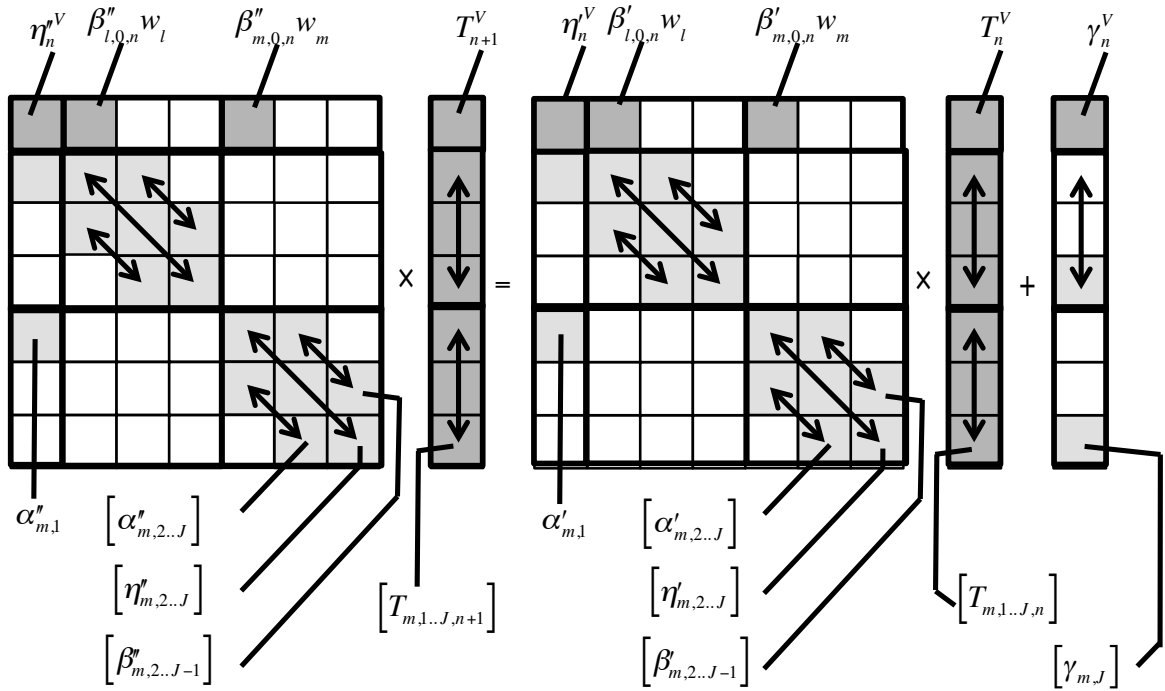

Fig 5-3. Schematic of an implicit time-step from time *n* to time *n*+1 for two locations, *l* and *m*, in Area IV (volatile-covered, interacting), using the Crank-Nicholson scheme. Dark gray elements (the temperatures and the elements of the upper row) change with each time step. Light gray elements are independent of time. White elements are zero. The variables in brackets refer to the vectors of length *J* or *J*–1 indicated by the double-arrowed lines.

The elements of the substrate arrays are derived from the discretation of the conductivity equation, Eq. (3.1-2), as before. The matrix elements for the substrate—the light gray elements in Figs 5-2 and 5-3—are unchanged from the previous cases. This holds even for the first layer, $j = 1$. The dependence of the temperature of the first layer at location *l*, $T_{l,1}$, depends only on the temperature below ($T_{l,2}$) and above ($T_{l,0}$). For Area IV, the assumption is that $T_{l,0} = T_V$ (that is, the upper surface of the substrate equals the volatile slab temperature, Fig 5-1). While this changes the format of the matrices (the line of $\alpha$'s in the left-most column in Fig 5-2 and 5-3), it does not change the value for the $\alpha$'s themselves. To find the elements for the implicit arrays $\alpha_{l,j}$, $\eta_{l,j}$, $\beta_{l,j}$ ($j = 1 .. J$) and the lower-boundary element $\gamma_{l,J}$, or their explicit counterparts (primed for the right-hand side and double-primed for the left) consult Tables 3 and 4.

The elements for the volatile slab—the dark gray elements on the top row of Figs 5-2 and 5-3—are related to, but different than, the corresponding elements for Area II (volatile-



covered, non-interacting). As before, I first solve for temperatures away from the solid phase transition ($T \neq T_C$). For Area IV, I integrate the conduction equation (Eq. 3.1-2) over the top layer, average that over Area IV, and add the result to Eq. 5.1-6a to replace the term with conduction at $z = 0$ (at the slab-substrate interface) with one at $-\Delta_0$ (at the bottom of the first substrate layer). Taking the time average from time $n$ to $n+1$ (indicated by overbars) yields Eq. 5.3-1, the areal averaged equivalent to Eq. 4.3-1. Compared with Eq. 4.3-1, Eq. 5.3-1 has areal averages for the thermophysical parameters (density, specific heat, mass of a slab, thermal conduction, emissivity), areal averages for the solar gain and heat lost by escape, and the inclusion of $f_V$, the fraction of the interacting area that is covered by volatiles, in the latent heat and escape terms.

$$\underbrace{\overline{\langle \rho_0 c_0 \rangle \int_{-\Delta_0}^{0} \frac{\partial T(z)}{\partial t} dz}}_{\text{Enthalpy, layer 0}} + \underbrace{\overline{\langle m_V c_V \rangle \frac{\partial T}{\partial t}}}_{\text{Enthalpy, volatile slab}} + \underbrace{\overline{\frac{L_S}{f_V g} \frac{dp_s(T)}{dT} \frac{dT}{dt}}}_{\text{Latent heat, volatile slab}} = \underbrace{\overline{\langle S_{n'} \rangle}}_{\text{Insolation}} - \underbrace{\overline{(\langle \varepsilon_n \rangle \sigma T^4)}}_{\text{Emission}} - \underbrace{\overline{\langle k \frac{dT}{dz} \big|_{z=-\Delta_0} \rangle}}_{\text{Conduction}} - \underbrace{\frac{L_S \overline{\langle E_{n'} \rangle}'}{f_V}}_{\text{Latent heat, escape}} \quad (5.3\text{-}1)$$

where $\langle \rho_0 c_0 \rangle$ is the areal average of the product of density and specific heat in layer 0, with cgs units of erg K$^{-1}$ cm$^{-3}$, and $\langle m_V c_V \rangle$ is the areal average of the product of volatile slab mass and specific heat in the volatile slab, with cgs units of erg K$^{-1}$ cm$^{-2}$.

The treatment of the first term is similar to that in the bare case; see the discussion near Eq. 3.3-2. As before, the temperature of layer 0 is sampled at the top of the layer. Because this is the slab-substrate interface, the temperature of layer 0 equals the volatile slab temperature within Area IV: $T_{l,0,n} = T_n^V$. With the assumption that we can sample the temperature at the top of layer 0, the enthalpy term depends only on the change in the volatile slab temperature:

$$\underbrace{\overline{\langle \rho_0 c_0 \rangle \int_{-\Delta_0}^{0} \frac{\partial T(z)}{\partial t} dz}}_{\text{Enthalpy, layer 0}} = \langle \Phi_0^H \rangle \left( T_{n+1}^V - T_n^V \right) \quad (5.3\text{-}2)$$

where $\langle \Phi_0^H \rangle$, like $\Phi_{l,j}^H$ (Eq. 3.3-5), has units of erg cm$^{-2}$ s$^{-1}$ K$^{-1}$, with the superscript $H$ representing heat or enthalpy. The discrete form for the areal average (cf. Eq. 5.1-3a) is simply the weighted average of the local values, summed over the locations within Area IV, $\{L_{IV}\}$:



$$\left\langle \Phi_0^H \right\rangle = \sum_{l \in \{L_{IV}\}} w_l \Phi_{l,0}^H \qquad (5.3\text{-}3)^1$$

The weights (Eq. 5.3-4) are simply the ratio of the solid angle of each location ($\Omega_l$) to the total solid angle of Area IV:

$$w_l = \frac{1}{\Omega_{IV}} \sum_{l \in \{L_{IV}\}} \Omega_l \qquad (5.3\text{-}4)$$

Continuing to treat Eq. 5.3-1 term-by-term, the change enthalpy of the volatile slab also depends on the change in volatile slab temperature; see discussion near Eq. 4.3-2 and 4.3-3.

$$\underbrace{\overline{\left\langle m_V c_V \right\rangle \frac{\partial T}{\partial t}}}_{\text{Enthalpy, volatile slab}} = \left\langle \Phi_n^V \right\rangle \left( T_{n+1}^V - T_n^V \right) \qquad (5.3\text{-}5a)$$

$$\left\langle \Phi_0^V \right\rangle = \sum_{l \in \{L_{IV}\}} w_l \Phi_{l,0}^V \qquad (5.3\text{-}5b)$$

The latent heat term is the same over all locations, but differs from the local equivalents (Eq. 4.3-4 to 4.3-8) by the factor of $f_V$:

$$\underbrace{\frac{L_S}{f_V g} \overline{\frac{dp_s(T)}{dT} \frac{dT}{dt}}}_{\text{Latent heat, volatile slab}} = \left\langle \Phi_n^A \right\rangle \left( T_{n+1}^V - T_n^V \right) + \left\langle F_n^A \right\rangle \qquad (5.3\text{-}6a)$$

$$\left\langle \Phi_n^A \right\rangle = \Phi_n^A / f_V \qquad (5.3\text{-}6b)$$

$$\left\langle F_n^A \right\rangle = F_n^A / f_V \qquad (5.3\text{-}6c)$$

The insolation terms is simply the areal average of the insolation at each location in Area IV:

$$\underbrace{\overline{\left\langle S_{n'} \right\rangle}}_{\text{Insolation}} = \sum_{l \in \{L_{IV}\}} w_l \overline{S_{l,n'}} \qquad (5.3\text{-}7)$$

The thermal emission depends on the areal average of the emissivity:

$$\underbrace{\overline{\left( \left\langle \varepsilon_n \right\rangle \sigma T_V^4 \right)}}_{\text{Emission}} = \left\langle \varepsilon_n \right\rangle \sigma \left( T_n^V \right)^4 + \left\langle \Phi_n^E \right\rangle \left( T_{n+1}^V - T_n^V \right) \qquad (5.3\text{-}8a)$$

---

[1] `avg = `**`vt3d_locavg`**`(val, weight)`



$$\langle \varepsilon_n \rangle = \sum_{l \in \{L_{IV}\}} w_l \varepsilon_{l,n} \tag{5.3-8b}$$

$$\langle \Phi_n^E \rangle = 2 \langle \varepsilon_n \rangle \sigma \left(T_n^V\right)^3 \tag{5.3-8c}$$

For explicit equations, the expression for the areal average of thermal conduction is found by taking the areal average of Eq. 3.3-9, and making the substitution that $T_{l,0,n} = T_n^V$:

$$\underbrace{\overline{\left\langle k \frac{dT}{dz}\bigg|_{z=-\Delta_0}\right\rangle}}_{\text{Conduction}} \approx \sum_{l \in \{L_{IV}\}} w_l \Phi_{l,0}^{K,B} \left(T_n^V - T_{l,1,n}\right) \tag{5.3-9}$$

where $\Phi_{l,0}^{K,B}$ is given by Eq. 3.3-10. Similarly, the expression for the implicit (Crank-Nicholson) equations takes the areal average of Eq. 3.3-13:

$$\underbrace{\overline{\left\langle k \frac{dT}{dz}\bigg|_{z=-\Delta_0}\right\rangle}}_{\text{Conduction}} \approx \frac{1}{2}\sum_{l \in \{L_{IV}\}} w_l \Phi_{l,0}^{K,B} \left(T_{n+1}^V - T_{l,1,n+1}\right) + \frac{1}{2}\sum_{l \in \{L_{IV}\}} w_l \Phi_{l,0}^{K,B} \left(T_n^V - T_{l,1,n}\right) \tag{5.3-10}$$

Finally, the escape rate is calculated by the average over all the interacting regions, Area III and Area IV:

$$\underbrace{\frac{L_S \langle E_n \rangle'}{f_V}}_{\text{Latent heat, escape}} = \frac{L_S}{f_V} \sum_{l \in \{L_{III}+L_{IV}\}} w'_l \overline{E_{l,n}} \tag{5.3-11a}$$

$$w'_l = \frac{1}{\Omega_{III} + \Omega_{IV}} \sum_{l \in \{L_{III}+L_{IV}\}} \Omega_l \tag{5.3-11b}$$

Substituting the expressions for the explicit equations gives

$$\underbrace{\langle \Phi_0^H \rangle \left(T_{n+1}^V - T_n^V\right)}_{\text{Enthalpy, layer 0}} + \underbrace{\langle \Phi_n^V \rangle \left(T_{n+1}^V - T_n^V\right)}_{\text{Enthalpy, volatile slab}} + \underbrace{\langle \Phi_n^A \rangle \left(T_{n+1}^V - T_n^V\right) + \langle F_n^A \rangle}_{\text{Latent heat, volatile slab}} =$$

$$\underbrace{\langle S_{n'} \rangle}_{\text{Insolation}} - \underbrace{\left[\langle \varepsilon_n \rangle \sigma \left(T_n^V\right)^4 + \langle \Phi_n^E \rangle \left(T_{n+1}^V - T_n^V\right)\right]}_{\text{Emission}} - \underbrace{\sum_{l \in \{L_{IV}\}} w_l \Phi_{l,0}^{K,B} \left(T_n^V - T_{l,1,n}\right)}_{\text{Conduction}} - \underbrace{\frac{L_S \langle E_n \rangle'}{f_V}}_{\text{Latent heat, escape}} \tag{5.3-12}$$

Collecting terms for the explicit equation gives



$$\left(\left\langle\Phi_0^H\right\rangle+\left\langle\Phi_n^E\right\rangle+\left\langle\Phi_n^V\right\rangle+\left\langle\Phi_n^A\right\rangle\right)T_{n+1}^V =$$
$$\left(\left\langle\Phi_0^H\right\rangle+\left\langle\Phi_n^E\right\rangle+\left\langle\Phi_n^V\right\rangle+\left\langle\Phi_n^A\right\rangle-\left\langle\Phi_0^{K,B}\right\rangle\right)T_n^V + \sum_{l\in\{L_{IV}\}} w_l \Phi_{l,0}^{K,B} T_{l,1,n} \qquad (5.3\text{-}13)$$
$$+\left(\left\langle\overline{S_{n'}}\right\rangle-\left\langle\varepsilon_n\right\rangle\sigma\left(T_n^V\right)^4-\left\langle F_n^A\right\rangle-\frac{L_S\left\langle\overline{E_n}\right\rangle'}{f_V}\right)$$

As in Section 3.3 and 4.3, divide by $\left\langle\Phi_n^T\right\rangle$, with cgs units erg cm$^{-2}$ s$^{-1}$ K$^{-1}$, where the superscript *T* represents *total*. The total "flux-per-temperature" includes terms for enthalpy of the slab and interaction with the atmosphere

$$\left\langle\Phi_n^T\right\rangle=\left\langle\Phi_0^H\right\rangle+\left\langle\Phi_n^E\right\rangle+\left\langle\Phi_n^V\right\rangle+\left\langle\Phi_n^A\right\rangle \qquad (5.3\text{-}14)$$

The resulting of dividing Eq. 5.3-13 by 5.3-14, and the resulting matrix elemens, are given in Table 8.

The implicit (Crank-Nicholson) equation (5.3-15) differs from equation (5.3-12) only with the substitution of the conduction term:

$$\underbrace{\left\langle\Phi_0^H\right\rangle\left(T_{n+1}^V-T_n^V\right)}_{\text{Enthalpy, layer 0}}+\underbrace{\left\langle\Phi_n^V\right\rangle\left(T_{n+1}^V-T_n^V\right)}_{\text{Enthalpy, volatile slab}}+\underbrace{\left\langle\Phi_n^A\right\rangle\left(T_{n+1}^V-T_n^V\right)+\left\langle F_n^A\right\rangle}_{\text{Latent heat, volatile slab}}=$$
$$\underbrace{\left\langle\overline{S_{n'}}\right\rangle}_{\text{Insolation}}-\underbrace{\left[\left\langle\varepsilon_n\right\rangle\sigma\left(T_n^V\right)^4+\left\langle\Phi_n^E\right\rangle\left(T_{n+1}^V-T_n^V\right)\right]}_{\text{Emission}} \qquad (5.3\text{-}15)$$
$$-\underbrace{\left[\frac{1}{2}\sum_{l\in\{L_{IV}\}} w_l \Phi_{l,0}^{K,B}\left(T_{n+1}^V-T_{l,1,n+1}\right)+\frac{1}{2}\sum_{l\in\{L_{IV}\}} w_l \Phi_{l,0}^{K,B}\left(T_n^V-T_{l,1,n}\right)\right]}_{\text{Conduction}}-\underbrace{\frac{L_S\left\langle\overline{E_n}\right\rangle'}{f_V}}_{\text{Latent heat, escape}}$$

Collecting terms for the implicit equation gives



$$\left(\left\langle\Phi_0^H\right\rangle+\left\langle\Phi_n^E\right\rangle+\left\langle\Phi_n^V\right\rangle+\left\langle\Phi_n^A\right\rangle+\frac{1}{2}\left\langle\Phi_0^{K,B}\right\rangle\right)T_{n+1}^V+\frac{1}{2}\sum_{l\in\{L_{IV}\}}w_l\Phi_{l,0}^{K,B}T_{l,1,n+1}=$$

$$\left(\left\langle\Phi_0^H\right\rangle+\left\langle\Phi_n^E\right\rangle+\left\langle\Phi_n^V\right\rangle+\left\langle\Phi_n^A\right\rangle-\frac{1}{2}\left\langle\Phi_0^{K,B}\right\rangle\right)T_n^V+\frac{1}{2}\sum_{l\in\{L_{IV}\}}w_l\Phi_{l,0}^{K,B}T_{l,1,n} \quad (5.3\text{-}16)$$

$$+\left(\left\langle\overline{S_{n'}}\right\rangle-\left\langle\varepsilon_n\right\rangle\sigma\left(T_n^V\right)^4-\left\langle F_n^A\right\rangle-\frac{L_S\left\langle\overline{E_n}\right\rangle'}{f_V}\right)$$

This equation is used to derive the elements for the matrix elements in Figs. 5-2 and 5-3, given in Table 8.

**Table 8. Matrix elements for $j = 0$, Area IV, $T \neq T_C$**

| Matrix equation | Matrix elements |
|---|---|
| Explicit<br>$T_{n+1}^V = \eta_n^V T_n^V + \sum_{l\in\{L_{IV}\}}w_l\beta_{l,0,n}T_{l,1,n} + \gamma_n^V$ | $\beta_{l,0,n} = \dfrac{\Phi_{l,0}^{K,B}}{\left\langle\Phi_n^T\right\rangle}$<br><br>$\eta_n^V = 1 - \sum_{l\in\{L_{IV}\}}w_l\beta_{l,0,n}$<br><br>$\gamma_n^V = \dfrac{\left\langle\overline{S_{n'}}\right\rangle-\left\langle\varepsilon_n\right\rangle\sigma\left(T_n^V\right)^4-\left\langle F_n^A\right\rangle-L_S\left\langle\overline{E_n}\right\rangle'/f_V}{\left\langle\Phi_n^T\right\rangle}$ |
| Implicit (Crank-Nicholson)<br>$\eta_n'^V T_{n+1}^V + \sum_{l\in\{L_{IV}\}}w_l\beta'_{l,0,n}T_{l,1,n} =$<br>$\eta_n''^V T_n^V + \sum_{l\in\{L_{IV}\}}w_l\beta''_{l,0,n}T_{l,1,n} + \gamma_n^V$ | $\beta'_{l,0,n} = \dfrac{\beta_{l,0,n}}{2}; \quad \beta''_{l,0,n} = -\dfrac{\beta_{l,0,n}}{2}$<br><br>$\eta_n'^V = 1 - \sum_{l\in\{L_{IV}\}}w_l\beta'_{l,0,n}; \quad \eta_n''^V = 1 - \sum_{l\in\{L_{IV}\}}w_l\beta''_{l,0,n}$ |

$\Phi_{l,0}^{K,B}$ is given by Eq. 3.3-9. $\left\langle\Phi_n^T\right\rangle$ is given by 5.3-14.

The discrete form of the equation for the change in temperature at a crystalline phase is trivial, since the volatile slab temperature does not change from time *n* to time *n*+1 in Eq. 5.1-6b.



**Table 9. Matrix elements for $j = 0$, Area IV, $T=T_C$**

| Matrix equation | Matrix elements |
|---|---|
| Explicit<br>$T^V_{n+1} = \eta^V_n T^V_n + \sum_{l \in \{L_{IV}\}} w_l \beta_{l,0,n} T_{l,1,n} + \gamma^V_n$ | $\beta_{l,0,n} = 0$<br>$\eta^V_n = 1$<br>$\gamma^V_n = 0$ |
| Implicit (Crank-Nicholson)<br>$\eta''_{0,n} T^V_{n+1} + \sum_{l \in \{L_{IV}\}} w_l \beta''_{l,0} T_{l,1,n} = \eta'_{0,n} T^V_n + \sum_{l \in \{L_{IV}\}} w_l \beta'_{l,0} T_{l,1,n} + \gamma_{l,0,n}$ | $\beta'_{l,0,n} = 0; \quad \beta''_{l,0,n} = 0$<br>$\eta'_{l,0} = 1; \quad \eta''_{l,j} = 1$ |

## 5.4 Matrix operations for interacting volatile-covered locations (Areas IV)

### 5.4a. Overview and explicit timesteps

In Section 3.4, I divided the temperature into the upper layer and the interior layers (Eq. 3.4-1), as a means to speeding up calculations in Areas I, II and III. In Area IV, this division is required, as the temperature of each of the upper layers ($T_{l,0,n}$) is equal to a single value for the volatile slab temperature, ($T^V_n$). With this, the matrix equation in Fig 5-2 can be written:

$$\begin{bmatrix} T^V_{n+1} \\ \mathbf{T}_{l,1..J,n+1} \\ \mathbf{T}_{m,1..J,n+1} \end{bmatrix} = \begin{bmatrix} \eta^V_n & \mathbf{b}_{l,n} & \mathbf{b}_{m,n} \\ \mathbf{a}_l & \mathbf{S}_l & 0 \\ \mathbf{a}_m & 0 & \mathbf{S}_m \end{bmatrix} \times \begin{bmatrix} T^V_n \\ \mathbf{T}_{l,1..J,n} \\ \mathbf{T}_{m,1..J,n} \end{bmatrix} + \begin{bmatrix} \gamma^V_n \\ \mathbf{g}_l \\ \mathbf{g}_m \end{bmatrix} \quad (5.4\text{-}1)$$

The matrix elements $\eta^V_n$ and $\gamma^V_n$ are defined in Table 8 or 9. The **b** arrays are similar to Eq. (3.4-3), except that the weighting factor is included: $\mathbf{b}_{l,n} = [w_l \beta_{l,0,n}, 0, \cdots, 0]$. The **a** arrays are defined in Eq. (3.4-4), the **S** is defined in the text between Eq. 3.4-4 and 3.4-5, and the **g** array is defined in Eq. 3.4-5. Eq. 5.4-1 is represented graphically in Fig 5-4.



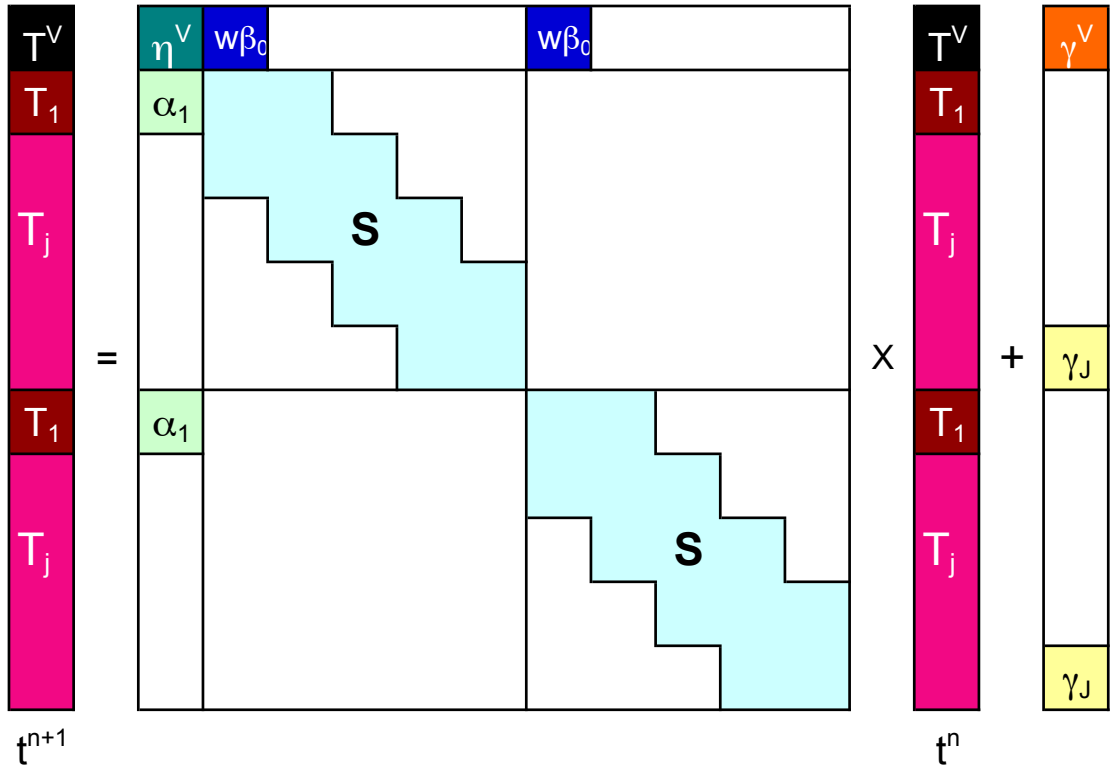

Fig 5-4. Graphical schematic of the implementation of an explicit time-step from time *n* to time *n+1* for multiple interacting locations in Area IV (Eq. 5.4-1). Compare with Fig 5-2. The temperature array is divided into the temperature of the volatile slab, $T^V$, (this is identical to the temperature in the uppermost layer, $T_0$), the next lower layer, $T_1$, and remaining layers for $j = 2..J$, $\mathbf{T}_j$. The elements of the substrate matrix *S* consist of the three arrays $\alpha_{2..J}$, $\eta_{1..J}$, and $\beta_{1..J-1}$. Darker elements with white lettering correspond to the dark gray elements in Fig. 5-2, and change with each time step. Lighter elements with black lettering correspond to the light gray elements in Fig. 5-2, and are independent of time. White elements are zero.

The new temperature of the volatile slab depends on the substrate (Eq. 5.4-2a); this is similar to Eq. 3.4-8a, but slightly simpler. The multi-location matrix equation for the temperatures of the substrate (Eq. 5.4-2b) is also similar to the non-interacting equivalent (Eq. 3.4-8b), differing only in that the topmost substrate temperature equals the temperature of the volatile slab.

$$T_{n+1}^V = \eta_n^V T_n^V + \left[w_l \beta_{l,0,n}, w_m \beta_{m,0,n}, \cdots\right] \cdot \mathbf{T}_{\{L\},1,n} + \gamma_n^V \qquad (5.4\text{-}2a)$$



$$\mathbf{T}_{\{L\},1..J,n+1} = \mathbf{S}_{\{L\}} \cdot \mathbf{T}_{\{L\},1..J,n} + \begin{Bmatrix} \alpha_{\{L\},1} T_n^V \\ 0 \\ \vdots \\ 0 \\ \gamma_{\{L\},J} \end{Bmatrix} \quad (5.4\text{-}2b)$$

Graphically, this is represented by Fig 5-5.

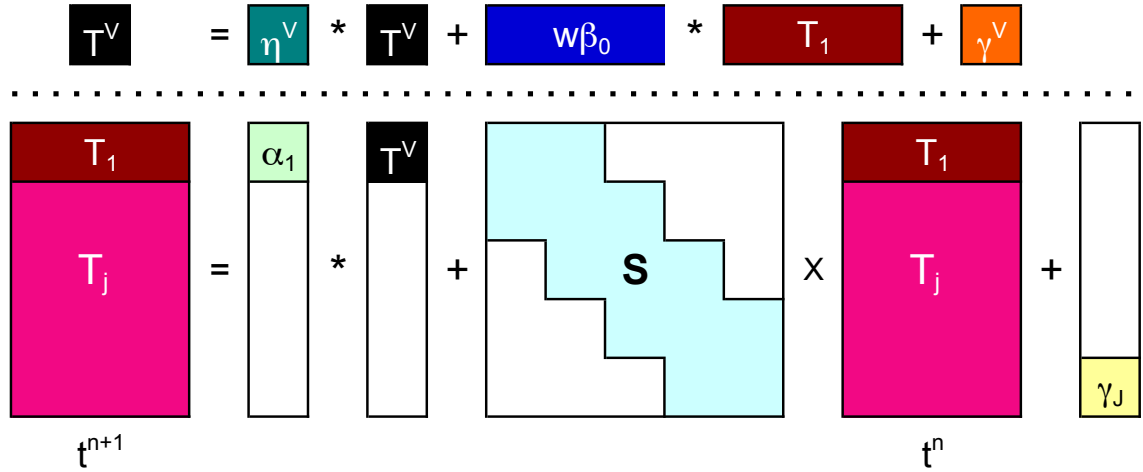

Fig 5-5. Graphical schematic of the implementation of an explicit time-step from time *n* to time *n+1* for multiple interacting locations (Eq. 5.4-2). Elements are labeled as in Fig 5-5. "*" indicates scalar multiplication (above the dotted line) or element-by-element multiplication of two arrays (above and below the dotted line). "X" indicates matrix multiplication.

### 5.4b. Implicit timesteps

For the implicit case, it is most straight-forward to write the Crank-Nicholson scheme in terms of intermediate temperatures for the volatile slab $\tilde{T}_n^V$ and substrate, $\tilde{T}_{l,1..J,n}$.

$$\begin{bmatrix} \tilde{T}_n^V \\ \tilde{\mathbf{T}}_{l,1..J,n} \\ \tilde{\mathbf{T}}_{m,1..J,n} \end{bmatrix} = \begin{bmatrix} \eta_n'^V & \mathbf{b}_{l,n}' & \mathbf{b}_{m,n}' \\ \mathbf{a}_l' & \mathbf{S}_l' & 0 \\ \mathbf{a}_m' & 0 & \mathbf{S}_m' \end{bmatrix} \times \begin{bmatrix} T_n^V \\ \mathbf{T}_{l,1..J,n} \\ \mathbf{T}_{m,1..J,n} \end{bmatrix} + \begin{bmatrix} \gamma_n^V \\ \mathbf{g}_l \\ \mathbf{g}_m \end{bmatrix} \quad (5.4\text{-}3a)$$



$$\begin{bmatrix} \eta_n''^V & \mathbf{b}_{l,n}'' & \mathbf{b}_{m,n}'' \\ \mathbf{a}_l'' & \mathbf{S}_l'' & 0 \\ \mathbf{a}_m'' & 0 & \mathbf{S}_m'' \end{bmatrix} \times \begin{bmatrix} T_{n+1}^V \\ \mathbf{T}_{l,1..J,n+1} \\ \mathbf{T}_{m,1..J,n+1} \end{bmatrix} = \begin{bmatrix} \tilde{T}_n^V \\ \tilde{\mathbf{T}}_{l,1..J,n} \\ \tilde{\mathbf{T}}_{m,1..J,n} \end{bmatrix} \qquad (5.4\text{-}3b)$$

For the other areas, the banded tridiagonal matrix was a computational convenience. For Area IV, it is the most direct way of solving Eq. 5.4-3b. The solution to Eq. (5.4-3b) can be written by defining two column vectors $\mathbf{y}_l$ and $\mathbf{z}_l$ of length $J$ (defined as in 3.4-16a, 16b), and two scalars $c_n$ and $d_n$:

$$c_n = \sum \mathbf{b}_{l,n}'' \cdot \mathbf{y}_l = \sum w_l \beta_{l,0,n}'' y_{l,0} \qquad (5.4\text{-}4a)$$

$$d_n = \sum \mathbf{b}_{l,n}'' \cdot \mathbf{z}_{l,n} = \sum w_l \beta_{l,0,n}'' z_{l,0,n} \qquad (5.4\text{-}4b)$$

with which the temperatures at the next time step for location $l$ are

$$T_{n+1}^V = \frac{\tilde{T}_n^V - d_n}{\eta_n''^V - c_n} \qquad (5.4\text{-}5a)$$

$$\mathbf{T}_{l,1..J,n+1} = \mathbf{z}_{l,n} - T_{n+1}^V \mathbf{y}_l \qquad (5.4\text{-}5b)$$

This solution can be confirmed by direct substitution into Eq. (5.4-3b). The solution is shown graphically in Fig 5-6. Note that only the time-independent substrate matrix needs to be inverted, and this can be done at the start of the computation, rather than for each time step. Furthermore, the array $y$ is also independent of time.



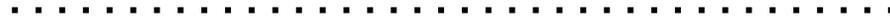

intermediate

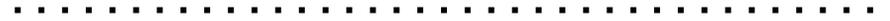

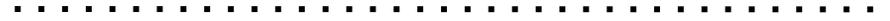



Fig 5-6. Graphical schematic of the implementation of an implicit time-step from time *n* to time *n+1* for multiple non-interacting locations (Eq. 3.4-8). Elements are labeled as in Fig 3-9. "*" indicates element-by-element multiplication of two arrays (above the dotted line). "X" indicates matrix multiplication (equivalent to the outer product of two arrays for the multiplication below the lowest dotted line).

## 5.5 Example: PNV9 from Young 2013

As a worked example, Fig 5-7 shows the results of the calculations used for case PNV9 (permanent northern volatile #9) from Young 2013. This example was illustrated in Fig 1 of Young (2013) and Fig 3 of Olkin et al. (2015). The format of the figure is a still from the movies that show the seasonal evolution, as shown in various talks (e.g., Young 2012a). The code is included in the supplemental materials as *vty16_fig5_7*. The code included here is taken from the code actually run for Young (2013), with only superficial changes, to allow myself or others to reproduce the results of Young (2013) and Olkin et al. (2015).

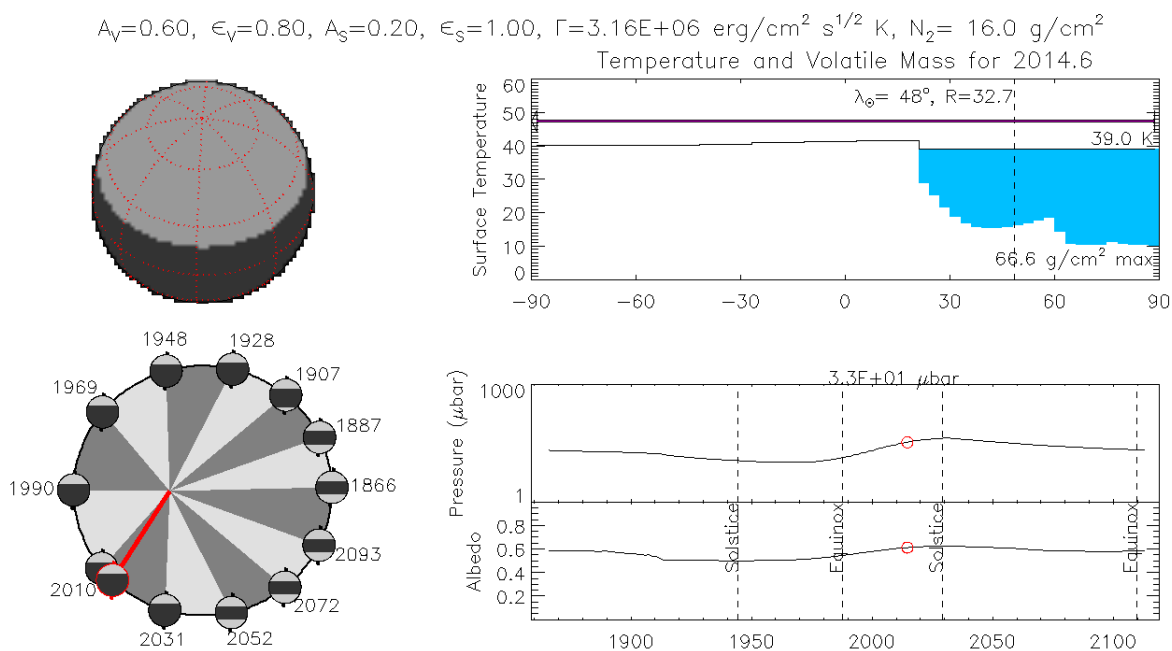



Fig 5-7.[1] Upper title: the Bond albedo and emissivity of the volatile ($A_V$ and $\varepsilon_V$), the Bond albedo and emissivity of the substrate ($A_S$ and $\varepsilon_S$), the thermal inertia of the substrate ($\Gamma$), and the globally averaged $N_2$ inventory ($N_2$). Top right: Pluto's temperature and volatile mass for the listed year (2014.6). The subsolar latitude and heliocentric radius are listed (48° and 32.7 AU). The purple line gives an indication of the direction and magnitude of the sublimation winds, running from the North to the South. Blue indicates the volatile mass, where volatiles are present; the thickness of the bars are proportional to the mass, and the maximum mass is indicated (66.6 g/cm$^2$). The thin solid line indicates the surface temperature, which is a uniform 39.0 K for volatile covered areas, and is just above 40 K for bare areas (south of ~20°). Top left: Pluto as seen from the sun. Volatiles and substrate are shaded by their respective albedos. Pluto is tilted by the subsolar latitude. When plotted as a movie, the size of Pluto varies with the inverse of the heliocentric distance. Bottom left: graphical depiction of the seasonal volatile evolution. The shape of the orbit is in scale with Pluto's eccentricity. The 12 light and dark gray "pie pieces" mark out equal durations in the orbit, with the sun at the vertex of the pie pieces. The circles represent Pluto as seen with a zero sub-observer latitude. The pole is a squat bar running behind the circles. The circles and the pole bar are oriented so that the pole is perpendicular to the Pluto-sun line at the two equinoxes and so the summer hemisphere is oriented toward the sun. The red line and the circle outlined in red represent Pluto's position and state at the listed year (2014.6). Within the circles, lighter gray shows the location of volatiles, and darker gray shows substrate. Lower right:

---

[1] **vty16_fig5_7,** which calls

res = **pluto_mssearch_func**(run, av, ev, as, es, ti, mvtot, n_off, res_all)

**vty16_plutostill_mssearch_func,** run, res, yr_still

**vty16_pluto_mssearch_resub_mat,** flag_frostslab, time_delta, n_loc, n_z, emis, temp_surf, eflux_sol, mass_slab, specheat_frost, z_delta, z_delta_bot, dens, specheat, thermcond, eflux_int, beta,alpha_bot, alpha_mid

**vty16_pluto_mssearch_resub_timestep,** flag_atm, freq, time_delta, gravacc, name_species, flag_stepscheme, n_loc, lat, n_z, z_delta,z_delta_bot, is_xport, angarea_delta, emis, temp_surf, eflux_sol, mass_slab, specheat_volatile, dens,specheat,thermcond, alpha_top, alpha_mid,alpha_bot,beta, denom, eflux_net, temp, temp_volatile, temp_next, temp_volatile_next, angarea_atm, mflux

**vty16_plutowrite_mssearch_func,** run, res_all



Surface pressure (log scale) and geometric albedo (linear) as a function of year, with the listed year marked by a small red circle. The pressure at the listed year is indicated (33 µbar), and the years of the equinoxes and solstices are marked.

In order to relate skin depth to depth with physical units, the substrate is assumed to have a density, $\rho$, of 0.93 g cm$^{-3}$ and the skin depth, $Z$, is assumed to be 15 m; from the specified thermal inertia, $\Gamma$, Eq. 3.2-5 and 3.2-6 define the specific heat, $c$, and the thermal conductivity, $k$. The specific heat of the volatile, $c_V$, (I remind the reader that the $V$ is for volatile, not volume) is assumed to be that of N$_2$ ($\beta$), or 1.3e7 erg g$^{-1}$ K$^{-1}$ (Spencer & Moore 1992).

The run is initialized with the entire surface of Pluto covered with N$_2$ at aphelion, and the initial surface and subsurface temperatures are calculated assuming that the entire surface was volatile-covered and interacting over the previous Pluto year. The solar forcing is calculated assuming orbital elements of eccentricity of 0.254, inclination of 23.439°, Longitude of Ascending Node of 43.960°, argument of perifocus of 183.994°, last periapsis at Julian date 2447899.597, mean motion of 0.00392581°/day, a semi-major axis of 39.79700 AU, and a pole with right ascension 132.993° and declination -6.163° (see code for full precision). The diurnally averaged absorbed solar flux was calculated at 240 time steps over Pluto's orbit, at each of 60 latitude bands, and expanded to $M = 2$ (constant and two sinusoidal terms). The initial temperature field is calculated from the sinusoidal expansion of the absorbed solar flux, assuming a flux from the interior of 6 erg cm$^{-2}$ s$^{-1}$. This follows the prescription of Section 5.2, except that the atmospheric "breathing" term is ignored (it is small on the seasonal timescales, Young 2013). The substrate uses a "medium" grid, with 19 layers of width 0.4 $Z$, where $Z$ is the skin depth. The top layer is half that, or 0.2 $Z$.

## 6. Conclusions

A variety of mathematical techniques for speeding up thermophysical models or volatile transport models have been presented. They include an improved initial condition, an implicit time-step step scheme, and a matrix formulation that allows for the calculation of several locations at once. These can be used separately or in combination.

This formulation described here has been previously applied to Pluto's diurnal cycle with volatile distributions and albedos that vary with both latitude and longitude (Young 2012a). The speed gains allowed me to perform a wide parameter-space search of Pluto's seasonal cycle in anticipation of New Horizons (Young 2013). This work has also been used to study



KBO seasons (Young and McKinnon 2013) and the first Pluto volatile transport models to include an $N_2$ reservoir (Young et al. 2015).

I hope you will find some of the techniques useful.

**Acknowledgement**

This work was supported, in part, by funding from NASA Planetary Atmospheres grant NNG06GF32G and the Spitzer project (JPL research support agreement 1368573). John Spencer and Candy Hansen were extremely generous with information about their thermophysical and volatile transport models. Andy Ingersoll, Melissa Brucker, Carly Howett, Angela Zalucha, Cathy Olkin, and Will Grundy all helped me birth this monster of a paper. Two thorough reviewers chose to be anonymous, reminding me of masked superheroes.

# Appendix A

## Table A1. Variable Names

| Variable | Name | Units (cgs) | Comments |
|---|---|---|---|
| $\alpha_{l,j}$ | Matrix element at location $l$, layer $j$ | unitless | Section 3.3, Fig 3-6, Table 4 & 5 |
| $\alpha_{\{L\},j}$ | Matrix element for locations in set $\{L\}$, layer $j$ | unitless | Section 3.4, Fig 3-10 |
| $\beta_{l,j}$ | Matrix element at location $l$, layer $j$, $j > 0$ | unitless | Section 3.3, Fig 3-6, Table 4 & 5 |
| $\beta_{l,0,n}$ | Matrix element at location $l$, layer 0, time $n$ | unitless | Section 3.3, Fig 3-6, Table 3 |
| $\beta'_{l,0,n}$ | Matrix element at location $l$, layer 0, time $n$ | unitless | Section 3.3, Fig 3-6, Table 3 |
| $\beta''_{l,0,n}$ | Matrix element at location $l$, layer 0, time $n$ | unitless | Section 3.3, Fig 3-6, Table 3 |
| $\Gamma$ | Thermal inertia | erg cm$^{-2}$ K$^{-1}$ s$^{-1/2}$ (tiu in MKS) | Eq. 3.2-5 |
| $\gamma_{\{L\},J}$ | Matrix element for locations in set $\{L\}$, layer $j$ | unitless | Section 3.4, Fig 3-10 |
| $\gamma_{l,0,n}$ | Matrix element at location $l$, layer 0, time $n$ | unitless | Section 3.3, Fig 3-6, Table 3 |
| $\Delta_j$ | Thickness of layer $j$ | cm | Fig 3-5, Section 3.3. |
| $\Delta^B_{l,j}$ | Distance to layer below | cm | Eq. 3.3-11 |
| $\Delta^B_{l,j}$ | Distance to layer above | cm | Eq. 3.3-25a,b |
| $\Delta t$ | Time step | s | Section 3.3. |
| $\delta_{l,j}$ | Unitless thickness of layer $j$ | unitless | Eq. 3.3-4 |
| $\delta^A_{l,j}$ | Unitless distance to layer above | unitless | Eq. 3.3-26 |
| $\delta^B_{l,j}$ | Unitless distance to layer below | unitless | Eq. 3.3-12 |
| $\varepsilon$ | Emissivity | unitless | Fig 2-1, Section 3.1 |
| $\varepsilon_{l,n}$ | Emissivity at location $l$ and time $n$, for discrete equations | unitless | Section 3.3 |
| $\langle \varepsilon_n \rangle$ | Emissivity, for discrete equations, averaged over Area IV | unitless | Eq. 5.3-8b |
| $\eta_{l,0,n}$ | Matrix element at location $l$, layer 0, time $n$ | unitless | Section 3.3, Fig 3-6, Table 3 |
| $\eta'_{l,0,n}$ | Matrix element at location $l$, layer 0, time $n$ | unitless | Section 3.3, Fig 3-6, Table 3 |
| $\eta''_{l,0,n}$ | Matrix element at location $l$, layer 0, time $n$ | unitless | Section 3.3, Fig 3-6, Table 3 |
| $\eta^V_n$ | Matrix element for interacting volatiles | unitless | Section 5-2, Fig 5-2, Table 8 & 9 |
| $\eta'^V_n$ | Matrix element for interacting volatiles | unitless | Section 5-2, Fig 5-3, Table 8 & 9 |
| $\eta''^V_n$ | Matrix element for interacting volatiles | unitless | Section 5-2, Fig 5-3, Table 8 & 9 |
| $\eta_{l,j}$ | Matrix element at location $l$, layer $j$, $j > 0$ | unitless | Section 3.3, Fig 3-6, Table 4 & 5 |
| $\eta'_{l,j}$ | Matrix element at location $l$, layer $j$, $j > 0$ | unitless | Section 3.3, Fig 3-6, Table 4 & 5 |
| $\eta''_{l,j}$ | Matrix element at location $l$, layer $j$, $j > 0$ | unitless | Section 3.3, Fig 3-6, Table 4 & 5 |



**Table A1. Variable Names, cont**

| Variable | Name | Units (cgs) | Comments |
|---|---|---|---|
| $\Theta_A$ | Thermal parameter, atmosphere | unitless | Eq. 4.2-8 |
| $\Theta_S$ | Thermal parameter, substrate | unitless | 3.2-11 |
| $\Theta_V$ | Thermal parameter, volatile slab | unitless | Eq. 4.2-7 |
| $\lambda$ | Latitude | radian | Section 3.1 |
| $\lambda_0$ | Sub-solar latitude | radian | Section 3.1 |
| $\mu_0$ | Cosine of solar incidence angle | unitless | Eq. 3.1-5 |
| $\zeta$ | Unitless depth | unitless | After Eq. 3.2-7. |
| $\rho$ | Density | g cm$^{-3}$ | Section 3.1 |
| $\rho_{l,j}$ | Density at location *l*, layer *j* | g cm$^{-3}$ | Section 3.3 |
| $\sigma$ | Stefan-Boltzmann constant | erg cm$^{-2}$ K$^{-4}$ s$^{-1}$ | Fig 2-1, Section 3.1 |
| $\tau$ | Unitless time step | unitless | Eq. 3.3-3 |
| $\phi$ | Longitude | radian | Section 3.1 |
| $\phi_0$ | Sub-solar longitude | radian | Section 3.1 |
| $\Phi_A$ | "Flux-per-temperature," atmosphere | erg cm$^{-2}$ s$^{-1}$ K$^{-1}$ | Eq. 4.2-4a |
| $\Phi^A_{l,n}$ | "Flux-per-temperature," atmosphere, for discrete equations | erg cm$^{-2}$ s$^{-1}$ K$^{-1}$ | Eq. 4.3-3 |
| $\langle \Phi^A_n \rangle$ | "Flux-per-temperature," atmosphere, for discrete equations, averaged over Area IV | erg cm$^{-2}$ s$^{-1}$ K$^{-1}$ | Eq. 4.3-3 |
| $\Phi_E$ | "Flux-per-temperature," emission | erg cm$^{-2}$ s$^{-1}$ K$^{-1}$ | Eq. 3.2-9b |
| $\Phi^E_{l,j}$ | "Flux-per-temperature," emission, for discrete equations | erg cm$^{-2}$ s$^{-1}$ K$^{-1}$ | Eq. 3.5-8 |
| $\langle \Phi^E_n \rangle$ | "Flux-per-temperature," emission, for discrete equations, averaged over Area IV | erg cm$^{-2}$ s$^{-1}$ K$^{-1}$ | Eq. 5.3-5b |
| $\Phi^H_{l,j}$ | "Flux-per-temperature," enthalpy of the substrate, for discrete equations | erg cm$^{-2}$ s$^{-1}$ K$^{-1}$ | Eq. 3.3-5 |
| $\langle \Phi^H_0 \rangle$ | "Flux-per-temperature," enthalpy of the top layer of the substrate, for discrete equations, averaged over Area IV | erg cm$^{-2}$ s$^{-1}$ K$^{-1}$ | Eq. 5.3-3 |
| $\Phi^{K,A}_{l,j}$ | "Flux-per-temperature," conduction from above, for discrete equations | erg cm$^{-2}$ s$^{-1}$ K$^{-1}$ | Eq. 3.3-24 |
| $\Phi^{K,B}_{l,j}$ | "Flux-per-temperature," conduction from below, for discrete equations | erg cm$^{-2}$ s$^{-1}$ K$^{-1}$ | Eq. 3.3-10 |
| $\Phi_S$ | "Flux-per-temperature," substrate | erg cm$^{-2}$ s$^{-1}$ K$^{-1}$ | Eq. 3.2-9a |
| $\Phi^T_{l,n}$ | "Flux-per-temperature," total, for discrete equations | erg cm$^{-2}$ s$^{-1}$ K$^{-1}$ | Eq. 3.3-18, 4.3-11. |



**Table A1. Variable Names, cont**

| Variable | Name | Units (cgs) | Comments |
|---|---|---|---|
| $\Phi_V$ | "Flux-per-temperature," volatile slab | erg cm$^{-2}$ s$^{-1}$ K$^{-1}$ | Eq. 4.2-4b |
| $\langle \Phi_0^V \rangle$ | "Flux-per-temperature," volatile slab, for discrete equations, averaged over Area IV | erg cm$^{-2}$ s$^{-1}$ K$^{-1}$ | Eq. 5.3-5b |
| $\omega$ | Frequency of solar forcing | s$^{-1}$ | Section 3.2 |
| $\Omega_{III}$ | Solid angle of Area III | ster | Section 5.1 |
| $\Omega_{IV}$ | Solid angle of Area IV | ster | Section 5.1 |
| $A$ | Albedo (approx. $A_h$ or $A_S$) | unitless | Discussion following Eq. 3.1-5 |
| $A_h$ | Hemispheric albedo | unitless | Eq 3.1-4. |
| $A_S$ | Spherical albedo (aka Bond albedo) | unitless | Discussion following Eq. 3.1-5 |
| $\mathbf{a}_l$ | $J$-element column vector with one non-zero element for location $l$ | unitless | Eq. 3.4-4 |
| $\mathbf{a}'_l$ | $J$-element column vector with one non-zero element for location $l$ | unitless | Eq. 3.4-11a |
| $\mathbf{a}''_l$ | $J$-element column vector with one non-zero element for location $l$ | unitless | Eq. 3.4-11b |
| $\mathbf{a}_{\{L\}}$ | $J$-element column vector with one non-zero element for location in set $\{L\}$ | unitless | Section 3-4 |
| $\mathbf{b}_{l,n}$ | $J$-element row vector with one non-zero element. | unitless | Eq. 3.4-3 |
| $\mathbf{b}_{l,n}$ | $J$-element row vector with one non-zero element. | unitless | Eq. 3.4-3 |
| $\mathbf{b}'_{l,n}$ | $J$-element row vector with one non-zero element. | unitless | Eq. 3.4-10a |
| $\mathbf{b}''_{l,n}$ | $J$-element row vector with one non-zero element. | unitless | Eq. 3.4-10b |
| $c_V$ | Specific heat of volatile | erg K$^{-1}$ g$^{-1}$ | Fig 2-1, Section 3.1. *V* for *Volatile*. |
| $c_l^V$ | Specific heat of volatile, , for discrete equations | erg K$^{-1}$ g$^{-1}$ | Eq. 4.3-2. *V* for *Volatile*. |
| $c$ | Specific heat of the substrate | erg K$^{-1}$ g$^{-1}$ | Section 3.1. |
| $c_{l,j}$ | Specific heat at location $l$, layer $j$ | erg K$^{-1}$ g$^{-1}$ | Section 3.3. |
| $c_{l,n}$ | Scalar for solving banded tri-diagonal matrix | unitless | Eq. 3.4-12c |
| $\mathbf{c}_n$ | Row vector for solving banded tri-diagonal matrix | unitless | Eq. 3.4-16c |
| $d_{l,n}$ | Scalar for solving banded tri-diagonal matrix | unitless | Eq. 3.4-12d |
| $\mathbf{d}_n$ | Row vector for solving banded tri-diagonal matrix | unitless | Eq. 3.4-16d |
| $E$ | Escape rate | g cm$^{-2}$ s$^{-1}$ | Section 4.1 |
| $\hat{E}_m$ | Sinusoidal coefficient of $E$ | g cm$^{-2}$ s$^{-1}$ | Eq. 4.2-2, Section 4.2. Complex |
| $F^E$ | Emitted thermal flux | erg cm$^{-2}$ s$^{-1}$ | Eq. 3.2-13 |
| $\hat{F}_m^E$ | Complex coefficients for $F^E$ | erg cm$^{-2}$ s$^{-1}$ | Eq 3.2-13 |
| $F$ | Heat flow at lower boundary | erg cm$^{-2}$ s$^{-1}$ | Fig 2-1, Section 3.1 |
| $F_{l,n}^A$ | Correction term for flux due to latent heat | erg cm$^{-2}$ s$^{-1}$ | Eq. 4.3-8 |
| $\langle F_n^A \rangle$ | Correction term for flux due to latent heat, averaged over Area IV | erg cm$^{-2}$ s$^{-1}$ | Eq. 5.3-6c |



**Table A1. Variable Names, cont**

| Variable | Name | Units (cgs) | Comments |
|---|---|---|---|
| $F_l$ | Heat flow at lower boundary for discrete equations | erg cm$^{-2}$ s$^{-1}$ | Eq. 3.3-20 |
| $f_V$ | Fraction of the interacting area that is volatile-covered | unitless | Eq. 5.1-4 |
| $g$ | Effective gravity | cm s$^{-2}$ | Section 4.1 |
| $\mathbf{g}_l$ | *J*-element column vector with one non-zero element for location *l* | unitless | Eq. 3.4-5 |
| $\mathbf{g}_{\{L\}}$ | J-element column vector with one non-zero element for locations in set *{L}* | unitless | Section 3.4 |
| $H_V$ | Enthalpy of the volatile | erg g$^{-1}$ | Fig 2-1, Section 3.1 |
| $h$ | Hour angle | radian | Section 3.1 |
| $h_0$ | Hour angle at time $t = 0$ | radian | Section 3.2 |
| $h_{max}$ | Maximum hour angle of sunlight | radian | Eq. 3.2-3 |
| $\mathbf{h}''_{0,n}$ | Row vector for solving banded tri-diagonal matrix | unitless | Eq. 3.4-16d |
| $H$ | Pressure scale height | cm | Section 4.1 only |
| $j$ | Index for layers | integer | Fig 3-5, Section 3.3. $j = 0 .. J$. |
| $J$ | Index of lowest layer | integer | Fig 3-5, Section 3.3. |
| $k$ | Thermal conductivity | erg K$^{-1}$ cm$^{-1}$ s$^{-1}$ | Fig 2-1, Section 3.1 |
| $k_B$ | Bolzmann's constant | erg K$^{-1}$ | Section 4.1 |
| $k_{l,j}$ | Thermal conductivity at location *l*, layer *j* | erg K$^{-1}$ cm$^{-1}$ s$^{-1}$ | Section 3.3 |
| $L_C$ | Latent heat of crystalline phase change | erg g$^{-1}$ | Section 4.1 |
| $L_S$ | Latent heat of sublimation | erg g$^{-1}$ | Fig 2-1, Section 3.1 |
| $l$ | Index for location | integer | Section 3.3 |
| $L$ | Number of locations | integer | Section 3.3 |
| $\{L\}$ | Set of locations with shared substrate properties & internal heat flux | Set | Section 3.4 |
| $\{L_{IV}\}$ | Set of locations in Area IV | Set | Section Section 5.3 |
| $M$ | Number of orders of sinusoidal expansion | integer | Section 3.2 |
| $m$ | Order of sinusoidal expansion | integer | Section 3.2 |
| $m$ | Another index for location | integer | Section 3.4, 5.3 |
| $m_A$ | Mass per area of the atmosphere | g cm$^{-2}$ | Section 4.1 |
| $m_{molec}$ | Mass per molecule | g molecule$^{-1}$ | Section 4.1 |
| $m_V$ | Mass per area of the volatile slab | g cm$^{-2}$ | Fig 2-1, Section 3.1 |
| $m^V_{l,n}$ | Mass per area of the volatile slab for discrete equations | g cm$^{-2}$ | Eq. (4.3-2) |
| $n$ | Index for time | integer | Section 3.3 |
| $P$ | Period of solar forcing | s | Section 3.2 |
| $p_S$ | Vapor pressure | μbar | Section 4.1 |



**Table A1. Variable Names, cont**

| Variable | Name | Units (cgs) | Comments |
|---|---|---|---|
| $r$ | Heliocentric distance | AU | Eq 3.1-4. |
| $R$ | Surface radius | cm | Section 4.1 only |
| $S$ | Absorbed insolation | erg cm$^{-2}$ s$^{-1}$ | Fig 2-1, Section 3.1; Eq 3.1-4. |
| $S_0$ | Constant term for $S$ in analytic expansion | erg cm$^{-2}$ s$^{-1}$ | Eq 3.2-1 |
| $\hat{S}_m$ | Complex coefficients for $S$ in analytic expansion | erg cm$^{-2}$ s$^{-1}$ | Eq 3.2-1 |
| $S_{1\,AU}$ | Solar flux at 1 AU | erg cm$^{-2}$ s$^{-1}$ | Section 3.1 |
| $\mathbf{S}_l$ | $J \times J$ tridiagonal matrix for location $l$ | unitless | Section 3.4 |
| $\mathbf{S}'_l$ | $J \times J$ tridiagonal matrix for location $l$ | unitless | Section 3.4 |
| $\mathbf{S}''_l$ | $J \times J$ tridiagonal matrix for location $l$ | unitless | Section 3.4 |
| $\mathbf{S}_{\{L\}}$ | $J \times J$ tridiagonal matrix for locations in set $\{L\}$ | unitless | Section 3.4 |
| $T$ | Temperature | K | Fig 2-1, Section 3.1 |
| $T_0$ | Constant term for $T$ in analytic expansion | K | Eq. 3.2-7 |
| $\hat{T}_m$ | Complex coefficients for $T$ in analytic expansion | K | Eq. 3.2-7 |
| $T_{l,0,n}$ | Discrete surface temperature at location $l$, time $n$ | K | Section 3.4 |
| $\mathbf{T}_{l,1..J,n}$ | Row vector of discrete substrate temperatures | K | Eq. 3.4-1 |
| $T_{l,j,n}$ | Temperature of location $l$, layer $j$, time $n$ | K | Fig 3-5, Section 3.3 |
| $\mathbf{T}_{\{L\},0,n}$ | Row array of length $L$ with temperatures of locations in set $\{L\}$, layer 0, time $n$ | K | Section 3.4 |
| $\mathbf{T}_{\{L\},1..J,n}$ | $J \times L$ matrix with temperatures of locations in set $\{L\}$, layers 1..$J$, time $n$ | K | Section 3.4 |
| $\tilde{T}_{l,0,n}$ | Intermediate temperature of location $l$, layer 0, time $n$ for Crank-Nicholson timesteps | K | Eq. 3.4-9a |
| $\tilde{\mathbf{T}}_{l,1..J,n}$ | $J$-element column vector of intermediate temperature of location $l$, layers 1..$J$, time $n$ for Crank-Nicholson timesteps | K | Eq. 3.4-9a |
| $T_V$ | Temperature of the volatile | K | Fig 2-1, Section 3.1. Constant over Area IV. |
| $T_n^V$ | Temperature of the volatile for discrete equations | K | Fig 2-1, Section 3.1. Constant over Area IV. |
| $t$ | Time | s | Fig 2-1, Section 3.1 |
| $w_l$ | Areal weight of location $l$ with respect to Area III | unitless | Eq. 5.3-4 |
| $w'_l$ | Areal weight of location $l$ with respect to Area III and IV | unitless | Eq. 5.3-11b |
| $\mathbf{y}_l$ | $J$ column vector, for solving banded tri-diagonal matrix | unitless | Eq. 3.4-12a |
| $\mathbf{y}_{\{L\}}$ | $J$ column vector, for solving banded tri-diagonal matrix | unitless | Eq. 3.4-16a |
| $z$ | Depth | cm | Fig 2-1, Section 3.1. Zero at top of substrate, negative at depth. |



**Table A1. Variable Names, cont**

| Variable | Name | Units (cgs) | Comments |
|---|---|---|---|
| $z_j$ | Depth of layer $j$ | cm | Fig 3-5, Section 3.3. Middle of layer $j$ for $j > 0$; $z_0 = 0$ (top of layer). |
| $\mathbf{z}_{l,n}$ | $J$ column vector, for solving banded tri-diagonal matrix | unitless | Eq. 3.4-12b |
| $\mathbf{Z}_{\{L\},n}$ | $J \times L$ matrix, for solving banded tri-diagonal matrix | unitless | Eq. 3.4-16b |
| $Z$ | Skin depth | cm | Eq. 3.2-6 |



## Appendix B

Table B2 and B3 are alphabetical lists of the IDL procedures and functions that directly implement the model described in this paper and are included in the supplementary material. Other routines are taken from elsewhere in the layoung IDL library, and from the astron library. As of the date of publication, these are accessible at http://www.boulder.swri.edu/~layoung, and http://idlastro.gsfc.nasa.gov. Figures were made with IDL Version 8.4.1, Mac OS X (darwin x86_64 m64).

**Table B2. IDL routines in layoung/vol_xfer/vty16/**

| Calling sequence | Notes |
| --- | --- |
| **vty16_fig3_1** | Fig 3.1 |
| **vty16_fig3_2** | Fig 3.2 |
| **vty16_fig3_3** | Fig 3.3 |
| **vty16_fig3_7a** | Fig 3.7a |
| **vty16_fig3_7b** | Fig 3.7b |
| **vty16_fig3_7c** | Fig 3.7c |
| **vty16_fig3_14** | Fig 3.14 |
| **vty16_fig4_2** | Fig 4.2 |
| **vty16_fig4_2_plot,** dist_sol_au, tod, theta_sva, temp_term_0, temp, p, $ <br>     name_therminertia, n_per_period | Fig 4.2 |
| **vty16_fig5_7** | Fig 5-7 |
| res = **vty16_fig5_7_func**(run, av, ev, as, es, ti, mvtot, n_off, res_all) | Fig 5-7 |
| **vty16_fig5_7_mat**, flag_frostslab, time_delta, n_loc, n_z, <br>   emis, temp_surf, eflux_sol, mass_slab, specheat_frost, <br>   z_delta, z_delta_bot, dens, specheat, thermcond, eflux_int, <br>   beta,alpha_bot, alpha_mid | Fig 5-7 |
| **vty16_fig5_7_still**, run, res, yr_still | Fig 5-7 |
| **vty16_fig5_7_timestep**, flag_atm, freq, time_delta, gravacc, name_species, <br> flag_stepscheme, n_loc, lat, n_z, z_delta,z_delta_bot, is_xport, angarea_delta, emis, <br> temp_surf, eflux_sol, mass_slab, specheat_volatile, dens,specheat,thermcond, <br> alpha_top, alpha_mid,alpha_bot,beta, denom, eflux_net, temp, temp_volatile, <br> temp_next, temp_volatile_next, angarea_atm, mflux | Fig 5-7 |
| **vty16_fig5_7_write**, run, res_all | Fig 5-7 |



**Table B3. IDL routines in layoung/vol_xfer/vt3d/**

| Calling sequence | Notes |
|---|---|
| **vt3d_1loc_diurnal_local**, constants, input, grid, program, output | Fig 4.2 |
| alpha_i = **vt3d_alpha_interior**(tau, del, del_a) | Table 4 & 5 |
| beta_i = **vt3d_beta_interior**(tau, del, del_b) | Table 4 & 5 |
| spp_tridc = **vt3d_cn_tridc**(alpha_i, beta_i, spp_indx) | Table 4 & 5, Fig 3-6, 3-8 |
| phi_a = **vt3d_dfluxdtemp_atm**(freq, temp_v, frac_varea, gravacc, name_species) | Eq. 4.1-5, 4.2-4b, 5.2-2d |
| phi_e = **vt3d_dfluxdtemp_emit**(emis, temp) | Eq. 3.2-9b |
| phi_v = **vt3d_dfluxdtemp_slab**(freq, mass_0, specheat) | Eq. 4.2-4b |
| phi_s = **vt3d_dfluxdtemp_substrate**(freq, therminertia) | Eq. 3.2-9a |
| flux_terms = **vt3d_eflux_terms_bare**(sol_terms, flux_int, emis, freq, therminertia, term_terms=temp_terms) | Eq. 3.2-17 |
| flux_terms = **vt3d_eflux_terms_local**(sol_terms, flux_int, emis, freq, therminertia, is_volatile, mass_volatile, specheat_volatile, gravacc, name_species, term_terms=temp_terms | Eq. 4.2-10 |
| avg = **vt3d_locavg**(val, weight) | Eq. 5.3-3 |
| z_skin = **vt3d_skindepth**(dens, specheat, thermcond, freq) | Eq. 3.2-6 |
| sol_terms = **vt3d_sol_terms_diurnal**(dist_sol_au, albedo, lat, h_phase0, lat_sol, n_terms) | Eq 3.2-4a, b, c |
| mu0 = **vt3d_solar_mu**(lat, lon, lat0, lon0) | Eq. 3.1-5 (Eq 3.2-4a for longitudinal averaged $\mu_0$) |
| sol = **vt3d_solwave**(sol_terms, phase) | Eq. 3.2-1 |
| **vt3d_step_cn_1loc**, alpha_i, beta_0, beta_i, gamma_0, gamma_J, temp_0, temp_i | Eq. 3.4-9a, 3.4-9b |
| **vt3d_step_cn_nloc**, alpha_i, beta_0, beta_i, gamma_0, gamma_J, temp_0, temp_i | Eq. 3.4-18 |
| **vt3d_step_cn_trisol_1loc**, alpha_i, beta_0, beta_i, gamma_0, gamma_J, spp_tridc, spp_indx, temp_0, temp_i | Eq. 3.4-9a, 3.4-9b |
| **vt3d_step_cn_trisol_nloc**, alpha_i, beta_0, beta_i, gamma_0, gamma_J, spp_tridc, spp_indx, temp_0, temp_i | Eq. 3.4-18 |
| **vt3d_step_expl_1loc**, alpha_i, beta_0, beta_i, gamma_0, gamma_J, temp_0, temp_i | Eq. 3.4-2 |
| **vty16_step_expl_nloc**, alpha_i, beta_0, beta_i, gamma_0, gamma_J, temp_0, temp_i | Eq. 3.4-8a, 3.4-8b |
| **vt3d_step_impl_1loc**, alpha_i, beta_0, beta_i, gamma_0, gamma_J, temp_0, temp_i | Eq. 3.4-9b |
| **vt3d_step_impl_nloc**, alpha_i, beta_0, beta_i, gamma_0, gamma_J, temp_0, temp_i | Eq. 3.4-18 |
| **vt3d_step_impl_trisol_1loc,** alpha_1, beta_0, gamma_0, gamma_J, spp_tridc, spp_indx, temp_0, temp_i | Eq. 3.4-9b |
| **vt3d_step_impl_trisol_nloc**, alpha_1, beta_0, gamma_0, gamma_J, spp_tridc, spp_indx, temp_0, temp_i | Eq. 3.4-18 |
| temp_0 = **vt3d_temp_term0_bare**(sol_0, flux_int, emis) | Eq. 3.2-8 |
| temp_0 = **vt3d_temp_term0_local**(sol_0, flux_int, emis, latheat, mflux_esc) | Eq. 4.2-3 |
| temp_terms = **vt3d_temp_terms_bare(**sol_terms,flux_int,emis,freq,therminertia) | Eq. 3.2-10 |



| | |
|---|---|
| temp_terms= **vt3d_temp_terms_bare_iter**(sol_terms,flux_int,emis,freq, therminertia, thermcond) | Eq. 3.2-10 |
| temp_terms = **vt3d_temp_terms_local**(sol_terms,flux_int,emis,freq, therminertia, is_volatile, mass_volatile, specheat_volatile, gravacc, name_species) | Eq. 4.2-5 |
| temp_terms = **vt3d_temp_terms_local_iter**(sol_terms,flux_int,emis,freq, therminertia, thermcond, is_volatile, mass_volatile, specheat_volatile, gravacc, name_species) | Eq. 4.2-5 |
| temp = **vt3d_thermwave**(temp_terms, phase, dtemp_dzeta, z_skin, zeta) | Eq. 3.2-7 |
| temp = **vt3d_thermwave_1loc_p0_nz**(temp_terms, dtemp_dzeta, z_skin, zeta) | Eq. 3.2-7 |
| temp = **vt3d_thermwave_nloc_p0_nz**(temp_terms, dtemp_dzeta, z_skin, zeta) | Eq. 3.2-7 |
| therminertia = **vt3d_thermalinertia**(dens, specheat, thermcond) | Eq. 3.2-5 |
| **vt3d_zdelta**, z, z_delta, z_delta_top, z_delta_bot | Eq. 3.3-11, 3.3-25a,b |
| z = **vt3d_zgrid**(skindepth,z_delta,n_z) | Section 3.3 |